\documentclass{pasj01}
\Received{$\langle$2019/11/08$\rangle$}
\Accepted{$\langle$2020/01/06$\rangle$}
\Published{$\langle$publication date$\rangle$}
\draft
\begin{document}

%\title{MAGNETIC FIELD STRUCTURE ON THE GALACTIC PLANE INVESTIGATED BY DIFFERENTIAL ANALYSIS OF INTERSTELLAR POLARIZATION}
\title{MAGNETIC FIELD STRUCTURE OF THE GALACTIC PLANE FROM DIFFERENTIAL ANALYSIS OF INTERSTELLAR POLARIZATION}
\author{Tetsuya \textsc{Zenko}\altaffilmark{1,}$^{*}$,  
Tetsuya \textsc{Nagata}\altaffilmark{1}, 
Mikio \textsc{Kurita}\altaffilmark{1},  
%Kino 
Masaru \textsc{Kino}\altaffilmark{2}, 
Shogo \textsc{Nishiyama}\altaffilmark{3}, 
Noriyuki \textsc{Matsunaga}\altaffilmark{4}, 
Yasushi \textsc{Nakajima}\altaffilmark{5}}%
\altaffiltext{1}{Department of Astronomy, Kyoto University, Kitashirakawa-Oiwake-cho, Sakyo-ku, Kyoto, Kyoto 606-8502, Japan}
\altaffiltext{2}{Astronomical Observatory, Kyoto University, Kitashirakawa-Oiwake-cho, Sakyo-ku, Kyoto, Kyoto 606-8502, Japan}
\altaffiltext{3}{Miyagi University of Education, Aoba-ku, Sendai, Miyagi 980-0845, Japan}   %長田1223
\altaffiltext{4}{Department of Astronomy, The University of Tokyo, 7-3-1 Hongo, Bunkyo-ku, Tokyo 113-0033, Japan}
\altaffiltext{5}{Center of Information and Communication Technology, Hitotsubashi University, 2-1 Naka, Kunitachi, Tokyo 186-8601, Japan}
\email{zenko@kusastro.kyoto-u.ac.jp}
%20191028 13:00 edited from 1027 12:50 edited from 1026 32:00 from 1026 20:30zen from 18:00nag from 20191025 ３１:４５ from 1025 8:50 from 1024 17:43}◎

\KeyWords{dust, extinction---Galaxy:center---infrared:ISM---ISM:magnetic fields---stars:distances---techniques:polarimetric} %update:2019/12/21(善光)}◎
\maketitle

\begin{abstract}
A new method for measuring the global magnetic field structure of the Galactic plane is presented. %これはすばらしい。
We have determined the near-infrared polarization of field stars around 52 Cepheids found in 
recent surveys toward the Galactic \textcolor{black}{plane}.%(善光_191127)center.
%, using the SIRPOL polarimetric camera.%<<削った>> ($\sim$ 8~arcmin field of view) 
%<<削った>>on the 1.4~m telescope IRSF. 
%We have determined the polarization of 52 Cepheids and field stars measured in $J$, $H$, and $K_{\mathrm S}$ bands using the near-infrared polarimetric camera SIRPOL on the 1.4, telescope IRSF. JHはあんまり使ってないからここには陽に書かない 
The Cepheids are located in the Galactic longitudes $-10^{\circ}\leq \, l\, \leq +10\fdg5$ and latitudes $-0\fdg22\leq \, b\, \leq +0\fdg45$, and 
their %distance is mainly in the range of 10 to 15~kpc from the Sun. %具体的に書こう ◆複数の方が自然やね
distances are mainly in the range of 10 to 15~kpc from the Sun. 
%<<削った>>The Galactic magnetic field orientation between the Sun and the far side of the Galactic center, 
%<<削った>>on the whole, 
%<<削った>>is almost parallel to the Galactic plane. %◎ここに持ってきた
Simple classification of the sightlines is made with the
polarization behavior vs. $H-K_{\mathrm S}$ color of field stars, and %◎
typical examples of three types of them are presented. 
Then, division of the field stars in each line of sight into
a) foreground, b) bulge, and c) background is made
with the $Gaia$ DR2 catalog,
the peak of the $H-K_{\mathrm S}$ color histogram, and
$H-K_{\mathrm S}$ colors consistent to be as distant as the Cepheid
in the center, respectively.
The differential analysis between them enables us to
examine the magnetic field structure more definitely than
just relying on the $H-K_{\mathrm S}$ color difference.
%Then, with the help of the $Gaia$ DR2 catalog, nearby field stars are identified;   the peak of $H-K_{\mathrm S}$ histogram is used to find bulge stars, which is the main population of the field stars, and the $H-K_{\mathrm S}$ color is also used to identify stars consistent to be as distant as the Cepheids. %, >>(善光_191027)もしかして繋げるつもりでした？◎いやいや。ありがとう。でも、上は今回つなげた。Such division of field stars into a) foreground, b) bulge, and c) background, and the differential analysis between them, enable us to examine the magnetic field structure more definitely than just relying on the $H-K_{\mathrm S}$ color.  
%We divided each field into three regions a) foreground, b) bulge, and c) background using the Cepheid and Gaia DR2 catalog, and calculated the differential polarization among the regions.
In one line of sight, 
%<<削った>>that 
the magnetic field is nearly parallel to the Galactic plane and  
well aligned all the way from the Sun to the Cepheid position in the other side of the Galactic center. 
Contrary to our preconceived ideas, however, sightlines having such well aligned magnetic fields in the Galactic plane is rather small in number.   
%Three example lines of sight are selected: toward a field with significantly increasing in the polarization degree, toward a field indicating that the differential position angle is more than $45^{\circ}$ between the bulge and the background, and toward a field with slightly increasing in the polarization degree.
%The field of views which have the magnetic field well ordered to the Galactic plane along the line of sight are in the minority, and 32 fields do not show any significant pattern. >>(善光_191026)
At least {36} Cepheid fields indicate %<<削った>>that
random magnetic field components are %dominant.   ◆ドミナントは強すぎる気がする
significant.  
Two Cepheid fields indicate %<<削った>>that, a fewをtwoに
the magnetic field orientation changes more than $45^{\circ}$ in the line of sight. %●間違ってたら指摘を。
The polarization increase per color change $\Delta P$$/$$\Delta (H-K_{\mathrm{S}})$ 
varies from region to region, 
reflecting the change in the ratio of the magnetic field strength and the turbulence strength.  
%
%The field of views which have the magnetic field along the line of sight aligns well in the same direction are in the minority, and 32 fields do not show any significant pattern.
%Therefore, the turbulence energy is higher than the magnetic field energy in our observation area.>>(善光_191026)Dflagだけでは根拠にならない。ちゃんと$\Delta P$ $/$ $\Delta (H-K_{\mathrm{S}})$を計算して議論できること
%Our results also indicate that the Galactic magnetic field orientation between the observer and the far side of Galactic center is approximately parallel to the Galactic plane.>>(善光_191026)議論のところを消すので同時にここも変える
%
%
%Our method probes the magnetic field orientation on a scale of several kiloparsecs in three fields, and the Galactic magnetic field orientation between the observer and the far side of Galactic center is almost parallel to the Galactic plane in most cases.
%We also calculate the polarization increase per color change $\Delta P$ $/$ $\Delta (H-K_{\mathrm{S}})$ among the fields, and we have shown that the ratio of the magnetic field and turbulence varies from region to region.
%>>(善光_191026)3つのサンプル領域に対して行った差分解析からはっきりわかることだけを記述する。
%>>1. に磁場の向きが3つの領域でどうだったのか
%>>2. $\Delta P$ $/$ $\Delta (H-K_{\mathrm{S}})$から、視野ごとに
\end{abstract}

\section{Introduction} 
%Polarized wave traces magnetic field structure. これは書かない
Interstellar magnetic fields of the Milky Way have been investigated for many years, %接続詞必要
and 
they are considered to play %an rolesが複数なら
important roles in astrophysics and astroparticle physics.
The large-scale structure of the Galactic magnetic field has been derived from studies of background starlight polarimetry (\cite{Mathewson}; Heiles \yearcite{aHeiles}, \yearcite{bHeiles}), dust emission polarimetry (Planck Collaboration et al. \yearcite{aPlanck}, \yearcite{bPlanck}), Faraday rotation (Han et al. \yearcite{aHan}, \yearcite{bHan}; \cite{Brown}; Mao et al. \yearcite{aMao}, \yearcite{bMao}; \cite{Pshirkov}; \cite{vanEck}), and synchrotron emission (Jansson \& Farrar \yearcite{aJansson}, \yearcite{bJansson}). 
These methods indicate that the large-scale magnetic field 
%inside the Galactic arms 
is parallel to the Galactic plane. %乱流と磁場のことも入れたい

Interstellar polarization, Faraday rotation and synchrotron emission are integral quantities between the observer and the target along the line of sight, and the observed polarization does not show the magnetic structure %which is 
at 
an arbitrary distance from the observer.
%Interstellar polarization, Faraday rotation and synchrotron emission are integral quantities along the line of sight, meaning that the observed properties depend on the conditions between the objects and the observer. 
Therefore, we %have 
would like 
to calculate the polarization as 
a
function of distance, %understanding じゃないかな
%understand ▼なんでこれが残ってるのかなあ
and understand 
the magnetic structure at a specific distance from the observer.
\textcolor{black}{This allows us to examine 3D tomography of large-scale magnetic field structures.}%(善光_191127)長田1220
%Therefore, we have to calculate the polarized wave and polarization as function of distance, understanding the magnetic structure at a specific distance from the observer.

Faraday rotation studies have combined observation of Galactic pulsars estimating distance with observation of polarized extragalactic sources, and detailed models of the Galactic magnetic field have been generated (e.g., \cite{Brown}; \cite{Sun}; \cite{vanEck}).
%Previous Faraday rotation studies combine observation of Galactic pulsars with distance estimate with observation of polarized extragalactic sources, and detailed models of the Galactic magnetic field have been generated (e.g., \cite{Brown}; \cite{Sun}; \cite{vanEck}). 
However, Faraday rotation only senses the line of sight component of magnetic field, and %this 要らぬでしょ
does not probe the sky-projected magnetic field. 
%However, Faraday rotation has only sense the line of sight component of magnetic field, and this does not reveal the sky-projected magnetic field. 
Also, pulsar distances have typical uncertainties of 0.5 kpc or greater, because few pulsars have stellar %▼ へんなアンド parallax and, H ${\mathrm I}\hspace{-.1em}$ kinematic distances typically only give upper or lower distance limits (\cite{Frail}; \cite{Verbiest}). %すこし調整
parallax, and H ${\mathrm I}\hspace{-.1em}$ kinematic distances typically only give upper or  lower distance limits (\cite{Frail}; \cite{Verbiest}). 
In addition, %▼この文って不自然だよねえ、私たちの観測エリアが急に出てくるのはthere are only 10 pulsars whose distance is more than 8 kpc toward our observed area ($|l|$ $\leq$ $10^{\circ}$, $|b|$ $\leq$ $0.2^{\circ}$) in the ATNF Catalogue (\cite{Manchester}). 
a small number of pulsars are known in the Galactic center region and beyond.  
For instance, there are only 10 pulsars whose distance is more than 8 kpc toward our observed area ($|l|$ $\leq$ $10^{\circ}$, $|b|$ $\leq$ $0\fdg2$) in the ATNF \textcolor{black}{Pulsar} Catalogue (\cite{Manchester}).%(善光_191127)
%Catalogue (\cite{Manchester}). 
This area %is が良いと思う 
suffers from %the stronger 
strong
confusion of diffuse emission from the Galactic disk at low Galactic latitudes, and Faraday rotation data are more difficult to %▼dataはメジャーするものかな measure in the far side of the Galactic center (Han et al. \yearcite{cHan}, \yearcite{dHan}).
obtain in the far side of the Galactic center (Han et al. \yearcite{cHan}, \yearcite{dHan}).
%Also, pulsar distances have typical uncertainties of 0.5kpc or greater, because few pulsars have stellar parallax and, H $I\hspace{-.1em}$ kinematic distances typically only give upper or lower distance limits(\cite{Frail}; \cite{Verbiest}). 
%In addition, Faraday rotation data are more difficult to measure in the far side of the Galactic center, because of the stronger confusion of diffuse emission from the Galactic disk at low Galactic latitudes (Han et al. \yearcite{cHan} \yearcite{dHan}).

%●●●ここまでの段落はきれいな英語だったのに。この段落、3章とか4章とかとは違って全く、ひどくはないけど、もうちょっと考えてみたら・・・。私はその後で直したい。
Starlight polarization %▼このあたりメジャーだらけmeasures the sky-projected magnetic field, but arises as an integral quantity of the line of sight between the %sun and the background stars.◆とういつしてみた
is sensitive to the sky-projected magnetic field, but arises as an integral quantity of the line of sight between the %sun and the background stars.◆とういつしてみた
Sun and the background stars.
To %▼measure change in the sky-projected magnetic field with distance, we need reliable stellar distance %makers.◆ですか？
examine changes in the sky-projected magnetic field with distance, we need reliable stellar distance %makers.◆ですか？
markers. 
%Starlight polarization senses the plane of sky magnetic field, and previous starlight polarization studies examine magnetic fields with distance.
%Starlight polarization has sense the plane of sky magnetic field, and previous starlight polarization studies disentangle magnetic fields with distance. 
\citet{bNishiyama} %have 過去で良いんじゃない
attempted to measure magnetic fields in the Galactic center with %▼この略号は使ってない NIR polarimetry by subtracting the polarization of ``bluer'' field %star at the close side in the Galactic bulge from the polarization of ``redder'' stars at the far side in the bulge.◆
near-infrared polarimetry by subtracting the polarization of ``bluer'' field 
stars at the %▼これはあんまり良くないと思う。Nishiyamaの中でもcloseだけじゃなくnearを使ってたclose side in the Galactic bulge from the polarization of ``redder'' stars at the far side in the bulge.
near side in the Galactic bulge from the polarization of ``redder'' stars at the far side in the bulge.  
%of small reddening field stars from the polarization of large reddening field stars located near the Galactic center.
This work was followed by \citet{cNishiyama} who used the same method to examine a transition from toroidal to poloidal magnetic field above and below the Galactic plane.
However, these works %◆
simply used $H$ $-$ $K_{\mathrm S}$ colors instead of actual distances, and kiloparsec-scale details of the morphology of the magnetic field remain %hidden.◆なんか弱い気が
unclear. 
\textcolor{black}{\citet{Pavel} attempted to decompose the line of sight structure of the Galactic magnetic field and provided photometrically identified red clump stars with the polarization data of the Galactic Plane Infrared Polarization Survey
% (e.g., は要らんと思うけどなあ。長田1222
(\cite{Clemens}).  
For real decomposition, however, spectroscopic follow-up to identify red clump stars is needed, as the author admitted in the paper. %長田1223
}

We %presents 
present a new method for examining the large-scale magnetic field geometry with distance by using classical Cepheids and $Gaia$ data in the Galactic plane.
%We presents a new method for resolving the large-scale magnetic field geometry with distance by using classical Cepheids and $Gaia$ data in the Galactic plane. 
Classical Cepheids are pulsating supergiants with the period luminosity relation (PLR) which enables us to estimate their distances %▼副詞または副詞句にしないとね！reasonably accurate, %assuming the extinction low (\cite{aNishiyama}).●law。それより、この形式（主語がassumeするわけじゃない）の分詞構文はなるべく避けよう
with reasonable accuracy, 
if we can assume an appropriate extinction law (e.g., \cite{aNishiyama}).
%Classical Cepheids are pulsating supergiants with the period luminosity relation (PLR) which enables us to estimate their distances reasonably accurate, assuming the extinction low (\cite{aNishiyama}). 
\citet{bMatsunaga} discovered 29 classical Cepheids between $-10^{\circ}$ and $+10^{\circ}$ in the Galactic longitude along the Galactic plane, using the Infrared Survey Facility (IRSF). 
D\'ek\'any et al. (\yearcite{aDekany}, \yearcite{bDekany}) %▼過去の遺物よ！have reported dozens of new Cepheids found in the VISTA Variables in the V\'ia L\'actea (VVV) survey (\cite{Minniti}). 　▲加えた　
reported dozens of new Cepheids found in the VISTA Variables in the V\'ia L\'actea (VVV) survey (\cite{Minniti}). 
Their new Cepheids are located %towards 
in the far side of %bulge region along the Galactic plane. 
the Galactic center.  
%ここの接続は自然か？OKと思いますよ。▼ここって、Gaiaが突然でてくるってところね。うーん、あんまり好かんけど入れますか、以下を。それから、theも必要かな。
On the other hand, 
the 
$Gaia$ astrometric mission (\cite{Gaia}) %will be %possible 事がpossible 何かはable ▼で、それよりもbe toの方がカッコいいかな。
%able 
is 
to produce a three-dimensional map of unprecedented precision, using parallaxes for billions of stars. 
Combining %the 
polarization with %▼parallaxは数えられそうだなあ 
parallaxes 
is a powerful tool to probe the 3D topography of the interstellar medium and magnetic field. 
%On the other hand, $Gaia$ astrometric mission (\cite{Gaia}) will be possible to produce a three-dimensional map of unprecedented accuracy, using precise parallaxes for billions of stars. 
Visual-wavelength % polarimetry 複数にするのなら
polarimetric studies 
have tried to resolve magnetic field structure in the diffuse ISM or dark globule using the $Gaia$ DR2 parallax (\cite{Panopoulou} and \cite{Eswaraiah}),  %▼カッコの後の空白なくした
%These works can probe the change of magnetic field orientation with distance. このニュアンスがよくわからんので下のようにしました。
and have made some success in detecting the change of magnetic field orientation with distance. 
%\citet{Panopoulou} and \citet{Eswaraiah} combined visual-wavelength polarimetry with parallaxes from $Gaia$ data release to find the cloud distance. 

In this paper, we divide field stars into three regions, %ここからしつこくa)b)c)と
a) foreground, b) bulge, and c) background, using %the classical Cepheids, $Gaia$ parallax data.
the $Gaia$ parallax data and near-infrared colors with the Cepheid distances.
%In this paper, we classify field stars according to foreground, bulge and background using classical Cepheids, $Gaia$ data and $H$ - $K_{\mathrm{S}}$ histograms. 
%Three fields are 
These three regions are 
then used as distance markers along a line of sight to probe 
the 
magnetic field structure among them via %▼NIR この略号使ってない
near-infrared starlight polarimetry. 

%These populations are then used as distance markers along a line of sight to probe magnetic field structure among them via NIR starlight polarimetry. 
In section 3, we classify 52 Cepheid fields %as 
into 
three types 
simply from 
polarization behavior vs. $H-K_{\mathrm S}$ color of field stars,  
choose a field of view from each of the three types, and 
discuss their polarimetric characteristics. 
%as three types and discuss the result of polarimetric observation to chosen three fields.
%The details of %a 
Our new method for measuring the Galactic magnetic field structure with distance is presented in Section 4. 
%The details of a new method for measuring the Galactic magnetic field structure with distance is presented in Section 3. 
%In addition, the method is applied to three lines of sight toward the inner Galaxy in Section 3. 
%Section 4 discussions the result on the morphology of the Galactic magnetic field toward three lines of sight.

\section{Observation}
We conducted near-infrared polarimetric observations of 52 classical Cepheids and the stars around them in 2016 and 2017 with the SIRPOL instrument. The 52 classical Cepheids are identified by D\'ek\'any et al. (\yearcite{aDekany}, \yearcite{bDekany}) and \citet{bMatsunaga}. Among them, 50 Cepheids are located very close to the Galactic %◆Plane ▲加えた
plane ($-10^{\circ}\leq \, l\, \leq +10\fdg5$、$-0\fdg2\leq \, b\, \leq +0\fdg2$), and the other two Cepheids are in $0\fdg2 \leq \, |b|\, \leq 0\fdg5$.
 
SIRPOL consists of a single-beam polarimeter (a half-wave plate rotator unit and a fixed wire-grid polarizer; \cite{Kandori}) and the near-infrared imaging camera SIRIUS (Simultaneous Infrared Imager for Unbiased Survey; \cite{Nagashima}; \cite{Nagayama}), and is attached to the 1.4 m telescope IRSF. The camera is equipped with three 1024 pixel $\times $ 1024 pixel HAWAII %$\rm I\hspace{-.1em}I$ ◆II とちゃう！！
arrays. %array▼This enables simultaneous observations in the $J$($\lambda_{J}=1.25\mu m$)、$H$($\lambda_{H}=1.63\mu m$), $K_{\mathrm S}$($\lambda_{K_{\mathrm S}}=2.14\mu m$)-bands by splitting the beam into the dichroic mirrors. 
SIRPOL provides images of a $7\farcm7$ $\times $ $7\farcm7$ area of sky, in the $J$ ($\lambda_{J}=1.25  \ \mu {\mathrm m}$), $H$ ($\lambda_{H}=1.63  \  \mu {\mathrm m}$), and $K_{\mathrm S}$ ($\lambda_{K_{\mathrm S}}=2.14 \ \mu {\mathrm m}$) bands, simultaneously. The image scale of the arrays is $0\farcs45$ pixel$^{-1}$. We obtained 10 dithered exposures, each 10 s or 20 s, at four wave-plate angles ($0^{\circ}$, $22\fdg5$, $45^{\circ}$, and $67\fdg5$ in the instrumental coordinate system) as one set of observations. %The total exposure time about half of the 52 Cepheids was 400 s, and the other were 800 s. 
The total exposure time was 400~s for most of the Cepheid fields, and 800~s %▼800s 
or more for some fields. The typical seeing was %▲　きまぐれ。ＪＨは全く要らんでしょ。$\sim 1''.4$ ($J$), $\sim 1''.3$ ($H$), and 
$\sim \, 1\farcs2$ in the $K_{\mathrm S}$ band during the observations. Twilight flat-field images were obtained at the beginning and end of the observations.

Standard procedures, dark subtraction, flat-fielding with twilight-flats, bad pixel subtraction, sky subtraction, and averaging of dithered images were applied with IRAF. In this paper, we discuss the result of $K_{\mathrm S}$-band polarimetry only. The Stokes parameters, $I$, $Q$, and $U$ for Cepheids and the field stars are determined from aperture photometry of combined images as follows. %\textcolor{black}{We calculated raw Stokes parameters} %(善光_191127)長田1220 いや、すでにThe Stokes parameters, $I$・・as followsと言っているのにこれを書くのはおかしくないですか。
$\it Q = \it I_{\rm 0}-\it I_{\mathrm 45}$, $\it U = \it I_{\rm 22.5}- \it I_{\rm 67.5}$, and $\it I = (\it I_{\rm 0}+\it I_{\rm 22.5}+\it I_{\rm 45}+\it I_{\rm 67.5})/{\rm 2}$, where $I_{\rm 0}, I_{\rm 22.5},  I_{\rm 45},$ and  $ I_{\rm 67.5}$ are intensities at four wave-plate angles. DAOFIND and PHOT tasks were used %◆コピペミスかな 　　　　長田1220 一か所65.5に戻っていたのが気になってました。
for 
point source identification and the aperture photometry %for ◆forで良いけど、先のforと重ならないようにしてみた
at 
each wave-plate angle ($ I_{\rm 0}, I_{\rm 22.5}, I_{\rm 45}$, and $I_{\rm 67.5}$). Since the aperture photometry gives a better result than PSF fitting photometry (\cite{Hatano}), aperture photometry was applied in the following procedure. Photometry measurements are greatly affected by the choice of aperture size. If the aperture radius is too large the obtained value suffers from background contamination in the %galactic ◆ 長田1221
Galactic plane and the signal to noise ratio is decreased. 
%On the other hand, ▼やっぱりこういう日本語的な英語はやめよう
If the aperture radius is too small, however, only a fraction of total flux is measured % ▼あまりにぶつ切りだねえ If the aperture is too small, only a fraction of total flux is measured.   If the fraction of the total flux measured is not the same, 
and the fraction inevitably changes, 
%the small size aperture introduces artificial polarization. 
resulting in artificial polarization.  
The aperture size was chosen after a search from 1.0 to 2.0 $\times$ FWHM at intervals of 0.1. We mostly adopted the aperture size of 2 $\times$ FWHM measured, but we adopted the aperture size of 1.5 or 1.0 $\times$ FWHM if the Cepheid was faint or the seeing was bad. The Two Micron All Sky Survey catalog (\cite{Skrutskie}) was used for absolute photometric calibration %▼セファイドは測光してないんだよねえ？(偏光を求める際にIを計算しているので測光はしていますが、使っていません)
of the field stars.  

The raw polarization degree $P_{\mathrm{raw}}$ and
the position angle $PA$ were calculated from %◎derived by
\begin{eqnarray*}
P_{\mathrm{raw}} = \sqrt{
\left(\frac{Q}{I}\right)^2+
\left(\frac{U}{I}\right)^2 }, \, \,
PA = \frac{1}{2}
\arctan \frac{U}{Q}.
\end{eqnarray*}
The debiased $P$ \citep{Wardle}
was finally derived from %◎
%◎by 
$P = \sqrt{P_{\mathrm{raw}}^2-\delta P_{\mathrm{raw}}^{2}}$
where $\delta P_{\mathrm{raw}}$ is the error of $P_{\mathrm{raw}}$, %given by
\textcolor{black}{
calculated from 
the propagation of errors in the four intensities at four wave-plate angles.  
}%
%独立した4つの波長板角の測光エラーを誤差伝播させて
%もしも式で書くなら
%\textcolor{black}{
%\begin{eqnarray*}
%\delta P_{\mathrm{raw}} &=&\sqrt{
%\left(\frac{\partial P_{\mathrm{raw}}}{\partial I_{0}}\right)^2\left(\delta I_{0}\right)^2
%+\left(\frac{\partial P_{\mathrm{raw}}}{\partial I_{22.5}}\right)^2\left(\delta I_{22.5}\right)^2
%+\left(\frac{\partial P_{\mathrm{raw}}}{\partial I_{45}}\right)^2\left(\delta I_{45}\right)^2
%+\left(\frac{\partial P_{\mathrm{raw}}}{\partial I_{67.5}}\right)^2\left(\delta I_{67.5}\right)^2}.
%\end{eqnarray*}%(善光_191127)
%}
%\textcolor{black}{
%\begin{eqnarray*}
%\delta P_{\mathrm{raw}} &=&\sqrt{
%\left(\frac{\partial P_{raw}}{\partial I_{0}}\delta I_{0}\right)^2
%+\left(\frac{\partial P_{raw}}{\partial I_{22.5}}\delta I_{22.5}\right)^2
%+\left(\frac{\partial P_{raw}}{\partial I_{45}}\delta I_{45}\right)^2
%+\left(\frac{\partial P_{raw}}{\partial I_{67.5}}\delta I_{67.5}\right)^2}.
%\end{eqnarray*}%(善光_191127)
%}
\textcolor{black}{
The typical error is $\delta P_{\mathrm{raw}}\, =\, 0.5 \%$ for $K_{\mathrm S}\, =\, 11.5$ mag. }%
%The typical magnitude for $\delta P_{\mathrm{raw}}\, =\, 0.5 \%$ is $K_{\mathrm S}\, =\, 11.5$ mag.%>>(善光)位置を前にする  長田1220

In this paper, we use the Galactic coordinate system,
and furthermore the position angle $PA_{\mathrm{GP}}$
is measured anti-clockwise
from the longitude-increasing direction of the Galactic plane
(i.e., the usual position angle in the Galactic coordinate system
(e.g., \cite{Appenzeller}),
which is measured from the north Galactic pole, minus $90^{\circ}$).
Lowercase $q_{\mathrm{GP}}$ and $u_{\mathrm{GP}}$ stand for 
$(Q/I)_{\mathrm{GP}}$ and $(U/I)_{\mathrm{GP}}$ in this system,
respectively.
\textcolor{black}{
Our zero-point of the $PA$ is estimated to be determined better than $\sim 3^{\circ}$ 
\citep{Kusune}.  %長田1223
}

\textcolor{black}{
We have checked our polarimetry by observing the unpolarized standard star WD~2539-434 
from the VLT list  %長田1222
\citep{Fos07} %長田1221
in order to estimate instrumentally induced polarization. %The value $p$ and $\theta$ of this target were derived by averaging normalized Stokes parameters resulting in a total exposure time of 1600 s. ◆ pとθがヘンだよなあ（Pにしようとか）と思ったけど、それ以前に、D論なら重要としても、投稿論文ではここまで書かなくても良いのでは？
We have confirmed that the instrumental polarization is small enough to neglect within the errors %in the $K_{\mathrm{S}}$ band.
of our polarimetry precision.   %長田1222 長田1223
Although it was found to vary with a period of 2.6950 hours and semiamplitude of 4 mmag in the $R$ band \citep{Gary}, 
variations at such a level do not affect our polarimetry 
because the average time difference is only about 30 s 
between the intensity measurements of different wave-plate angles.  }

\textcolor{black}{
We have also examined 
how the measured polarization of WD~2539-434 %長田1222
changes as a function of aperture size 
from 1.0 to 4.0 $\times$ FWHM 
to check whether our chosen aperture is too small and introduces artificial polarization.  
Its debiased polarization stays zero between %1.6 and 2.6 $\times$ FWHM;善光191221 
1.5 and 2.4 $\times$ FWHM;
its calculated $\delta P_{\mathrm{raw}}$ 
stays below 0.28\% between 1.0 and 2.2 $\times$ FWHM, but 
it increases to 4.0 $\times$ FWHM at a rate of nearly 0.01\% per 0.1 $\times$ FWHM, 
probably due to too much background contamination.  
We have examined several field stars of different magnitude as a function of aperture size also, and 
many of them tend to be stable near 2 $\times$ FWHM 
although some stars fainter than 11~mag have a variety of fluctuations.  }%長田1221

\begin{figure}[H]
\begin{center}
\includegraphics[width=80mm]{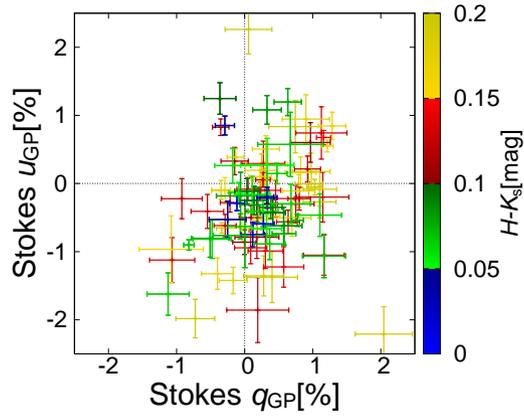}
\end{center}
\vspace{1cm}
\caption{Stokes parameters $q_{\mathrm{GP}}$ and $u_{\mathrm{GP}}$ of 
all the field stars that are brighter than $K_{\mathrm{S}}=11$~mag and 
bluer than $H$ - $K_{\mathrm{S}} = 0.2$~mag 
%field stars which are bluer than $H$ - $K_{\mathrm{S}}$ $\leq$ 0.2 mag 長田1222
in the 52 fields.}
\label{fig:Fig0.2mag}
\end{figure}

\textcolor{black}{
In a similar vein, we have examined the polarization of all the stars 
that are brighter than $K_{\mathrm S} = 11$~mag 
and bluer than $H-K_{\mathrm S} = 0.2$~mag in the 52 fields, 
to check the level of instrument polarization.  %長田1221 長田1222
%In a similar vein, we examined the polarization of all the stars 
%that are bluer than $H-K_{\mathrm S} \leq 0.2$~mag in the 52 fields, %善光191221(\leq)
%to check the level of instrument polarization.  %長田1221
We have observations of only a few fields which suffer interstellar extinction small enough, 
so polarimetry of such blue stars less affected by extinction in the 52 fields can serve %長田1222
to estimate the level of instrument polarization of our observation system.  
The mean of $q$, its standard deviation, $u$, and its standard deviation of these 99 
stars are $0.28$\%, $(\pm)0.58$\%, $-0.27$\%, $(\pm)0.73$\%, respectively (see Figure \ref{fig:Fig0.2mag}).  
Field stars have intrinsic $H-K_{\mathrm S} \sim 0.1$~mag on average according to \citet{Wainscoat} (see below), and %長田1222
the selected stars will have at most 0.9\% 
if they follow the upper limit $P / E(H-K_{\mathrm S}) = 9.0\% \mathrm{mag}^{-1}$ \citep{Hatano}.  
The polarization  $q \sim 0.28$\%, $u \sim -0.27$\% is also consistent with the results 
for disk star candidates ($PA \sim 0^\circ$ in the equatorial coordinates, 
which is $PA_\mathrm{GP} \sim -30^\circ$) in \citet{Hatano} %長田1223 長田1223
although the disk star candidates in \citet{Hatano} include reddened stars 
up to $H-K_{\mathrm S} = 0.4$~mag and their mean $P$ is 0.8\%.  %長田1222
\citet{Kandori} states that the stability of SIRPOL is better than 0.3\%, and %長田1222
they were unable to detect instrument polarization.  
Our results are consistent with it, and %長田1222
we regard our instrument polarization as being smaller than 0.3\%.  %長田1222
Also, the stability of SIRPOL in a long term has been demonstrated, 
but we have to be careful because in a recent paper 
\citet{Kan19} mention that the instrument polarization of SIRPOL can reach 
0.26\% due to inappropriate handling of skyflat frames.  
We will investigate this problem in a future paper.  %長田1223
} %長田1221

\section{Results}
In %Table 1 ◆
table~1 
we list the %galactic ◆◎能動態にした
Galactic 
coordinates, $H$ and $K_{\mathrm{S}}$ band mean magnitudes adopted from D\'ek\'any et al. (\yearcite{aDekany}, \yearcite{bDekany}) and \citet{bMatsunaga}, extinctions $A_{K_{\mathrm{S}}}$, distances $D$, polarization degrees $P$ and position angles $PA_{\mathrm{GP}}$ in the $K_{\mathrm{S}}$ band of the 52 Cepheids. %●ピリオドの前に空白は不要 　▲加えた　
%We use $H$ and $K_{\mathrm S}$ band mean magnitudes for estimating distances to Cepheids as done by \citet{bDekany} and \citet{bMatsunaga}. 
\textcolor{black}{We use distance modulus for estimating distances to Cepheids as done by Matsunaga et al. (2016). %However, 
Since D\'ek\'any et al. (\yearcite{bDekany}, \yearcite{bDekany}) estimated the distances using their own extinction law, we have re-calculated the distance of the Cepheids reported by D\'ek\'any et al. (\yearcite{bDekany}, \yearcite{bDekany}) in accordance with \citet{bMatsunaga}.}%
%D\'ek\'any et al.  (\yearcite{bDekany},\yearcite{bDekany}) estimated the distances using the differential extinction relation, and we re-calculated the distance of the Cepheids reported by D\'ek\'any et al.  (\yearcite{bDekany},\yearcite{bDekany}) in the \citet{bMatsunaga} way.}%(善光_191127)長田1220  Howeverはおかしくないですか・・・等。
%◎(善光)先に観測していたD\'ek\'anyの$H$と$K_{\mathrm{S}}$をつかっており、D\'ek\'anyで発見されていない天体に関しては松永のを用いた.
We adopt the PLR of classical Cepheids from \citet{aMatsunaga}:
\textcolor{black}{
\begin{eqnarray}
M(H) &=&-3.256 (\log Pd-1.3)-6.562,\\
M(K_{\mathrm S})&=&-3.295 (\log Pd-1.3)-6.685,
\end{eqnarray}
}%長田1220 長田1221
\textcolor{black}{
where $Pd$ is their variation period in days.  } %長田1223 
%These relations were calibrated based on {\it Hubble Space Telescope} Fine Guidance Sensor parallaxes of Cepheids in the solar neighborhood (Benedict et al. 2007; van Leeuwen et al. 2007). 
These PLRs and the extinction relation of $A_{K_{\mathrm S}}/E_{H-K_{\mathrm S}}=1.44$ (\cite{aNishiyama}) %●タイポ修正
are combined with the observed magnitudes $H$ and $K_{\mathrm S}$ %●カンマは不要
to estimate the distance and the foreground extinction $A_{K_{S}}$.

\begin{figure}[H]
\begin{center}
\includegraphics[width=80mm]{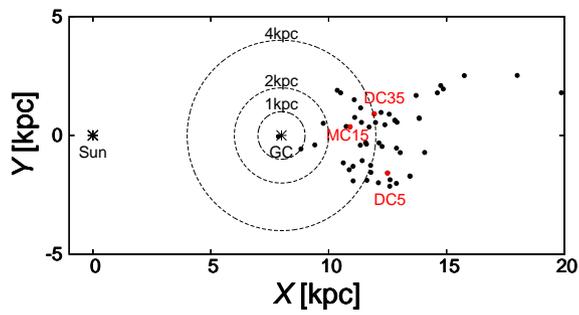}
\end{center}
\vspace{1cm}
\caption{Distribution of the 52 Cepheids 
% , indicates by black filled circles %, respectively, on the face-on view of the Galactic disc. Red filled circles mean that we ●respectivelyはこんな使い方をしません、A, B, C, by a, b, c, respectivelyというように１対１で「対応」するときだけです。
in the face-on view of the Galactic disc. %, ▲
%except for the two Cepheids (MC22 and MC26) at the distances more than 18~kpc from the Sun.  
Red dots are the three Cepheids analyzed in detail  (Section 3. and 4.). 
%analyze the targets in detail (Section 3. and 4.). 
%The sun is indicated by the asterisk at the origin, and also the Galactic Center is indicated by the asterisk at an assumed distance of 8 kpc. 
The Sun - Galactic Center (asterisks) distance is assumed to be  8~kpc. }
%Dashed circles at 1, 2 and 4 kpc from the Galactic Centre are drawn for readers's convenience. We do not plot MC22 and MC26, because the distances of these stars are more than 18 kpc.}●キャプションでは普通、Weがプロットしたとは言いません。もっと客観的な言い方をします
\label{fig:Fig1}
\end{figure}

Figure \ref{fig:Fig1} shows %figure  plots●
the distribution of the 52 Cepheids with filled circles. Only one Cepheid (MC10) is located close to the Galactic center, and all the others are located on the far side of the Galactic center. 
Although %●やっぱりSinceはヘンだと思う
the polarization of these Cepheids is 
\textcolor{black}{an} %長田1220
integral quantity between the observer and the Cepheids along the line of sight, 
we can probe the magnetic field structure in the far side of the Galactic center 
if we perform some differential analysis.  %using the polarization of the Cepheids. 
%●Therefore, we can probe the magnetic field which Faraday rotation data can not reveal. (e.g. Han \yearcite{cHan}, \yearcite{dHan}).
The information about 
\textcolor{black}{the} %長田1220
magnetic field structure in the far side was quite unclear even with Faraday rotation and other data (e.g. Han \yearcite{cHan}, \yearcite{dHan}).

\begin{figure*}[H]
\begin{center}
\includegraphics[width=160mm]{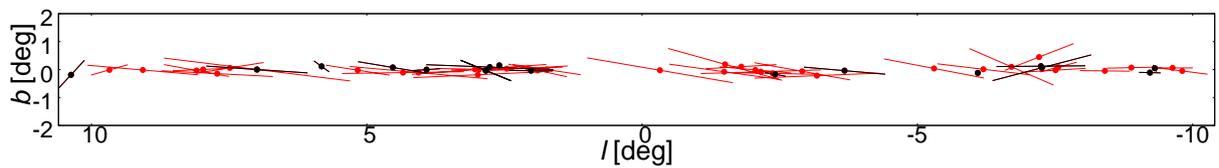}
\end{center}
\vspace{1cm}
\caption{Polarization of the 52 Cepheids.
The red vectors are measurements with small errors $P/\delta P \, >\, 3$, but the black vectors are those with $P/\delta P \, <\, 3$.}
\label{fig:eFig1}
\end{figure*}

\begin{figure}[h]
\begin{center}
\includegraphics[width=80mm]{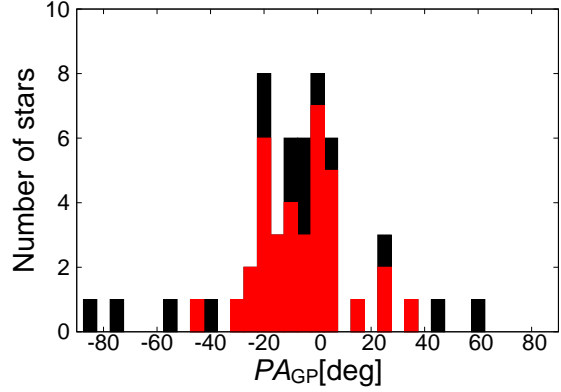}
\end{center}
\vspace{1cm}
\caption{Histogram of $PA_{\mathrm GP}$ for the 52 Cepheids.  
The red filled boxes are measurements with small errors $P/\delta P \, >\, 3$, 
%but ▲やっぱりandが自然な気がしてきた。気まぐれナガタ
\textcolor{black}{
and the black filled boxes are those with $P/\delta P \, <\, 3$.%▲vectors are those with $P/\delta P <3$.
}
}
\label{fig:eFig2}
\end{figure}

The magnetic field component horizontal to the Galactic plane is dominant over the vertical magnetic field component. %●, because 47 Cepheids out of the 52 Cepheids have the position angles of $|PA_{\mathrm{GP}}| < 45^{\circ}$. ◎horizontalを後ろに持って来た
\textcolor{black}{38 Cepheids out of the 52 Cepheids have %the 長田1220 
position angles of $|PA_{\mathrm{GP}}|\, <\, 20^{\circ}$ (in figure~\ref{fig:eFig1} and figure~\ref{fig:eFig2}).} %(善光_191127)
%47 Cepheids out of the 52 Cepheids have the position angles of $|PA_{\mathrm{GP}}|\, <\, 45^{\circ}$ (in figure~\ref{fig:eFig1} and figure~\ref{fig:eFig2}).
%●5 Cepheids out of the 52 Cepheids have the large
\textcolor{black}{Five Cepheids} have large position angles of $|PA_{\mathrm{GP}}|\, \geq \, 45^{\circ}$, %●and %(善光_191127)
but their position angle errors $\delta PA_{\mathrm{GP}}$ are more than $10^{\circ}$, which means that the polarization signal to noise ratios are smaller than 3. %●The position angle error $PA_{\mathrm{err}}$ of more than $10^{\circ}$ do not represent correctly the position angle, and the position angles $PA_{\mathrm{GP}}$ of 5 Cepheids do not indicate the vertical magnetic field component is dominant to the horizontal magnetic field component.
We do not regard them representing the magnetic field correctly.  

\begin{figure}[h]
\begin{center}
\includegraphics[width=80mm]{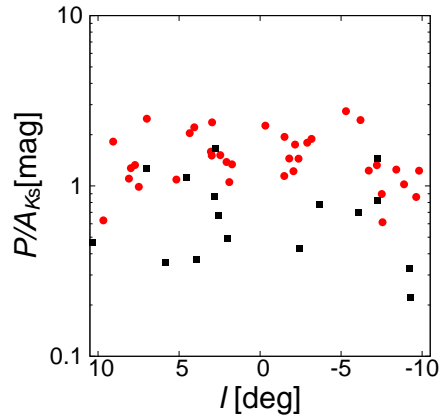}
\end{center}
\vspace{1cm}
\caption{Polarization efficiency $P/A_{K_{\mathrm{S}}}$ vs. Galactic longitude $l$ for the 52 Cepheids. The red filled circles %indicate the detected polarization 
are measurements with small errors $P/\delta P \, >\, 3$, %at exceeding 3,●これは文法的におかしいよね 
but the black filled squares %indicate the detected polarization $P/\delta P$ at not exceeding 3.
are those with $P/\delta P \, <\, 3$. }
\label{fig:Fig2}
\end{figure}

Kobayashi et al. (\yearcite{aKobayashi} \& \yearcite{bKobayashi}) observed the Galactic longitudes of $0^{\circ}$, $20^{\circ}$, and $30^{\circ}$, and found difference in the polarization efficiency $P_{K_{\mathrm{S}}}/A_{K_{\mathrm{S}}}$. 
The polarization efficiency at the Galactic longitudes of $20^{\circ}$ and $30^{\circ}$ %●
is smaller than 
\textcolor{black}{the} %長田1220
one %●
at the Galactic center of $0^{\circ}$. 
Since the Galactic longitudes $l$
of the 52 Cepheids 
range %●動詞
from $-9\fdg8$ to $+10\fdg4$, 
%●from $-9.81^{\circ}$ to $+10.372^{\circ}$, 
we examine if the polarization efficiency $P_{K_{\mathrm{S}}}/A_{K_{\mathrm{S}}}$ is %difference ●
different from field to field.  %●in each field. 
Figure %●
\ref{fig:Fig2} shows the relationship between the Galactic longitude and the polarization efficiency of the 52 Cepheids. 
The polarization efficiency with $|l|\, \leq$ $5^{\circ}$ is 1.39 $\pm$ 0.55 $\% / \mathrm{mag}$, %●絶対値はｌにつける。 ●mag$, 
and the polarization efficiency with $|l|$ $>$ $5^{\circ}$ is 1.15 $\pm$ 0.63 $\% / \mathrm{mag}$; 
the Cepheids %●at the outer of observed field 
in the outer part 
($|l|$ $>$ $5^{\circ}$) have %●the 
polarization efficiency similar to 
the Cepheids %which are 
close to the Galactic center ($|l|\, \leq$ $5^{\circ}$),  %◎やっぱりthereforeは大げさやねえ
and the polarization efficiency is not %significant difference 
significantly different in our observed region.  %(7/29)

%●段落を変えた
%●新
It is believed that interstellar polarization depends on the magnetic field 
of constant and random components similar in strength \citep{Jones}.  
Then it might be natural to assume that 
the interstellar polarization degree gradually increases 
as a more distant and therefore more reddened star is observed.  
Also, the position angle of polarization is expected to be rather constant, and 
possibly along the Galactic plane ($|PA_{\mathrm{GP}}|\, \sim \, 0^{\circ}$).  
We ask if such a picture of undisturbed magnetic structure holds in most of the observed 
Cepheid fields.  
%●●●分類　前景バルジ後景はあとで
%We %divides 
%divide each field %of view ●
%into the three regions %●
%a) foreground, b) bulge, 
%and c) background, and %●calculates 
%calculate the differential polarization between them. 
We %probe 
examine polarization efficiency and change %of ◆inの方が自然な気が
in 
differential position angles $PA_{\mathrm{GP}}$ %to ◆なんでtoだっけ
as the reddening increases %▲要るよねえ
in the %48◆ここはまだ51でしょ。だけどくどいから 
Cepheid fields, 
and we %can classify them into three types. (7/29)
have classified them into three types %●新
of 
1) fairly constant and regular magnetic field, 
2) containing an abrupt change in the field direction, and 
3) of rather complicated pattern. 
%●段落の最後へ持って来てつなげた、
We tried to %
conduct a new polarimetric differential analysis with the observed 52 Cepheid fields, %▲Howeverは強すぎるかな
%However, 
but it was %● 
difficult to %calculate the polarization of field stars in %4 
derive accurate polarization of field stars in %4 
four Cepheid %●field
fields due to poor signal to noise ratios.  %,▲２文に 
%and 
Thus, we looked into the other 48 Cepheid fields. %▲theかな、やっぱり%●

%●Hatanoの段落をここへ持ってきた
When we plot the field stars in a polarization degree $P$ vs. $H$ $-$ $K_{\mathrm S}$ diagram, 
stars with larger polarization extend redward, and the slope 
$\Delta P$/$\Delta (H$ $-$ $K_{\mathrm S})$ %◆統一しよう
reflects the polarization efficiency. 
%●Hatanoの説明は短くしてしまおう、私たちのやってることじゃないのだから。
Toward the Galactic center, 
\citet{Hatano} divided the field stars into two groups 
%at $H$ - $K_{\mathrm S}$ $\sim$ 0.4 mag; disk sources have $H$ - $K_{\mathrm S}$ $<$ 0.4 mag and bulge sources have $H$ - $K_{\mathrm S}$ $\geq$ 0.4 mag. 
of disk sources with $H$ $-$ $K_{\mathrm S}$ $<$ 0.4 mag 
consisting of A/F dwarfs and G/K giants, 
and bulge sources with $H$ $-$ $K_{\mathrm S}$ $\geq$ 0.4 mag
consisting of K/M giants.   
%Toward th Galactic center, 
%stellar types of the disk sources are $A/F$ dwarfs and $G/K$ giants, but stellar type of the bulge sources are $K/M$ giants. 
%Therefore, the color excess $E(H-K_{\mathrm{S}})$ for the disk sources is smaller than the color excess $E(H-K_{\mathrm{S}})$ for the bulge sources. 
Similarly, we %defined the field stars with $H$ - $K_{\mathrm S}$ $\geq$ 0.5 mag as the bulge stars and calculate the gradient. 
define the field stars with $H$ $-$ $K_{\mathrm S}$ $\geq$ 0.5 mag as the bulge stars and calculate the gradient. 

If the magnetic field along the line of sight %well aligns the same direction, 
aligns well in the same direction, 
in the figure of polarization degree $P$ vs. $H$ $-$ $K_{\mathrm S}$ color, %>>(善光191104)冠詞は不要では? the $H$ - $K_{\mathrm S}$ color, %
the polarization slope should be great.  
%the increase of the polarization degree with %
%$H$ - $K_{\mathrm S}$ color is large. 
We make linear fitting to the polarization of 
field stars with $H$ $-$ $K_{\mathrm S}$ $\geq$ 0.5 mag 
and calculate %a slope of polarization efficiency in each field. 
the slope in each field. 
If the slope of polarization %●
per color %$P_{K_{\mathrm S}} / (H$ - $K_{\mathrm S})$ %◆統一しよう
$\Delta P$/$\Delta (H$ $-$ $K_{\mathrm S})$
is more than %
1.2~$\% / \mathrm{mag}$ in both the color ranges of  
0.5~mag $\leq \, H$ $-$ $K_{\mathrm{S}}\, \leq$ 1.5~mag and  
0.5~mag $\leq \, H$ $-$ $K_{\mathrm{S}}\, \leq$ 3.0~mag (table~1), 
we define %●
%1 $\% / \mathrm{mag}$, we define %●
the field as Dflag = 1.   %and classify 13 Cepheid fields into Dflag = 1.
%◎で、この範囲でこう私が書いたので良いんだっけ？　(善光)色$H$ - $K_{\mathrm{S}}$の範囲を指定して、2つのパターンで傾きを計算している(表1).
Although this slope break is somewhat arbitrary, 
it corresponds to a fairly random magnetic field geometry 
in the compilation of infrared polarimetry by \citet{Jones} 
if we adopt the extinction law by \citep{aNishiyama} and assume 
$H$ $-$ $K_{\mathrm{S}}\,  \sim$ 1-2~mag.  %※ここで言ってることが正しいかな？JonesのFig.2にプロットしてみて。%▲１-２に戻しました。ドッペルゲンガーで１-２というのが消えていた。
Only \textcolor{black}{10} Cepheid fields are classified to have Dflag = 1.
%when we conduct a differential analysis to 35 Cepheid fields, 
In the remaining \textcolor{black}{38} Cepheid fields,
two fields have %the differential position angles $PA_{\mathrm{GP}}$ are more than $45^{\circ}$ between the bulge and the background. 
turned out to be populated by field stars 
whose position angles change greatly as the $H$ $-$ $K_{\mathrm S}$ color increases, and 
%$PA_{\mathrm{GP}}$ more than $45^{\circ}$ 
%between the bulge and the background.  %●
we have defined these %▲three 
two fields as Dflag = 2.  % 
%This result indicates the magnetic field orientation greatly change between the bulge and the background. 
As we show in the next section, the magnetic field orientation 
$PA_{\mathrm{GP}}$ seems to change more than $45^{\circ}$ 
between the bulge and the background in these fields of view. 
%%%%%%%%こういう感じ?
%In the remaining 39 Cepheid fields, %In the remaining 35 Cepheid fields,
%two fields have %three fields have %the differential position angles $PA_{\mathrm{GP}}$ are more than $45^{\circ}$ between the bulge and the background. turned out to be populated by field stars whose position angles change greatly %more than $45^{\circ}$ as the $H$ - $K_{\mathrm S}$ color increases%between the bulge and the background, and %$PA_{\mathrm{GP}}$ more than $45^{\circ}$ %between the bulge and the background.  %●twoだよね
%一つでしか示していないなあ。
%other にはtheが要るね。
%具体的に書かないと形容詞はアカンなあ
%andの後にスペースを打ってからパーセントを入れないとだめ。
%we have defined these two fields as Dflag = 2.  % %This result indicates the magnetic field orientation greatly change between the bulge and the background. As we show in the next section, the magnetic field orientation $PA_{\mathrm{GP}}$ seems to change more than $45^{\circ}$ between the bulge and the backgroundin one of these fields of view. %Therefore, we defined these three field of views as Dflag = 2 %and other 32 field of views as Dflag = 3
%%%%%%%%%%%%
In the other \textcolor{black}{36} Cepheid fields, the polarization degree does not increase very much ($\Delta P$/$\Delta (H$ $-$ $K_{\mathrm S}) \leq 1.2$), 
and %In other 31 Cepheid fields, the polarization degree does not increase very much, and 
we do not notice any significant pattern.    
Their polarization is generally parallel to the Galactic plane, but 
some of them show broad distribution of position angle 
in the polarization of field stars.  
We have defined these \textcolor{black}{36} filelds as Dflag = 3.%We have defined these 31 filelds as Dflag = 3.

%The purpose of this paper tests the usefulness of the new differential analysis method. Therefore, 
We choose a field of view from each of the three types, %discussing 
and discuss the change of magnetic field structure %
in detail. 
%Regarding residual 49 field of views, w
We will present the analysis of the other %▲49   48-3じゃない？
45 Cepheid fields elsewhere (Zenko et al. in preparation).   
%prepare the next paper. 
%Since we perform the statistical processing to the new method, w
We have selected Cepheid %
fields with large number of %
field stars in all of the foreground, %▲なんか不自然the 
bulge, and background categories.  
%We determine the MC28 from Dflag =1, the DC5 from Dflag = 2, and MC15 from Dflag = 3. 
%MC28 
DC35 from Dflag = 1, DC5 from Dflag = 2, and MC15 from Dflag = 3 have been selected. 
These Cepheid locations %
are indicated with red points in figure~\ref{fig:Fig1}. 
%In this section, w
We discuss the result of polarimetric observation of each field. %of view. (7/29)%

%%%%%%2019/8/14(善光191105)ここから下は私が修正したものを挿入
\subsection{DC35}
\begin{figure*}[h]
\begin{center}
\begin{minipage}{0.5\hsize}
\begin{center}
\includegraphics[width=90mm]{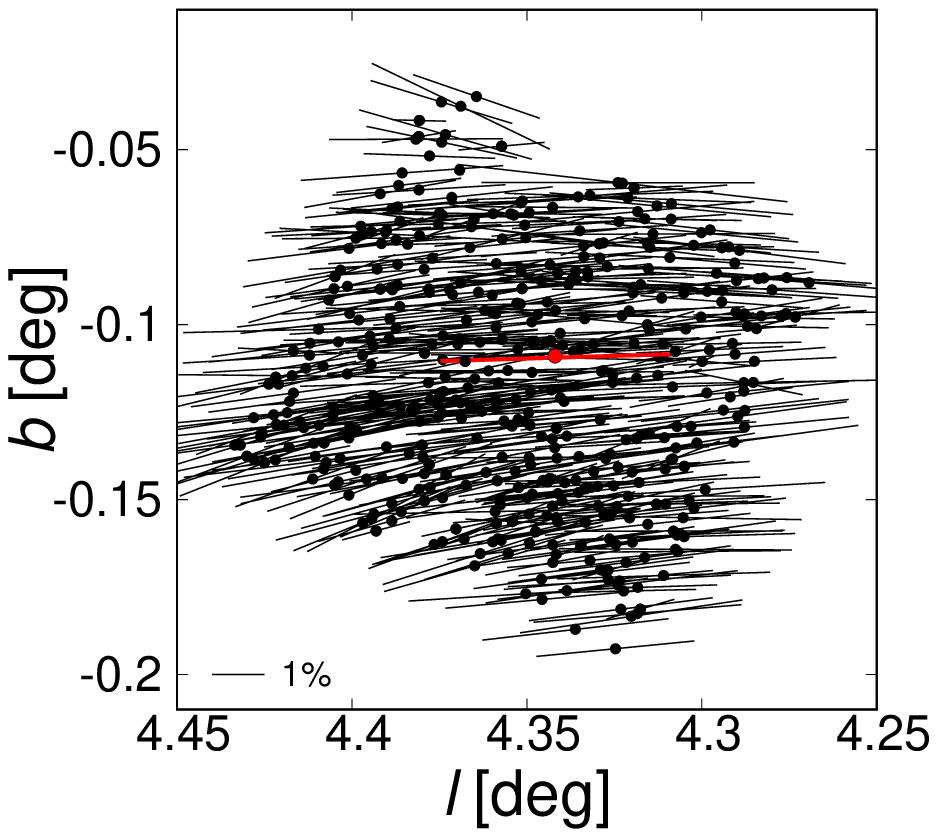}
\end{center}
\vspace{0cm}
\end{minipage}
\hspace{-3cm}
\begin{minipage}{0.5\hsize}
\begin{center}
\includegraphics[width=55mm]{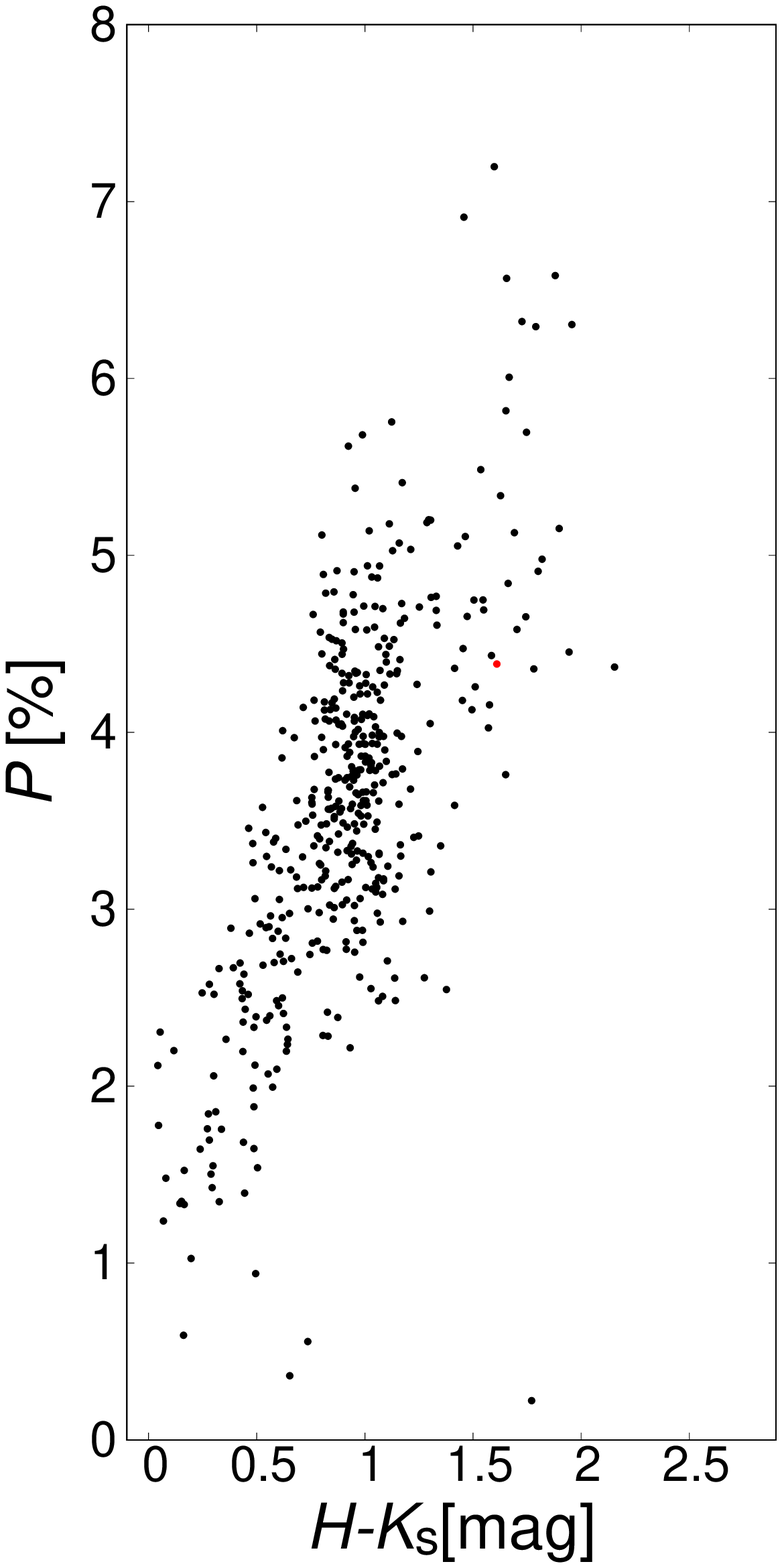}
\end{center}
\end{minipage}\\
\hspace{5.755cm}
\vspace{0cm}
\begin{minipage}{0.5\hsize}
\begin{center}
\includegraphics[width=55mm]{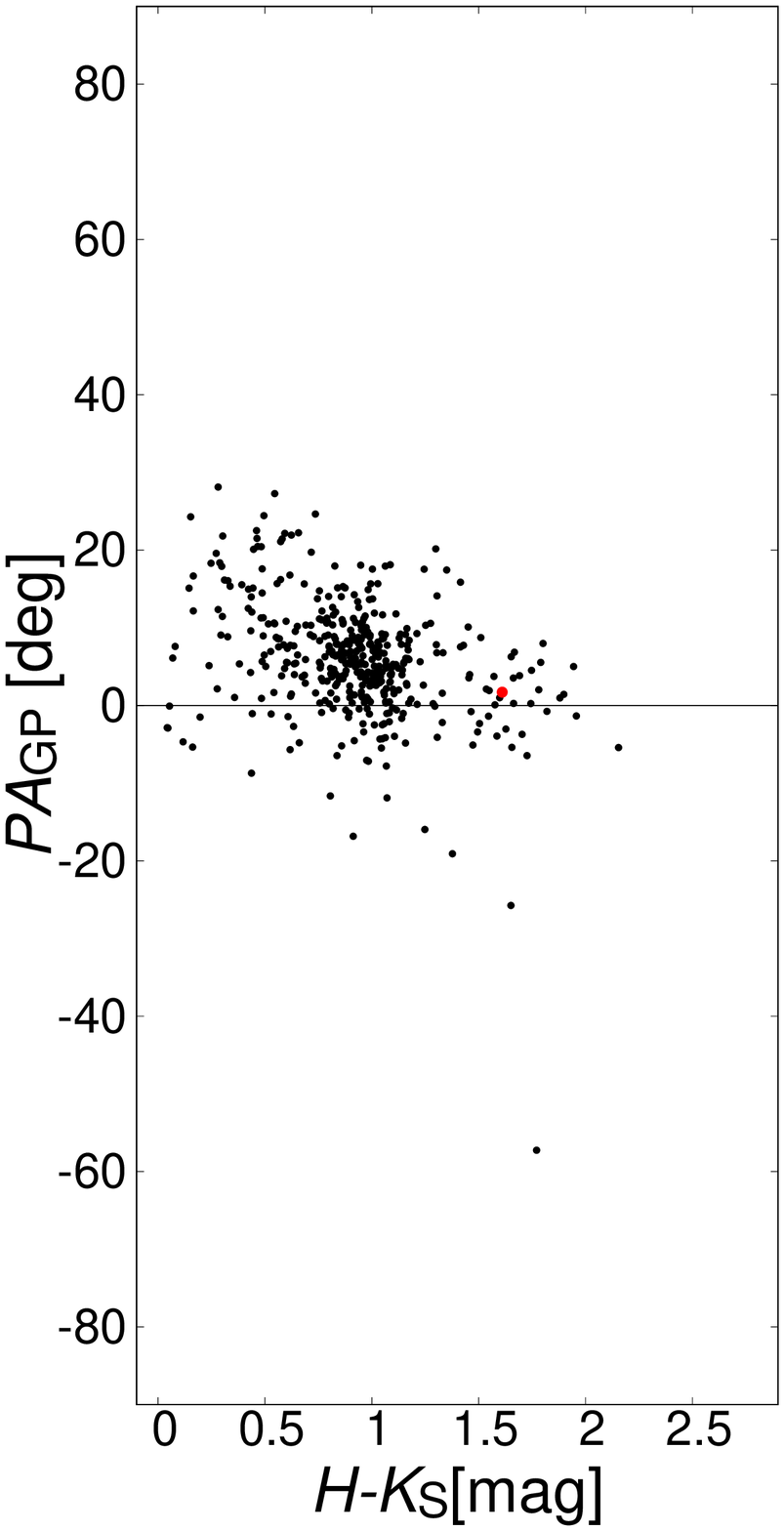}
\end{center}
\end{minipage}
\vspace{1cm}
\end{center}
\caption{Polarization of the DC35 field.  %We plot ●題名なんだから、文にはしない。だからWe plotとはしない、という言い方もできるかな。439 stars and DC35 on the three figure, and these polarizations are detected with $\delta P_{K_{\mathrm{S}}} < $ 0.5 \%. The red dot is located at the DC35. Upper left: Polarization detected stars are shown as black or red vectors centered on their stars. Vector length represent percentage linear polarization- a 1\% reference bar is shown in lower left. Vector orientations represent polarization position angle $PA_{\mathrm{GP}}$.これらは自明で書かなくて良いでしょ。 Upper right: Degree of debiased polarization P vs. $H$ - $K_{\mathrm S}$ colors for the field stars. Bottom: Position angle $PA_{\mathrm{GP}}$ vs. $H$ - $K_{\mathrm S}$ colors for the field stars. The black horizontal line represents the orientation of the Galactic plane.
The Cepheid DC35 (red dot) and 439 field stars whose polarization is detected with $\delta P_{K_{\mathrm{S}}}\, <\, $ 0.5 \%. 
Upper left: Polarization map in the Galactic coordinates.  
A 1\% reference bar is shown in lower left. 
Upper right: Degree of debiased polarization $P$ vs. $H$ $-$ $K_{\mathrm S}$ colors.
Bottom: Position angle $PA_{\mathrm{GP}}$ vs. $H$ $-$ $K_{\mathrm S}$ colors. 
The black horizontal line represents the orientation of the Galactic plane.}
\label{fig:Fig7}
\end{figure*}

DC35 %is found by ●ここはまあハッキリと過去で
was found by 
%\citet{bMatsunaga}, and this period is determined 21.41 d. 
\citet{bDekany}, and %this period is determined 13.450 d. 
its period was determined to be 13.5 d. 
%This polarization degree is more than 4\%, and the position angle $PA_{\mathrm{GP}}$ is less than $|20^{\circ}|$. 
Its polarization degree is more than 4\%, and 
the 9th largest in the 52 Cepheids.  
Its position angle $PA_{\mathrm{GP}}$ is %●less than $|20^{\circ}|$. 
%▲$1.71^{\circ}$, and 
$2^{\circ}$, 
and this $|PA_{\mathrm{GP}}|$ is the 5th smallest, 
%Therefore, the polarization vector of the DC35 is approximately parallel to the Galactic plane. 
approximately parallel to the Galactic plane.  
%Extinction $A_{K_{\mathrm S}}$ of the DC35 is relatively small among the 52 Cepheids. The polarization efficiency of the DC35 is relatively large among the 52 Cepheids. 
The extinction $A_{K_{\mathrm S}}$ to %the 
DC35 is relatively small %among the 52 Cepheids. 
among the 52 Cepheids (35th largest), 
so the polarization efficiency $P_{K_{\mathrm{S}}}/A_{K_{\mathrm{S}}}$ 
to %the 
DC35 is large.  
%The polarization efficiency of the DC35 is relatively large 「among the 52 Cepheids くどいかな.」
In the upper left panel of %figure\ref{fig:Fig7}, the field stars with $P_{\mathrm{err}}$ $\leq$ 0.5\% show the polarization vectors. 
figure \ref{fig:Fig7}, the field stars with $\delta P_{\mathrm{GP}}$ $\leq$ 0.5\% %●show the polarization vectors. 
are shown.  
%Field 
The field stars around %the 
DC35 have similar polarization %vectors 
to %the 
DC35. 
Most of 
the field stars have polarization %vectors 
approximately parallel to the Galactic plane, and %some 
only a few field stars have %the 
polarization %vectors 
%perpendicular 
slightly tilted from the Galactic plane with small polarization %degree. 
degrees.  
%
%
%
%これはアステリスクの説明とともに図のキャプションにしてしまおう
%Also, there are 18 field stars with the %debias 
%debiased polarization degrees $P_{\mathrm{deb}}$ = 0, 
%that is, such stars show no significant polarization with our $\delta P$ polarimetry.   
%
%
%
Overall, these stars are evenly distributed in the field of view, and 
we find no significant local change in the magnetic field.
%These field stars are widely distributed to the field of view, and we cannot find significant local change of the magnetic field. (8/3)
Also, the field of view toward DC35 has constant polarization efficiency (1.8~$\% / \mathrm{mag}$) over the 
0.5~mag $\leq \, H - K_{\mathrm S} \leq$ 3.0~mag range, as shown in table~1.   
We %divided 
divide 
the $H$ $-$ $K_{\mathrm S}$ data set in bins of equal size %(0.5 mag)カッコは要らんでしょう・・・ 
0.5 mag, and draw the polarization maps (figure~\ref{fig:DC35map}). 
\textcolor{black}{
Since 
most field stars have intrinsic $H-K_{\mathrm S}$ of 0 - 0.2~mag \citep{Wainscoat}, 
the $H-K_{\mathrm S}$ color can be regarded as similar to the color excess $E(H-K_{\mathrm S})$.} %長田1222
From the smallest reddening %polarization 
map %at 
a) 0.0~mag $\leq \, H - K_{\mathrm S}\, <$ 0.5~mag 
to the largest reddening map
e) 2.0~mag $\leq \, H - K_{\mathrm S}\, <$ 2.5~mag,  
coherent patterns of polarization parallel to the Galactic plane is always dominant, and 
the polarization degree $P$ grows greater 
as the color excess $E(H$ $-$ $K_{\mathrm S})$ becomes larger. %%善光191221ここは修正しなくてもいいのか？ 
In the map a), some stars have polarization slightly tilted to the Galactic plane from lower left to upper right, but 
in the maps b) - e), fewer and fewer stars have such polarization direction.  
This is apparent in the bottom panel of figure~\ref{fig:Fig7}, 
where very slow gradient of overall distribution of 
smaller and smaller position angle $PA_{\mathrm{GP}}$ 
toward  larger $H$ $-$ $K_{\mathrm S}$.  
%●少なくともfigure~\ref{fig:DC35map}では、1%の大きさをaからeの図で統一するのが良いと思います
%●●●第4段落と思っていたけど、まとめよう
%The correlation between the position angle $PA_{\mathrm{GP}}$ and $H$ - $K_{\mathrm S}$ color is shown in the bottom panel of figure\ref{fig:Fig7}. 
%To examine the dependence of the position angle $PA_{\mathrm{GP}}$ on $H$ - $K_{\mathrm S}$, we divided the $H$ - $K_{\mathrm S}$ data set in bins of equal size (0.5 mag) and calculated the mean and the standard deviation of the position angle $PA_{\mathrm{GP}}$ in each bin. 
Also, the standard deviation of position angle $PA_{\mathrm{GP}}$ 
becomes smaller from a)
%We can see a change of standard deviation position angle $PA_{\mathrm{GP}}$ at $H$ - $K_{\mathrm S}$ $\sim$ 1.0 mag. 
%On the small reddening side, the mean of position angles $PA_{\mathrm{GP}}$ are 
%$11.06^{\circ} \pm 
$8\fdg63$ to
%(0 mag $<$ $H$ - $K_{\mathrm S}$ $\leq$ 0.5 mag) and $6.62^{\circ} \pm 
b) \textcolor{black}{$5\fdg94$} and  
%(0.5 mag $<$ $H$ - $K_{\mathrm S}$ $\leq$ 1.0 mag). 
%On the other hand, the mean position angle $PA_{\mathrm{GP}}$ is $4.34^{\circ} \pm 
c) $6\fdg21$.   %%%and ◎もうここでやめましょか。論理がつながらん。, and then becomes larger again in  
%at 1.0 mag $<$ $H$ - $K_{\mathrm S}$ $\leq$ 1.5 mag, and it is $-1.51^{\circ} \pm 
%%%%%% d) $12.05^{\circ}$.   
This might be understood very well if the magnetic field in the distance corresponding to 
0.5 mag $\leq$ $H$ $-$ $K_{\mathrm S}$ $<$ 1.5 mag is well ordered 
in the Galactic plane direction. 
The position angles of these reddened stars are very close to that of the cepheid DC35.   
%at 1.5 mag $<$ $H$ - $K_{\mathrm S}$ $\leq$ 2.0 mag. There is a field stars at 2.0 mag $<$ $H$ - $K_{\mathrm S}$ $\leq$ 2.5 mag, the position angle $PA_{\mathrm{GP}}$ of this star is $-5.42^{\circ} \pm 0.68^{\circ}$. This change of standard deviation position angle $PA_{\mathrm{GP}}$ indicates the dust well aligns the same direction along the line of sight.(8/3)
%We can see a change of standard deviation position angle $PA_{\mathrm{GP}}$ at $H$ - $K_{\mathrm S}$ $\sim$ 1.0 mag. On the small reddening side, the mean of position angles $PA_{\mathrm{GP}}$ are $19.13^{\circ} \pm 43.86^{\circ}$ (0 mag $<$ $H$ - $K_{\mathrm S}$ $\leq$ 0.5 mag) and $6.59^{\circ} \pm 35.16^{\circ}$ (0.5 mag $<$ $H$ - $K_{\mathrm S}$ $\leq$ 1.0 mag). On the other hand, the mean position angle $PA_{\mathrm{GP}}$ is $-3.05^{\circ} \pm 11.62^{\circ}$ at 1.0 mag $<$ $H$ - $K_{\mathrm S}$ $\leq$ 1.5 mag, it is $-4.48^{\circ} \pm 9.40^{\circ}$ at 1.5 mag $<$ $H$ - $K_{\mathrm S}$ $\leq$ 2.0 mag, and it is $-5.56^{\circ} \pm 4.12^{\circ}$ at 2.0 mag $<$ $H$ - $K_{\mathrm S}$ $\leq$ 2.5 mag.

%%%%%%%%%%%%%%%%%%%%%%%%%%%%%%%%%%%%%%%%%%%%%%%%%%%%%%%%%%%%%%%%%%%%%%%%%%%%%%%%%%
%%%%%%%%%%%%%%%%%%%%%%%%%%%%%%%%%%%%%%%%%%%%%%%%%%%%%%%%%%%%%%%%%%%%%%%%%%%%%%%%%%
%%%%%%%%%%%%%%%%%%%%%%%%%%%%%%%%%%%%%%%%%%%%%%%%%%%%%%%%%%%%%%%%%%%%%%%%%%%%%%%%%%

\subsection{DC5}
\begin{figure*}[h]
\begin{center}
\begin{minipage}{0.5\hsize}
\begin{center}
\includegraphics[width=90mm]{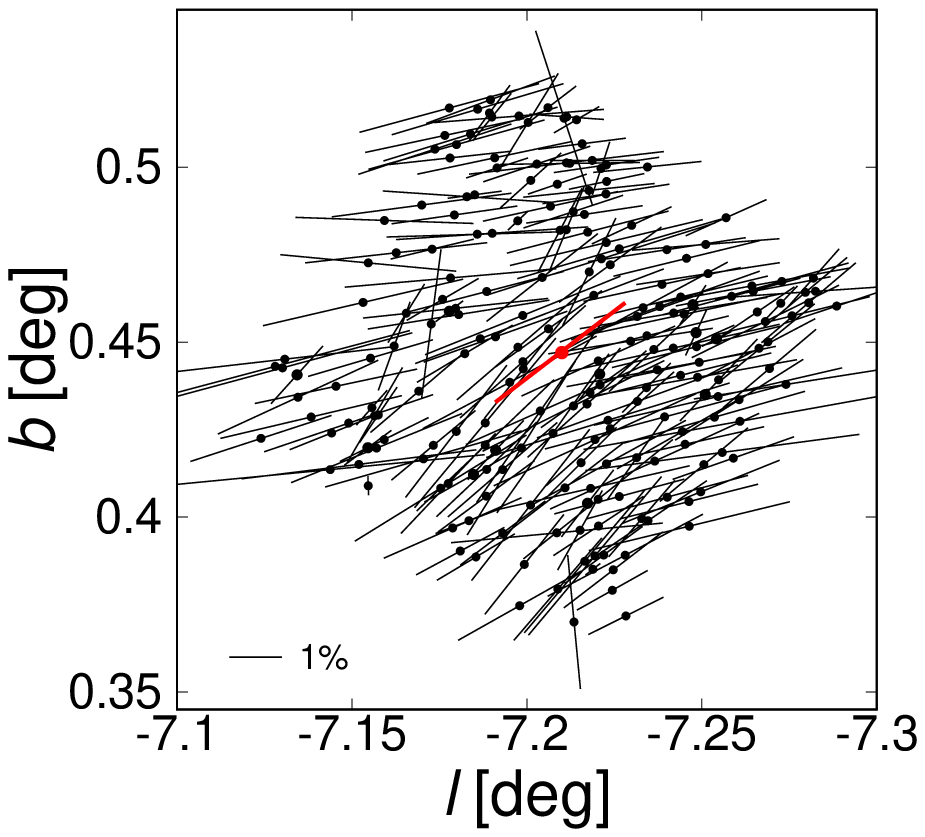}
\end{center}
\vspace{0cm}
\end{minipage}
\hspace{-3cm}
\begin{minipage}{0.5\hsize}
\begin{center}
\includegraphics[width=55mm]{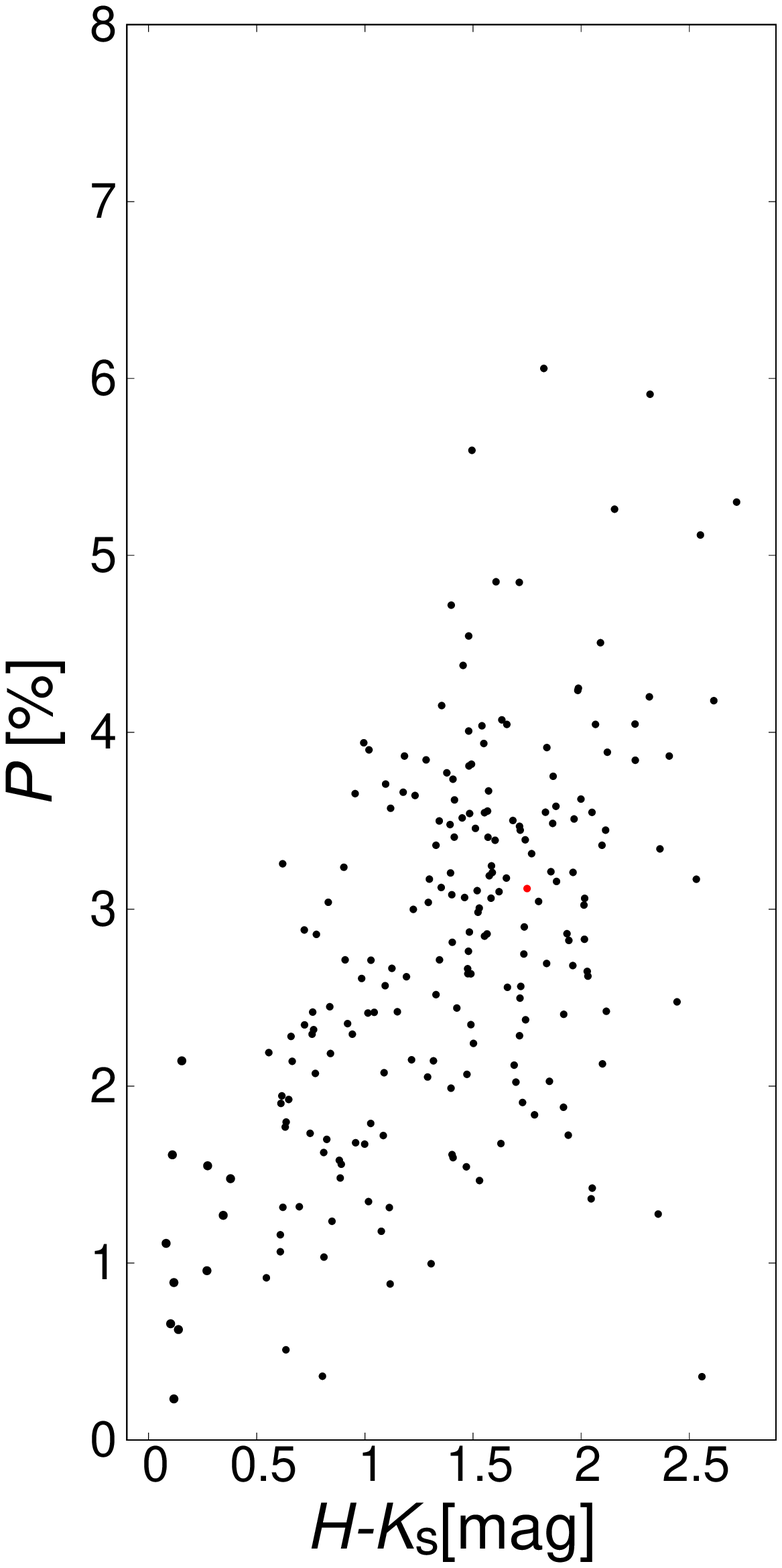}
\end{center}
\end{minipage}\\
\hspace{5.755cm}
\vspace{0cm}
\begin{minipage}{0.5\hsize}
\begin{center}
\includegraphics[width=55mm]{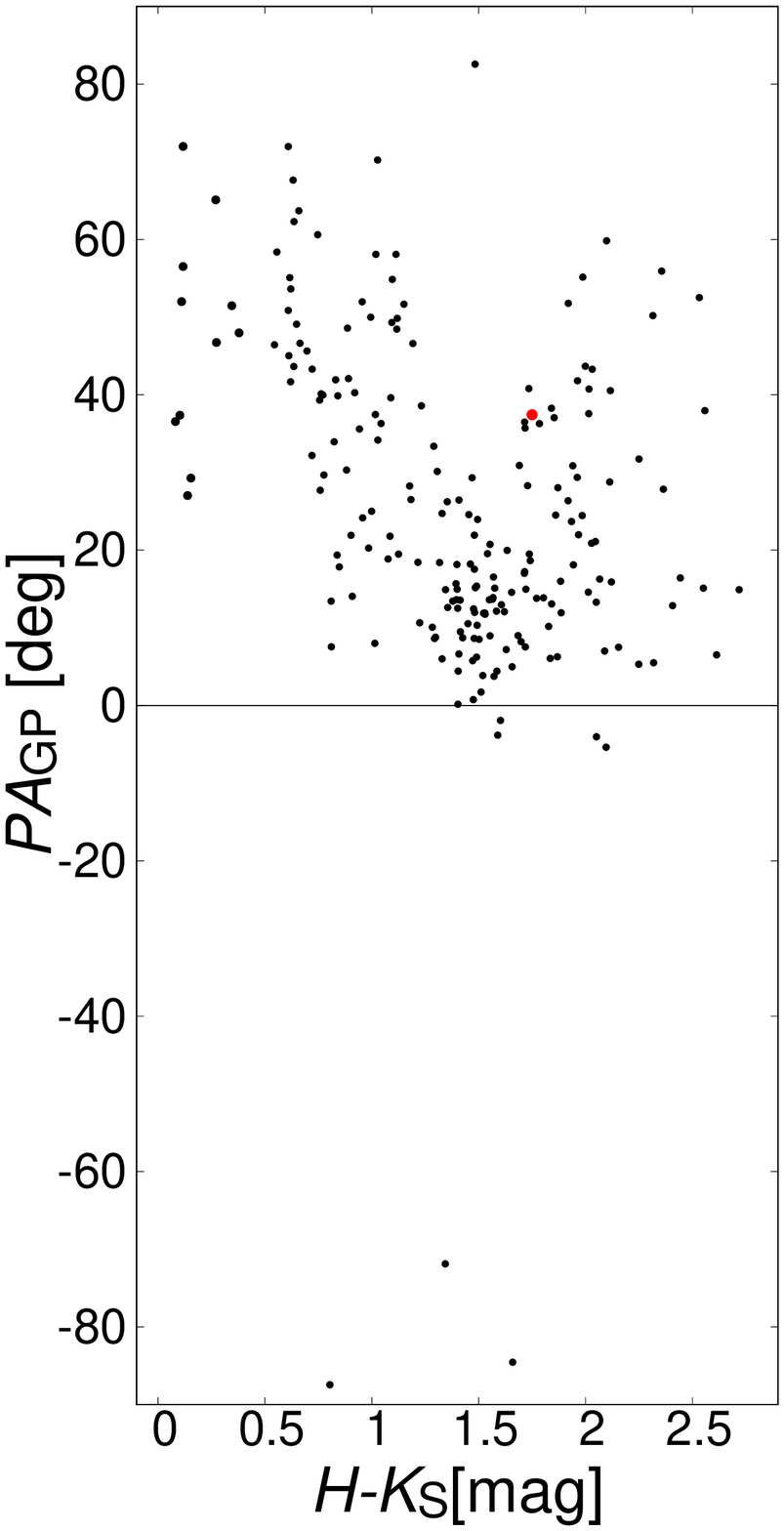}
\end{center}
\end{minipage}
\vspace{1cm}
\end{center}
\caption{DC5 (red dot) and \textcolor{black}{211}%210%(善光_191127)
 field stars. Same as figure~\ref{fig:Fig7}. }
%We plot 210 field stars and DC5 on the three figure, and these polarizations are detected with $\delta P_{K_{\mathrm{S}}} < $ 0.5 \%. The red dot is located at the DC5. See figure.~\ref{fig:Fig7} for vector and dot descriptions
\label{fig:Fig5}
\end{figure*}

DC5 was found by \citet{bDekany}, and its period was determined to be 12.3 d. 
%DC5 is found by \citet{bDekany}, and this period is determined 12.3267 d. 
%The 
Its Galactic latitude $0\fdg4$
%47^{\circ}$ %of DC5 
is largest in our sample, but 
it is moderately extincted among the 52 Cepheids with $A_{K_{\mathrm S}}$ of 2.4~mag, %2.35 mag, 
and also moderately polarized with $P$ of 3.1\%. %●26番目と22番目
%The galactic latitude of the DC5 is the farthest away from the Galactic plane. 
%This polarization degree is more than 3\%, and the position angle $PA_{\mathrm{GP}}$ is more than $20^{\circ}$. 
However, its position angle $PA_{\mathrm{GP}}$ is $37^{\circ}$, and 
its $|PA_{\mathrm{GP}}|$ is among the largest 
in the  Cepheids with 
$P/\delta P\, >\,3$ and therefore 
well determined $PA_{\mathrm{GP}}$, 
%along with DC9 and DC21.●これって良くないなあ、 DC21はS/Nが3ない  
along with DC9.  
This indicates %there is %the 
a magnetic field oblique to the Galactic plane %along 
in this line of sight. 
%Extinction $A_{K_{\mathrm S}}$ of the DC28 is relatively small among the 52 Cepheids. 
In the upper left panel of %figure\ref{fig:Fig5}, the field stars with $P_{\mathrm{err}}$ $\leq$ 0.5\% show the polarization vectors. 
figure~\ref{fig:Fig5}, many field stars around DC5 have polarization 
similar to DC5, 
both in the degree and position angle.  
At the same time, 
some other field stars have polarization parallel to the Galactic plane. 
%figure~\ref{fig:Fig5}, many field stars around DC5 have similar polarization to DC5, both in the degree and position angle. At the same time, some other field stars have polarization parallel to the Galactic plane. 
%Field stars around the DC5 have similar polarization vectors to the DC5. 
%Some field stars have the similar position angle $PA_{\mathrm{GP}}$ $\sim$ $40^{\circ}$ to the DC5, but other field stars have the position angle $PA_{\mathrm{GP}}$ parallel to the Galactic plane. 
Furthermore, a few field stars have small polarization perpendicurar to the Galactic plane.  
%Several field stars have the position angle $PA_{\mathrm{GP}}$ perpendicular to the Galactic plane and relatively small polarization vectors compared to the surrounding field stars. 
The magnetic field in this field of view is complicated.  
%This polarization map indicates there is a magnetic field which is not parallel to the Galactic plane.(8/4)
The gradient of the field stars with 
0.5~mag $ \leq \, H - K_{\mathrm S}\, \leq$ 1.5~mag is \textcolor{black}{1.8} %1.9
$\pm$ 0.3~\%/mag, but 
that of the field stars with 
0.5~mag $ \leq \, H - K_{\mathrm S}\, \leq$ 3.0~mag is 1.0 $\pm$ 0.1~\%/mag, %◎
much smaller than the former. 
%There is a positive correlation between the polarization degree and $H$ - $K_{\mathrm S}$ in the upper right panel of figure\ref{fig:Fig5}, we revealed low or high polarization efficiency to calculate the gradient. The gradient of the field stars with 0.5 mag $\leq$ $H$ - $K_{\mathrm S}$ $\leq$ 1.5 mag is 1.84, but the gradient of the field stars with 0.5 mag $\leq$ $H$ - $K_{\mathrm S}$ $\leq$ 3.0 mag is 0.97. The gradient with 0.5 mag $\leq$ $H$ - $K_{\mathrm S}$ $\leq$ 3.0 mag is smaller than the gradient with 0.5 mag $\leq$ $H$ - $K_{\mathrm S}$ $\leq$ 1.5 mag, and we consider whether the dust alignment is bad or the magnetic field orientation greatly change. (8/4)

In the bottom panel of figure~\ref{fig:Fig5}, the position angles $PA_{\mathrm{GP}}$ decrease to $\sim  \, 10^{\circ}$ %with 
in the range of field stars 0.5~mag $ \leq \, H - K_{\mathrm S}\, <$ 1.5~mag, but 
%these star to 
they increase again from $H$ $-$ $K_{\mathrm S}\, \sim \, 1.5$ mag . 
If we look at the detail, 
the mean \textcolor{black}{and standard deviation} of position angle $PA_{\mathrm{GP}}$ is 
$34^{\circ}\, \pm \, 15^{\circ}$ (0.75 mag $\leq$ $H$ $-$ $K_{\mathrm S}$ $<$ 1.25 mag), 
$15^{\circ}\, \pm \, 9^{\circ}$ (1.25 mag $\leq$ $H$ $-$ $K_{\mathrm S}$ $<$ 1.75 mag; except three field stars whose position angles $PA_{\mathrm{GP}}$ are more than 3$\sigma$ away from the mean position angle $PA_{\mathrm{GP}}$), and 
$24^{\circ}\, \pm \, 15^{\circ}$ (1.75 mag $\leq$ $H$ $-$ $K_{\mathrm S}$ $<$ 2.25 mag). 
The magnetic field orientation changes greatly at $H$ - $K_{\mathrm S}$ $\sim$ 1.5 mag %◎
to produce $10^{\circ}$ or $20^{\circ}$ polarization increase.  
We note that DC5 is more reddened ($H$ - $K_{\mathrm S}\, =\, 1.75$~mag) and has the position angle $PA_{\mathrm{GP}}\, \sim$ 37$^{\circ}$.    
%To define the $H$ - $K_{\mathrm S}$ = 1.5 mag as the base point, we divided the field stars with 0.75 mag $\leq$ $H$ - $K_{\mathrm S}$ $\leq$ 2.25 mag into three subgroups in bins of equal size (0.5 mag) and calculated the mean and the standard deviation of the position angles $PA_{\mathrm{GP}}$ in each bin. The mean of position angles $PA_{\mathrm{GP}}$ are $33.73^{\circ} \pm 15.16^{\circ}$ (0.75 mag $<$ $H$ - $K_{\mathrm S}$ $\leq$ 1.25 mag), $14.97^{\circ} \pm 9.40^{\circ}$ (1.25 mag $<$ $H$ - $K_{\mathrm S}$ $\leq$ 1.75 mag; except 3 field stars whose position angles $PA_{\mathrm{GP}}$ are more than 3 $\sigma$ away from the mean position angle $PA_{\mathrm{GP}}$), and $24.23^{\circ} \pm 15.31^{\circ}$ (1.75 mag $<$ $H$ - $K_{\mathrm S}$ $\leq$ 2.25 mag). These results indicate the magnetic field orientation change at $H$ - $K_{\mathrm S}$ $\sim$ 1.5 mag. (8/4)

The polarization maps divided into five $H$ $-$ $K_{\mathrm S}$ ranges are in 
figure~\ref{fig:DC5map}.  
In the small  $H$ $-$ $K_{\mathrm S}$ ranges a) and b), many field stars show 
$\sim \, 50^{\circ}$ $PA_{\mathrm{GP}}$, and 
in the maps c), d), and e) are mixture of such field stars and 
those polarized nearly parallel to the Galactic plane.  
\subsection{MC15}
MC15 %is 
was found by \citet{bMatsunaga}, and its period was determined to be 16.5 d.
%The MC15 is located near the Galactic center. 
%Its galactic longitude $1.9^{\circ}$ is smaller than our sample. ●どういう意味？>>意味がないので削除しておきます●意味がないとは？　それとも何か間違った文？
%>>(善光)間違ったわけではなく、こういう文章を入れて何か主張したい(この天体は銀河系中心に近いとか)かというと意味がないというつもりで書きました●こだわるようですが、「1.9度だと私達のサンプルよりも銀経が小さい」ってやっぱりどういう意味？　含めるべきでなかったということ？>>(善光_191027)52個のセファイドの中でも銀河系に近い。つまり銀経が小さいと書きたかったのです(どうしてこんなふうに書いたのがわからない、羞恥心でいっぱいだったのですが)。ですが、この内容を含めてもふくめなくても問題無いので、じゃあも書かなくてよいと思ったので削除すると書いたのです
%This polarization degree is more than 2\%, and the position angle $PA_{\mathrm{GP}}$ is more than $20^{\circ}$. 
Its polarization degree and extinction are both modest, and 
the position angle $PA_{\mathrm{GP}}$ is $-23^{\circ}$, 
having a large $|PA_{\mathrm{GP}}|$ after DC5 and a few other Cepheids.  
%It's ●It isではないpolarization degree is less than 3\%, and the 18th smallest in the 52 Cepheids.
%Its position angle $PA_{\mathrm{GP}}$ is $-22.90^{\circ}$, and this $|PA_{\mathrm{GP}}|$ is slightly large.
%This indicates a magnetic field oblique to the Galactic plane in this line of sight. 
%Extinction $A_{K_{\mathrm S}}$ of the MC28 is relatively small among the 52 Cepheids.
%The extinction $A_{K_{\mathrm{S}}}$ to MC15 is relatively small among the 52 Cepheids, so the polarization efficiency $P_{K_{\mathrm{S}}}/A_{K_{\mathrm{S}}}$ to MC15 is small.●前にも書いた気がするけど書いてなかったかな、小さいのを小さいので割ると小さいという意味がわからんです。
%
%
%In the bottom left panel of figure~\ref{fig:Fig2}, %>>(善光)ここのレファレンスは別の図になっています。
%In the bottom left panel of figure~\ref{fig:fig2},
%>>(善光_191026)修正するならば上の通りになりますが、この図はFig11の年周視差vs偏光度or位置角の図ですよね？なぜここに含める必要があるのでしょうか?この下の3つの文もそうですけど。
%most of field stars have %the 
%position angles $PA_{\mathrm{GP}}$ $\sim -30^{\circ}$. 
%Therefore, the magnetic field direction seem slightly oblique to the Galactic plane.
%
%
%
%Field stars around the MC15 have similar polarization vectors to the MC15. In the upper left panel of figure\ref{fig:Fig6}, the polarization vectors are parallel or incline to the Galactic plane in this field of view. Some field stars have the polarization vector parallel to the Galactic plane in the upper field of view. In addition, other field stars have the small polarization degree ($\leq$ 1\%) such as the left side of field. The MC15 has the similar polarization vector to the surrounding field stars. 
%In the upper left panel of figure \ref{fig:Fig6}, 
Most field stars around MC15 have similar polarization to MC15, both in the degree and position angle.
Therefore, the magnetic field direction seem to be well aligned and slightly oblique to the Galactic plane in this line of sight.

\begin{figure*}[h]
\begin{center}
\begin{minipage}{0.5\hsize}
\begin{center}
\includegraphics[width=90mm]{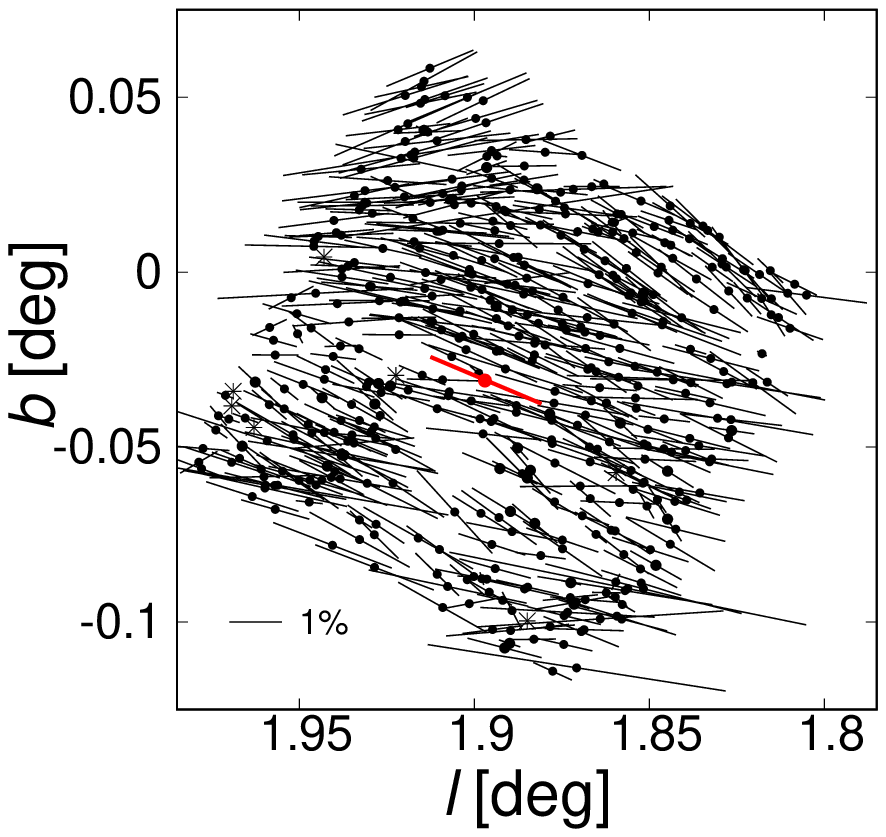}
\end{center}
\vspace{0cm}
\end{minipage}
\hspace{-3cm}
\begin{minipage}{0.5\hsize}
\begin{center}
\includegraphics[width=55mm]{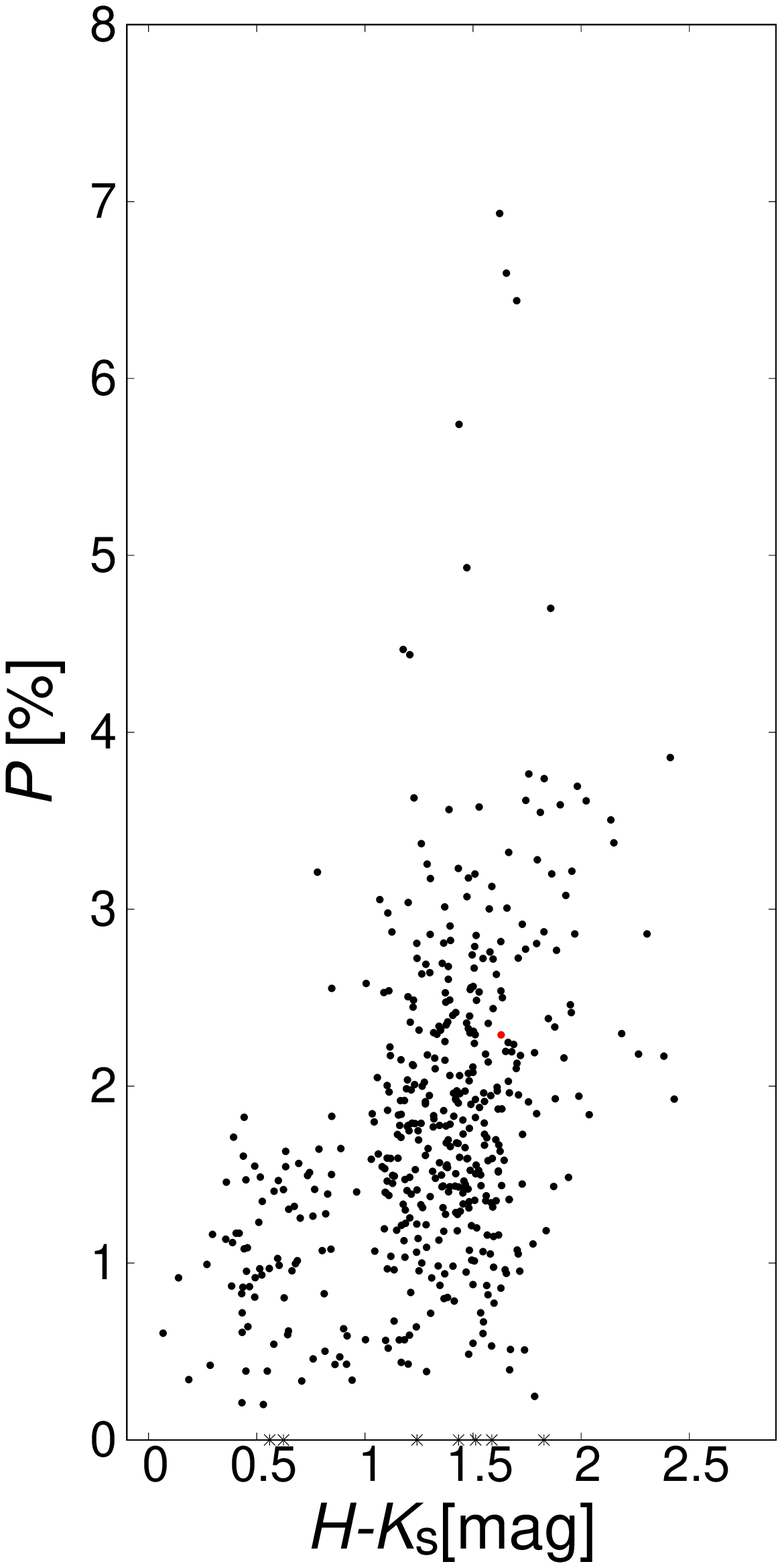}
\end{center}
\end{minipage}\\
\hspace{5.755cm}
\vspace{0cm}
\begin{minipage}{0.5\hsize}
\begin{center}
\includegraphics[width=55mm]{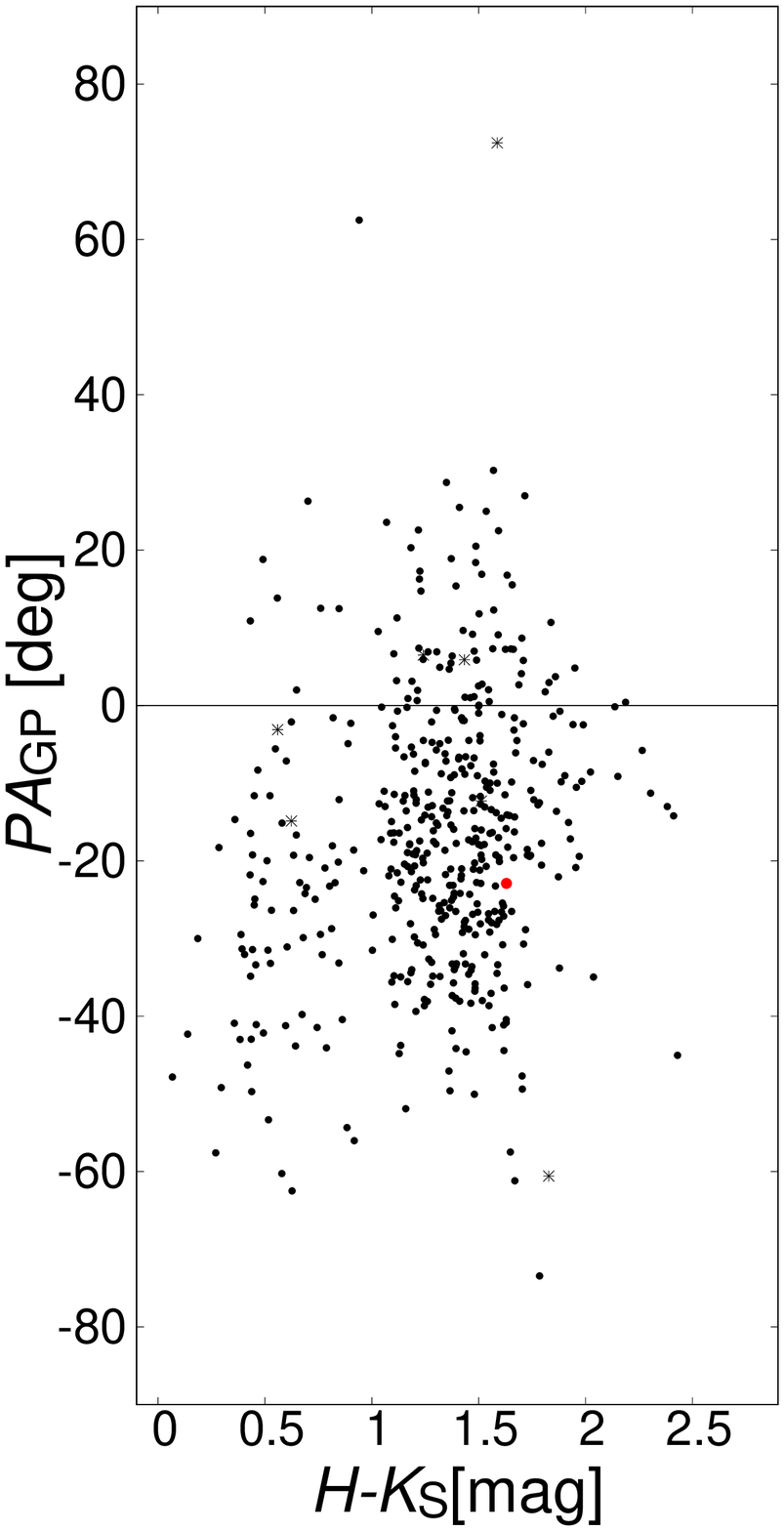}
\end{center}
\end{minipage}
\vspace{1cm}
\end{center}
\caption{MC15 (red dot) and \textcolor{black}{467}%464%(善光_191127)
 field stars. Same as figure~\ref{fig:Fig7}. }
%We plot 464 stars and MC15 on the three figure, and these polarizations are detected with $\delta P_{K_{\mathrm{S}}} < $ 0.5 \%. The red dot is located at the MC15. Black 7 asterisk means debiased polarization degree $P_{\mathrm{deb}}$ = 0\%. See figure.~\ref{fig:Fig7} for vector and dot descriptions}
\label{fig:Fig6}
\end{figure*}

\begin{figure*}[h]
\begin{minipage}{0.3\hsize}
\begin{center}
\includegraphics[width=70mm]{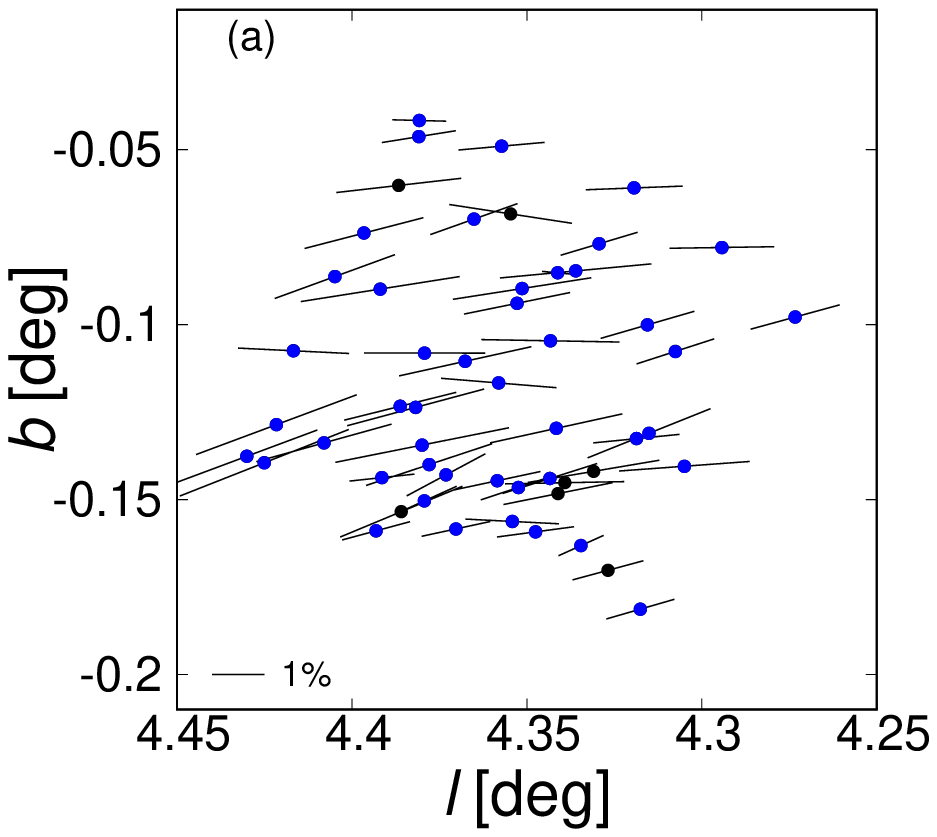}
\end{center}
\end{minipage}
\begin{minipage}{0.3\hsize}
\begin{center}
\includegraphics[width=70mm]{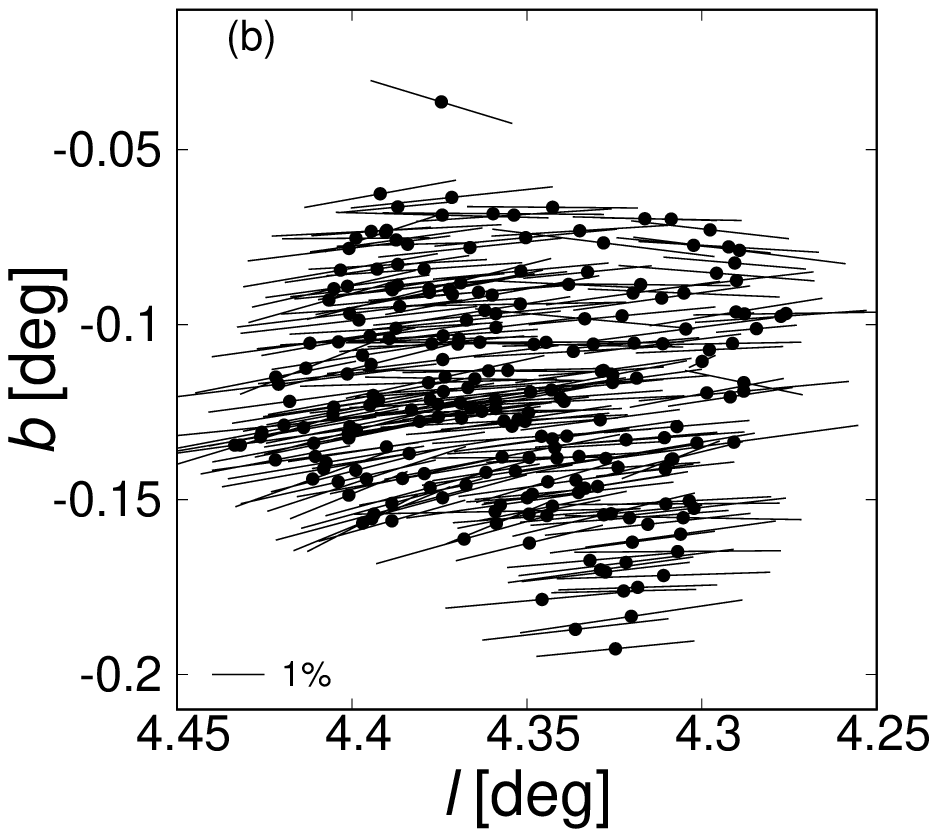}
\end{center}
\end{minipage}
\begin{minipage}{0.3\hsize}
\begin{center} 
\includegraphics[width=70mm]{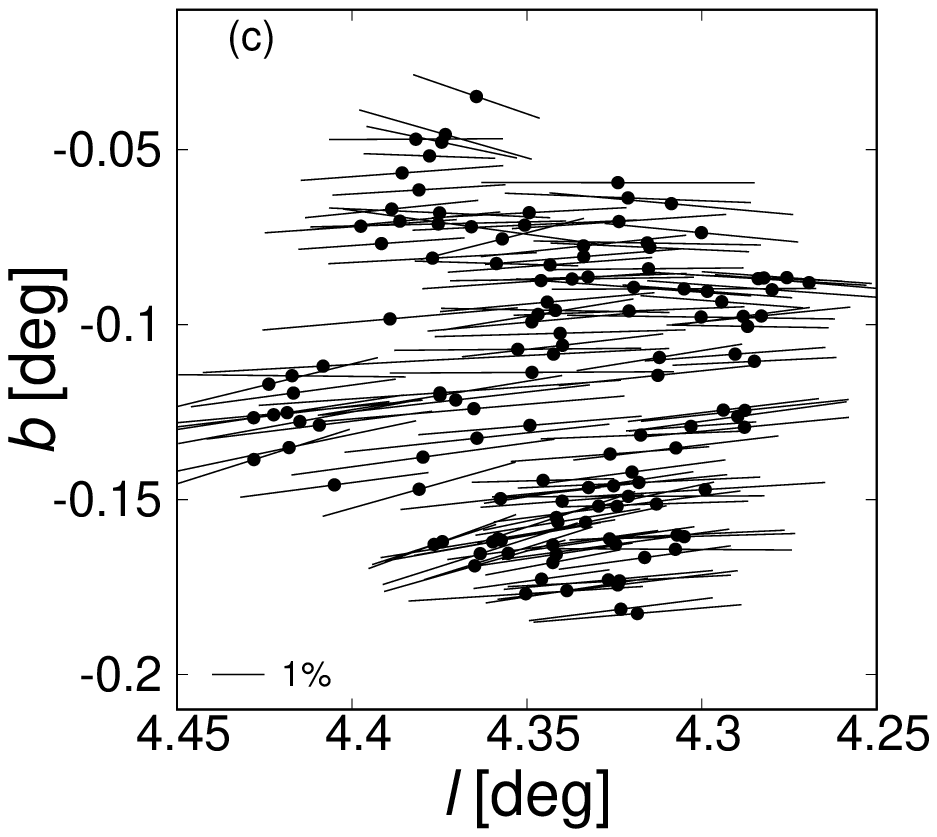}
\end{center}
\end{minipage}\\
\begin{minipage}{0.3\hsize}
\begin{center}
\includegraphics[width=70mm]{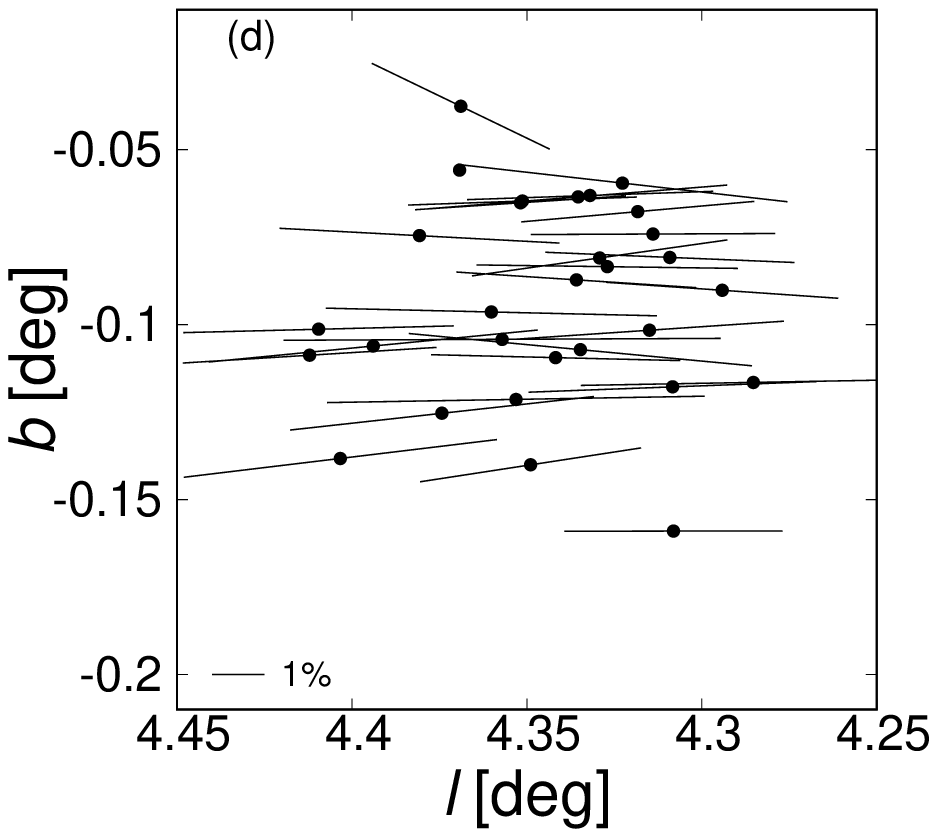}
\end{center}
\end{minipage}
\begin{minipage}{0.3\hsize}
\begin{center}
\includegraphics[width=70mm]{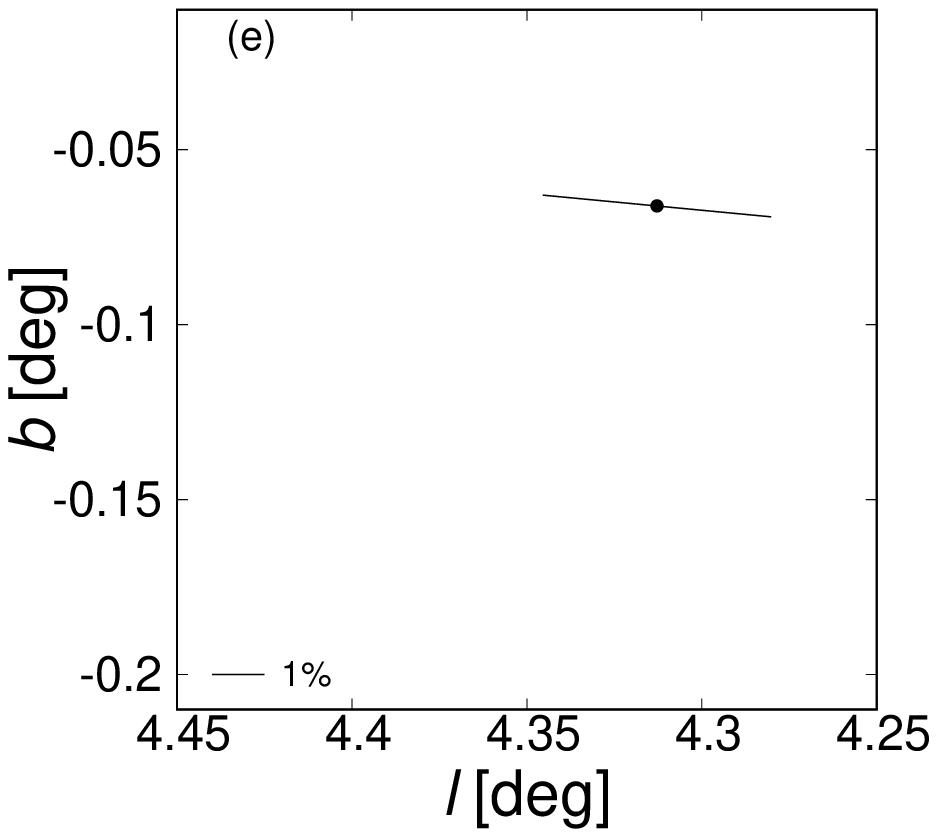}
\end{center}
\end{minipage}
\vspace{1cm}
\caption{Polarization maps of DC35 field %are ●題名だから文じゃない
divided into bins of 0.5 mag of
%the 
$H$ $-$ $K_{\mathrm S}$.  %respectively. ●対応しないrespective
%Polarization detected stars are shown as black vectors centered on their stars. Vector length represent percentage linear polarization- a 1\% reference bars are shown in lower left. ●1%の大きさを変えていたりしたら特筆すべきだろうけど、ここでは前のFigと同じで自明として不要でしょう。
(a) %In the upper left panel,●これを書かなくて良いように(a)と名付けた気がする 
$H$ $-$ $K_{\mathrm S}$ range %is 
from 0.0 to 0.5 mag. 
Blue dots %mean 
are foreground stars, 
described in Section 4. 
(b) %In the upper middle panel, 
%$H$ - $K_{\mathrm S}$ range is from 0.5 to 1.0 mag.●等号をどちらに付けるか、受験数学を思い出すような。 
$0.5\, \mathrm{mag}\, \leq \, H$ $-$ $K_{\mathrm S}\, <\, 1.0\, \mathrm{mag}$. 
(c) %In the upper right panel, 
%$H$ - $K_{\mathrm S}$ range is from 1.0 to 1.5 mag. 
$1.0\, \mathrm{mag}\, \leq \, H$ $-$ $K_{\mathrm S}\, <\, 1.5\, \mathrm{mag}$. 
(d) %In the bottom right panel, 
%$H$ - $K_{\mathrm S}$ range is from 1.5 to 2.0 mag. 
$\textcolor{black}{1.5}\, \mathrm{mag}\, \leq \, H$ $-$ $K_{\mathrm S}\, <\, 2.0\, \mathrm{mag}$. 
(e) %In the bottom middle panel, 
%$H$ - $K_{\mathrm S}$ range is from 2.0 to 2.5 mag.
$2.0\, \mathrm{mag}\, \leq \, H$ $-$ $K_{\mathrm S}\, <\, 2.5\, \mathrm{mag}$. }
\label{fig:DC35map}
\end{figure*}

\begin{figure*}[h]
\begin{minipage}{0.3\hsize}
\begin{center}
\includegraphics[width=70mm]{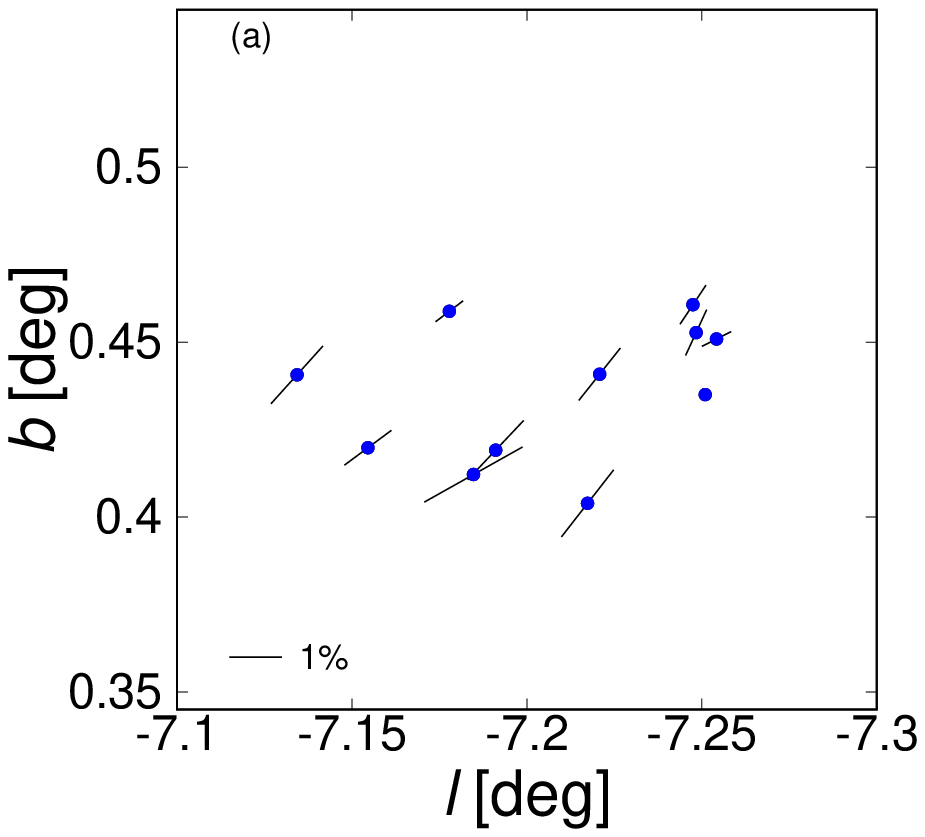}
\end{center}
\end{minipage}
\begin{minipage}{0.3\hsize}
\begin{center}
\includegraphics[width=70mm]{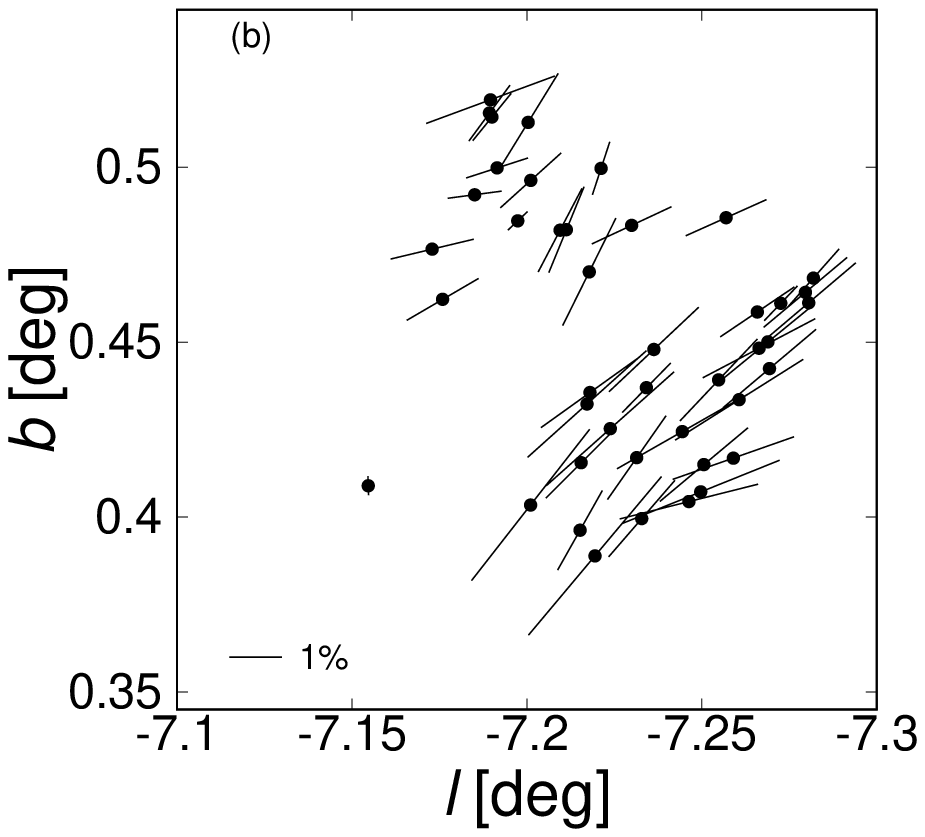}
\end{center}
\end{minipage}
\begin{minipage}{0.3\hsize}
\begin{center}
\includegraphics[width=70mm]{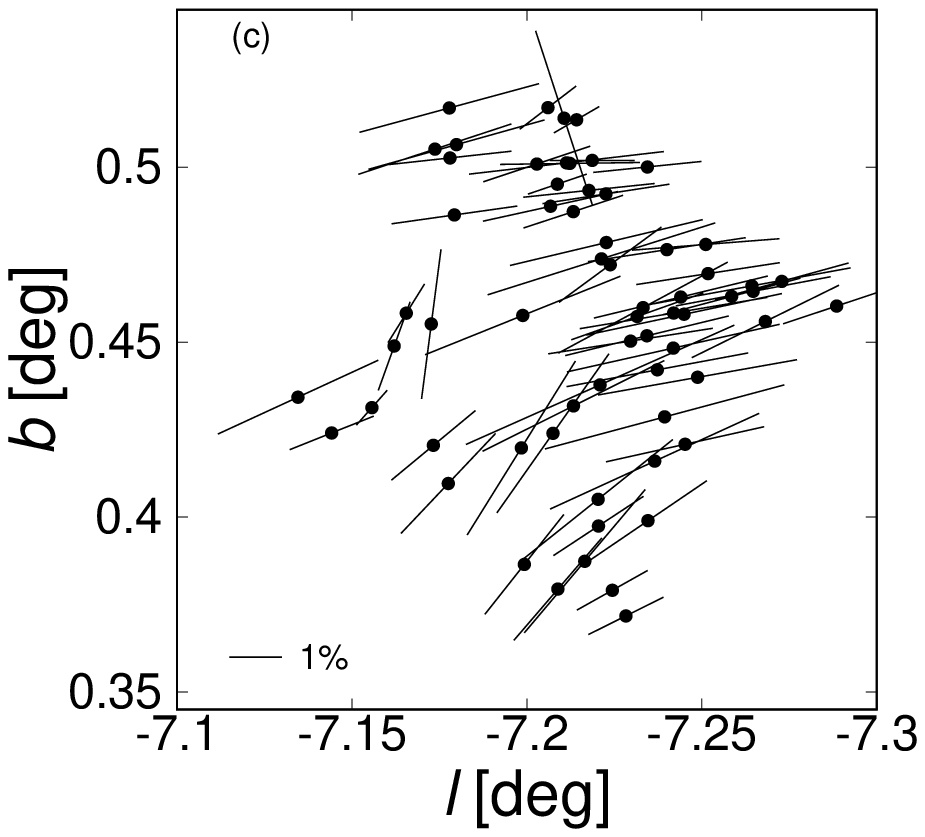}
\end{center}
\end{minipage}\\
\begin{minipage}{0.3\hsize}
\begin{center}
\includegraphics[width=70mm]{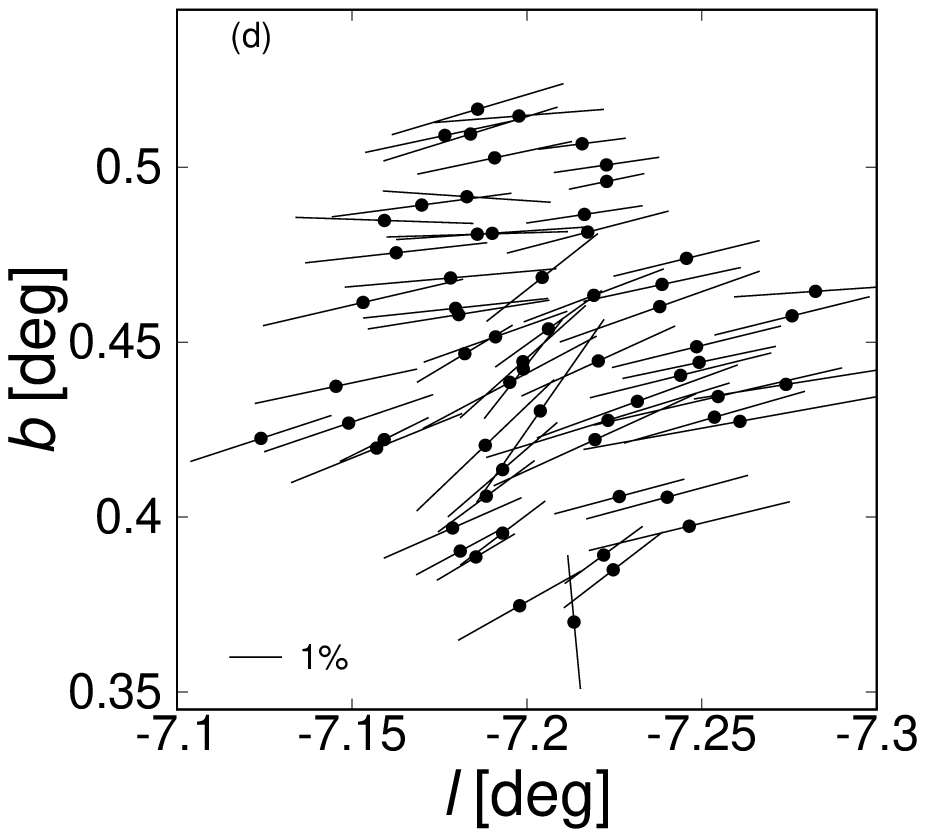}
\end{center}
\vspace{-0cm}
\end{minipage}
\begin{minipage}{0.3\hsize}
\begin{center}
\includegraphics[width=70mm]{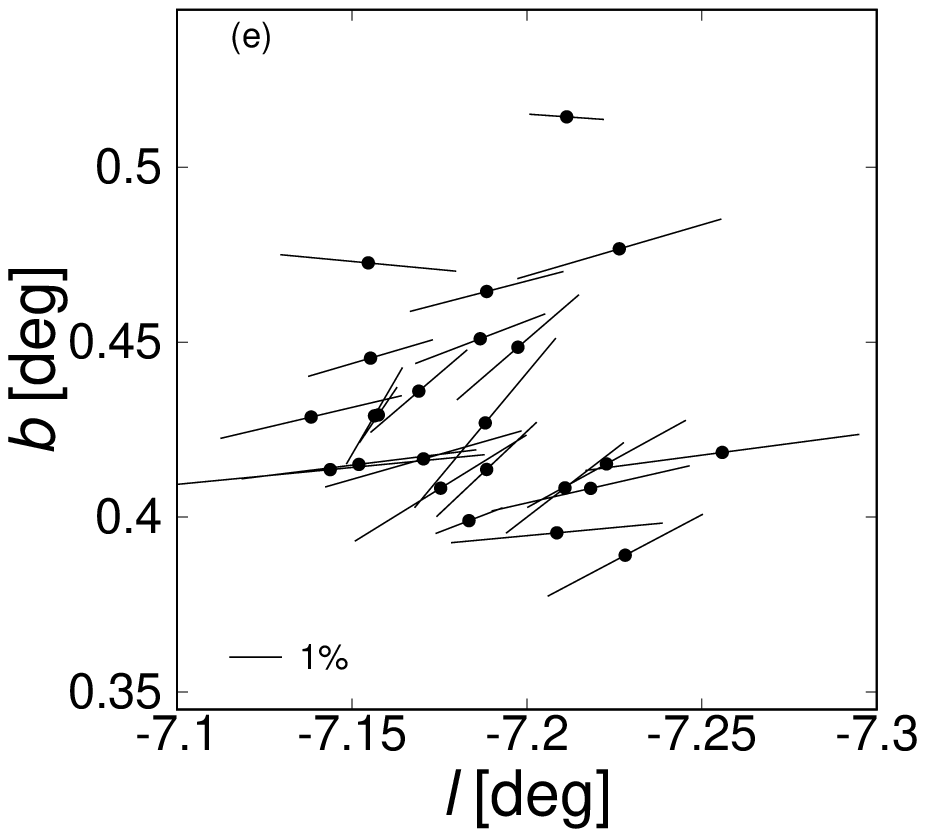}
\end{center}
\end{minipage}
\vspace{1.5cm}
\caption{Same as figure~\ref{fig:DC35map}, but for DC5.}
%Polarization maps of DC5 field are divided the $H$ - $K_{\mathrm S}$ data set in bins of 0.5 mag size, respectively. See figure\ref{fig:DC35map} for vector descriptions
\label{fig:DC5map}
\end{figure*}

However, we notice a group of field stars having polarization nearly parallel to the Galactic plane ($PA_{\mathrm{GP}}\, \sim \, 0^{\circ}$) 
in the uppermost small area in figure~\ref{fig:Fig6}.  
The line of sight to this small area seems to be different from the other part of this field.    
%At the same time, some other filed stars have polarization parallel to the Galactic plane. 
%However, %only 
We also notice one small ($<$ 2~arcmin) region of large extinction exists, with the center coordinates of ($l, b$) = ($1\fdg9$, $-0\fdg05$), 
where we have very few stars whose polarization has been measured accurately.   
%where we have very few stars deteced. 
%●なのか、polが受かるほど明るい星が少ないのか？>>2massのイメージで見たら全く星が見当たらないところと明るい天体が無いのが半々になっているようでした。周囲にある明るい天体は偏光が受かっています。●ならば○の文かなあ。
%○where we have very few stars whose polarization has been measured accurately.   
%this region radius is smaller than 2 arcmin.
%We also have checked the dark cloud catalog.There is no dark cloud nearby the MC15 field, but only one large dust extinction region exists, with the center coordinates of (l, b) = ($1.9^{\circ}$, $-0.05^{\circ}$); this region radius is smaller than 2 arcmin.
%This region does not seem to affect the polarization of the field, because there are no polarized stars within this region. ●なんでそう言えるのん？>>(善光)その領域内で天体が受かっているなら、その天体はその領域内の磁場構造をトレースしているはずだからとおもいましたけど、それを理由にするくらいならたかだか2$arc^{2}$ぐらいのサイズのdark cloudのようなものがそれよりもはるかに大きい範囲の差分解析の結果に影響を及ぼさないというふうに言うのが自然な気がしてきました。
This small region does not seem to influence the MC15 field analysis as a whole.  
%Therefore, this region does not seem to the polarization of the field.

\begin{figure*}[h]
\begin{minipage}{0.3\hsize}
\begin{center}
\includegraphics[width=70mm]{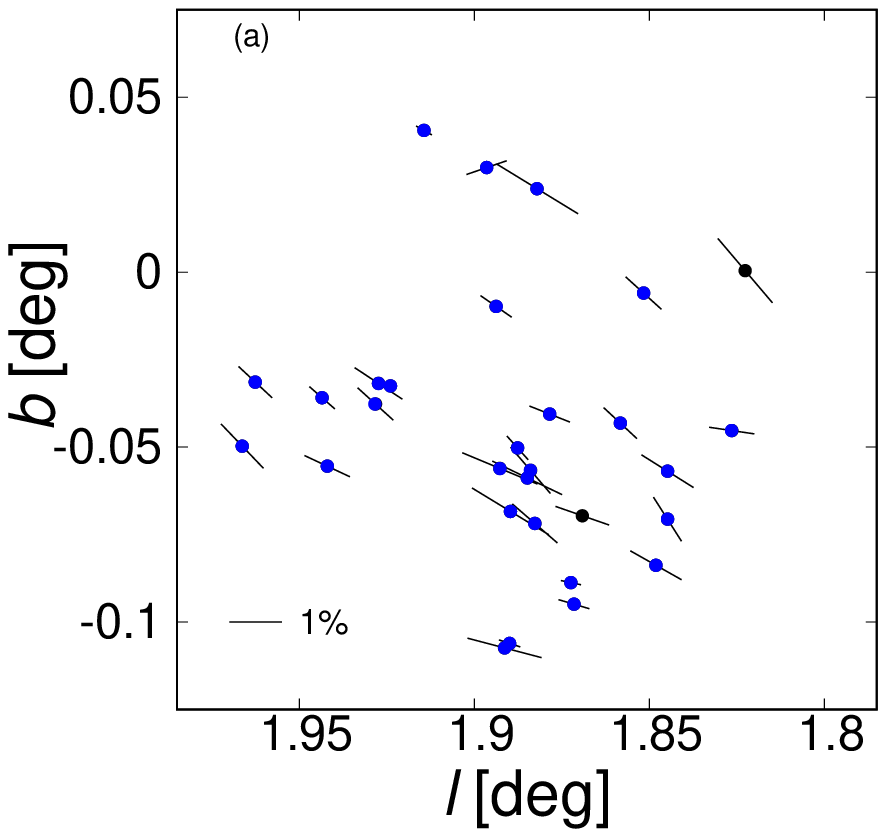}
\end{center}
\end{minipage}
\begin{minipage}{0.3\hsize}
\begin{center}
\includegraphics[width=70mm]{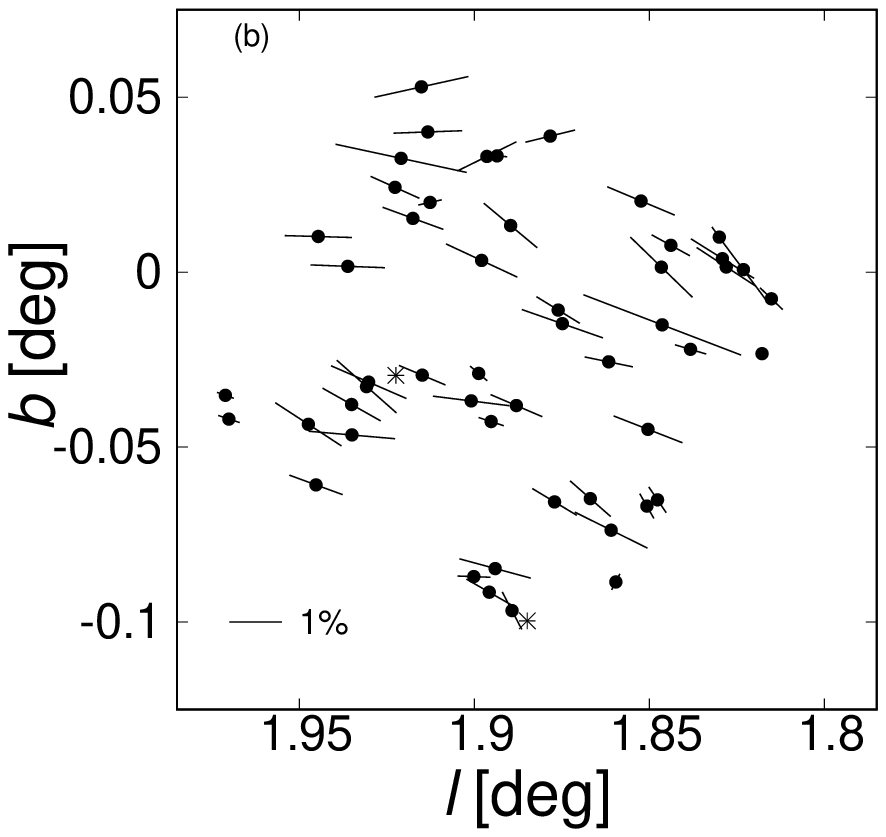}
\end{center}
\end{minipage}
\begin{minipage}{0.3\hsize}
\begin{center}
\includegraphics[width=70mm]{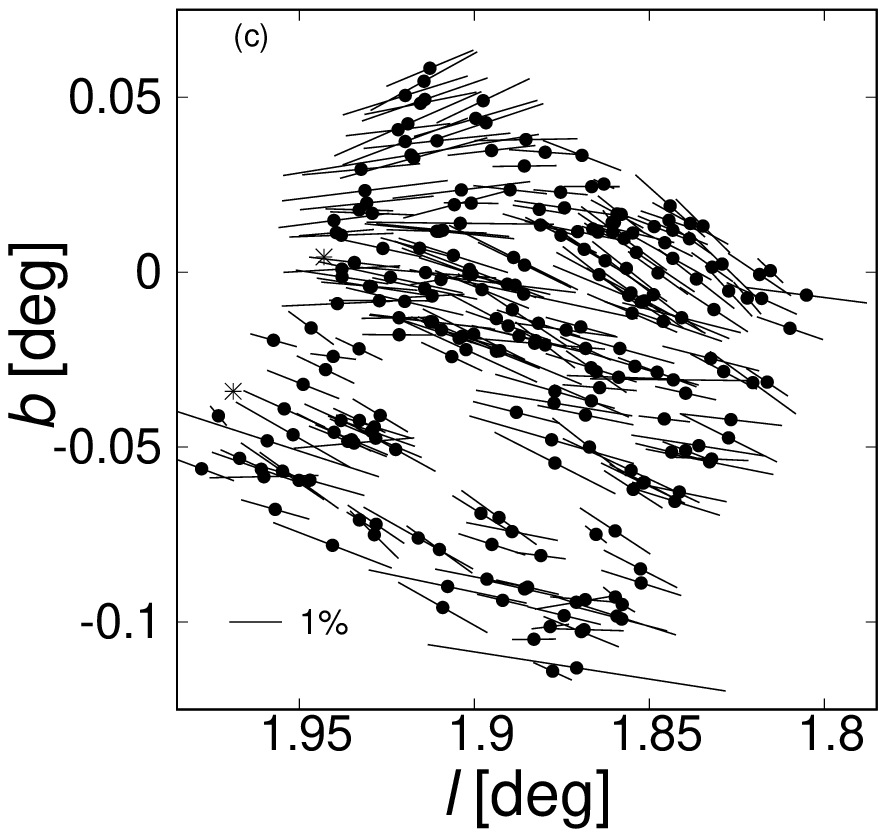}
\end{center}
\end{minipage}\\
\begin{minipage}{0.3\hsize}
\begin{center}
\includegraphics[width=70mm]{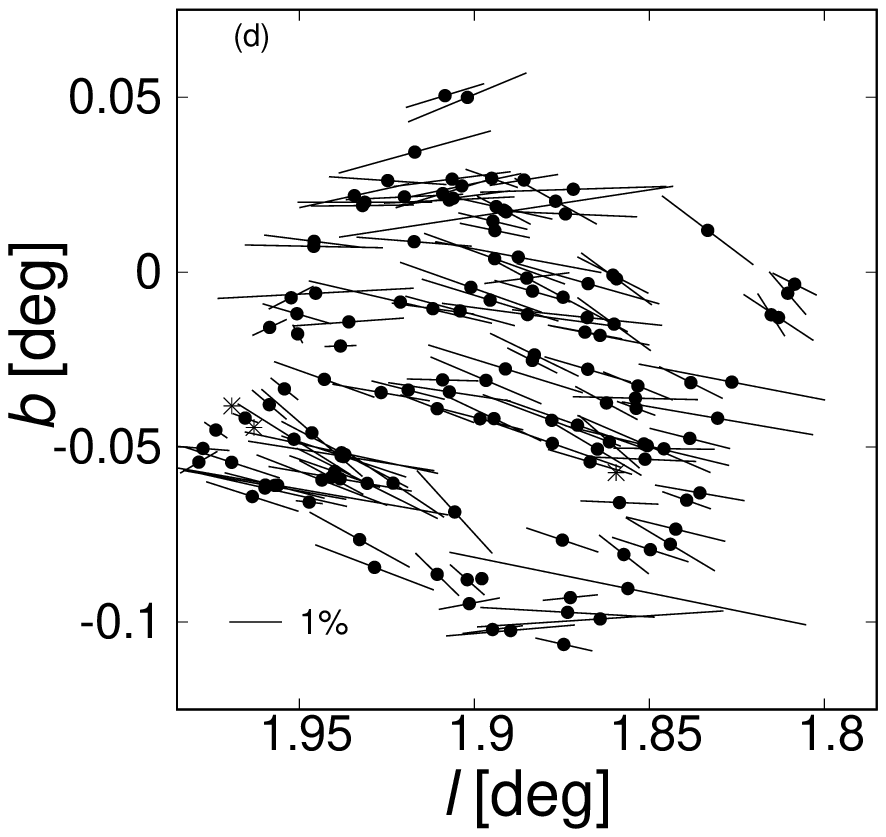}
\end{center}
\end{minipage}
\begin{minipage}{0.3\hsize}
\begin{center}
\includegraphics[width=70mm]{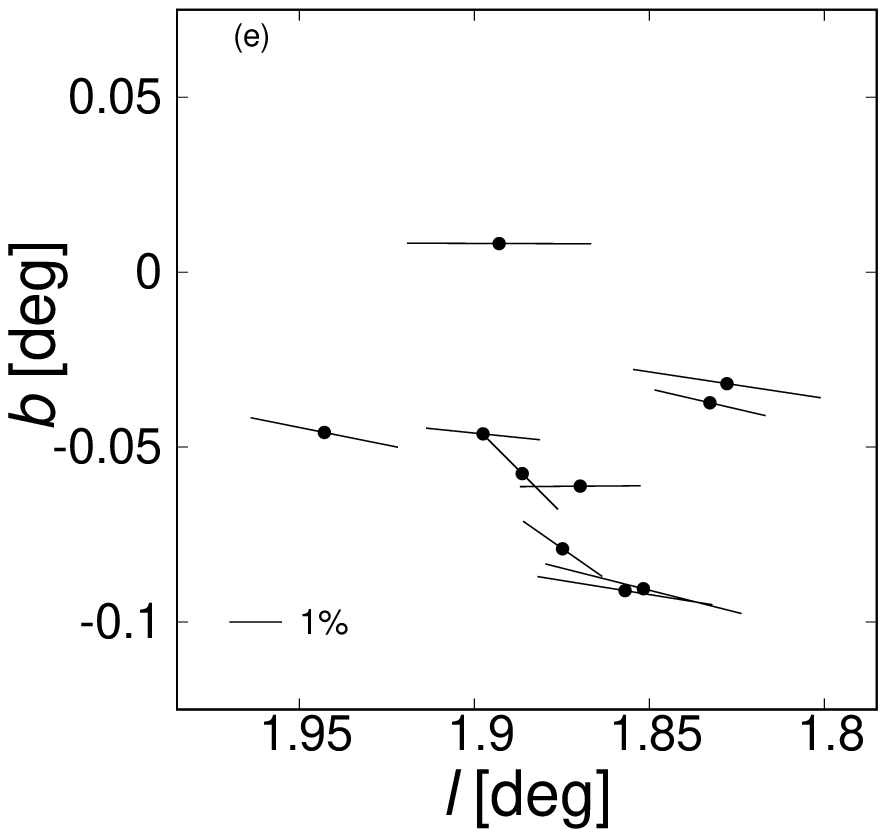}
\end{center}
\end{minipage}
\vspace{1.5cm}
\caption{Same as figure~\ref{fig:DC35map}, but for MC15.}
%{Polarization maps of MC15 field are divided the $H$ - $K_{\mathrm S}$ data set in bins of 0.5 mag size, respectively. Black asterisk means debits polarization degree $P_{\mathrm{deb}}$ = 0\%.  See figure\ref{fig:DC35map} for vector descriptions. }
\label{fig:MC15map}
\end{figure*}

In the other part, most field stars have small polarization degree $P$ of less than 3\% and are evenly distributed.  %>>(善光)ざっくり計算してみて2\%ではなく3\%というのを確認しました。全部平均偏光度は1.77で標準偏差は0.90
The gradient of the field stars with 0.5~mag $\leq \, H - K_{\mathrm S}\, \leq$ 1.5~mag  is 1.0 $\pm$ 0.2~\%/mag, 
and that of the field stars with 0.5~mag $\leq \, H - K_{\mathrm S}\, \leq$ 3.0~mag is 1.0 $\pm$ 0.1~\%/mag. %◎
The field of view toward MC15 thus has constant polarization efficiency over the 0.5~mag $< H - K_{\mathrm S}\, \leq$ 3.0~mag range.
%Also, there are y field stars with the debias polarization degrees $P_{\mathrm{deb}}$ = 0. (8/4)
%
%There is a positive correlation between the polarization degree and $H$ - $K_{\mathrm S}$ in the upper right panel of figure\ref{fig:Fig6}, and we check the polarization efficiency to calculate the gradient. The gradient of the field stars with 0.5 mag $\leq$ $H$ - $K_{\mathrm S}$ $\leq$ 1.5 mag  is 1.04, but the gradient of the field stars with 0.5 mag $\leq$ $H$ - $K_{\mathrm S}$ $\leq$ 3.0 mag is 0.9. The MC15 gradient at 0.5 mag $\leq$ $H$ - $K_{\mathrm S}$ $\leq$ 3.0 mag is similar to the DC5, but the gradient at 0.5 mag $\leq$ $H$ - $K_{\mathrm S}$ $\leq$ 1.5 mag is different from the DC5. The cause for the deteriorated polarization efficiency differ between the MC15 and the DC5. (8/4)
%
%
%In the bottom panel of figure\ref{fig:Fig6}, the position angles $PA_{\mathrm{GP}}$ decrease to $\sim$ $15^{\circ}$ in the range of field stars 0 mag < $H$ - $K_{\mathrm S}$ $\leq$ 1.0 mag. 
%The standard deviation of the position angles $PA_{\mathrm{GP}}$ is approximately constant in the range of field stars 0.0 mag < $H$ - $K_{\mathrm S}$ $\leq$ 2.5 mag.
%If we look at the detail, the mean of position angle $PA_{\mathrm{GP}}$ is $-25^{\circ} \pm 21^{\circ}$ (0.0 mag $<$ $H$ - $K_{\mathrm S}$ $\leq$ 1.0 mag), and $-15^{\circ} \pm 16^{\circ}$ (1.0 mag $<$ $H$ - $K_{\mathrm S}$ $\leq$ 2.5 mag).
We find no significant global change in the magnetic field.
%In the bottom panel of figure\ref{fig:Fig6}, the dispersion of the position angles $PA_{\mathrm{GP}}$ are shown approximately constant increasing the reddening, and the position angles $PA_{\mathrm{GP}}$ have $\sim$ $20^{\circ}$. 
%To research the dispersion of the position angles $PA_{\mathrm{GP}}$, we divided the $H$ - $K_{\mathrm S}$ data set in bins of equal size (0.5 mag) and calculated the mean and the standard deviation of the position angle $PA_{\mathrm{GP}}$ in each bin. We can see a small change of mean position angle at $H$ - $K_{\mathrm S}$ $\sim$ 1.0 mag:the mean position angle $PA_{\mathrm{GP}}$ is $-29.62^{\circ} \pm 16.84^{\circ}$ at 0 mag $<$ $H$ - $K_{\mathrm S}$ $\leq$ 0.5 mag, and it is $-21.57^{\circ} \pm 22.46^{\circ}$ at 0.5 mag $<$ $H$ - $K_{\mathrm S}$ $\leq$ 1.0 mag. On the other hand, the mean position angle is $-16.03^{\circ} \pm 15.89^{\circ}$ at 1.0 mag $<$ $H$ - $K_{\mathrm S}$ $\leq$ 1.5 mag, it is $-14.57^{\circ} \pm 17.48^{\circ}$ at 1.5 mag $<$ $H$ - $K_{\mathrm S}$ $\leq$ 2.0 mag, and it is $-14.17^{\circ} \pm 17.48^{\circ}$ at 2.0 mag $<$ $H$ - $K_{\mathrm S}$ $\leq$ 2.5 mag. These results indicate the magnetic field orientation do not change along the line of sight.(8/5)
%
The polarization maps divided into five $H$ $-$ $K_{\mathrm S}$ ranges are in figure~\ref{fig:MC15map}. 
%In the small $H$ - $K_{\mathrm S}$ ranges a) and b), many field stars show $\sim$ $25^{\circ}$ $PA_{\mathrm{GP}}$, and in the maps c), d), and e) are mixture of such field stars and those polarized nearly parallel to the Galactic plane.
%In the MC15 field of view, we also divided the $H$ - $K_{\mathrm S}$ data set in bins of equal size (0.5 mag) and draw the polarization maps (figure\ref{fig:DC5map}). In the smallest reddening polarization maps (0.0 mag $\leq$ $H$ - $K_{\mathrm S}$ $\leq$ 0.5 mag), most of polarization vectors look to be aligned oblique to the Galactic plane. In the relatively reddening polarization maps (0.5 mag $\leq$ $H$ - $K_{\mathrm S}$ $\leq$ 1.0 mag), most of polarization vectors are oblique to the Galactic plane, but few polarization vectors are vertical to the Galactic plane. The polarization map at 1.0 mag $\leq$ $H$ - $K_{\mathrm S}$ $\leq$ 1.5 mag shows an overlap of coherent patterns. In the polarization maps at 1.5 mag $\leq$ $H$ - $K_{\mathrm S}$ $\leq$ 2.0 mag, some field stars show a coherent pattern parallel to the Galactic plane, and other field stars have very small polarization degree. The polarization maps at 2.0 mag $\leq$ $H$ - $K_{\mathrm S}$ $\leq$ 2.5 mag shows few detectable polarizations.(8/15)
Some of the most reddened field stars  $H$ $-$ $K_{\mathrm S}\, > $ 2.0~mag 
might be polarized nearly parallel to the Galactic plane (see the bottom panel of figure~\ref{fig:Fig6} also), but it is evident that 
many field stars reddened similarly to MC15 have polarization similar to MC15 ($PA_\mathrm{GP}\, =\, -23^\circ$).

%%&&&&&&&&&&&&&&&&&&&&&&&&&&&&&&&&&
%%&&&&&&&&&&&&&&&&&&&&&&&&&&&&&&&&&
\begin{longtable}[h]{p{16mm}ccccccccccc}%◎にしてみました
\caption{Cepheids found by D\'ek\'any et al. (\yearcite{aDekany}, \yearcite{bDekany}) and \citet{bMatsunaga}, and 
their polarization measured in our survey. 
ID (DC and DCC stand for D\'ek\'any Cepheid and MC stands for Matsunaga Cepheid, and the number %◎
or name in their lists), Galactic longitude, Galactic latitude, $H$ and $K_{\mathrm{S}}$ mean magnitude, $K_{\mathrm{S}}$ extinction, distance, debiased polarization degree, its error, position angle $PA_{\mathrm{GP}}$ in Galactic coordinates, its error, gradients $G1$ (0.5~mag $ \leq \, H - K_{\mathrm S} \leq$ 1.5~mag), gradients $G2$ (0.5~mag $ \leq \, H - K_{\mathrm S} \leq$ 3.0~mag), and Dflag (Section 3). 
The mean magnitudes are from D\'ek\'any et al. (\yearcite{aDekany}, \yearcite{bDekany}) and from \citet{bMatsunaga} for the Cepheids discovered by them.} %●合ってる？とにかく等級なんぞは論文から取ってきたものならば
%{Catalog of classical Cepheids are calculated ●この文（主語、動詞）はどうなっとる？　で、そもそも文にはしない、図のキャプションと同様。polarization in our survey. Then listed are names●複数にしても良いかも知れないけど、単数にしました。そして最初はCepheidではなくIDとしました。, galactic longitude, galactic latitude, $HK_{\mathrm{S}}$ mean magnitudes, ●違うよねえ　distances, debiased polarization degrees, polarization degree errors, position angles $PA_{\mathrm{GP}}$, position angle $PA_{\mathrm{GP}}$ errors, and Dflag (Section 3.). The mean magnitudes are intensity-scale means of maximum and minimum, and the amplitudes refer to peak to valley variation. The definition of the names DC is given in D\'ek\'any et al. (2015), and the definition of the names MC is given in Matsunaga et al. (2013). }\\
  \hline \footnotesize
 ID &  $l$ & $b$ & $H$ & $K_{\mathrm S}$ & $A_{K_{\mathrm S}}$ & $D$ & $P$ & $PA_{\mathrm{GP}}$ & $G1$ & $G2$ & Dflag \\
  \hline
 & deg & deg & mag & mag & mag &  kpc &  \%  & deg & $\% \,/ \, \mathrm{mag}$ & $\% \,/ \, \mathrm{mag}$ \\
  \hline
    \endfirsthead
    \multicolumn{12}{c}{\footnotesize}\\ \hline
 ID &  $l$ & $b$ & $H$ & $K_{\mathrm S}$ & $A_{K_{\mathrm S}}$ & $D$ & $P$ & $PA_{\mathrm{GP}}$ & $G1$ & $G2$ & Dflag\\
  \hline
 & deg & deg & mag & mag & mag &  kpc &  \%  & deg &$\% \,/ \, \mathrm{mag}$ &$\% \,/ \, \mathrm{mag}$ \\
  \hline
  \endhead
  %----- ページの表の最下部 --------
  \hline
  \multicolumn{12}{c}{\footnotesize} \\
  \endfoot
  %----- 最終ページの表の最下部 --------
  \hline
  \multicolumn{12}{c}{\footnotesize} \\
  \endlastfoot
  \hline   \footnotesize 
DC1  &  -8.886 &  0.068  &  13.32 &  12.01  & 1.73 &  13.0  &  1.77 $\pm$  0.51 &    2.3 $\pm$  7.8 & \textcolor{black}{0.18 $\pm$ 0.13} & \textcolor{black}{0.30 $\pm$ 0.12} & 3 \\
DC2  &  -9.312 &  0.047  &  14.19 &  12.57  & 2.17 &  12.3  &  0.48 $\pm$  1.02 &  -74.9 $\pm$  25.9 & \textcolor{black}{0.33 $\pm$ 0.47} & \textcolor{black}{0.05 $\pm$ 0.19}  & 3 \\
DC3/MC1  &  -9.810 &  -0.050 &  13.20 &  12.04  & 1.53 &  11.2  &  1.88 $\pm$  0.61 &  -14.2 $\pm$  8.8 & \textcolor{black}{0.35 $\pm$ 0.36} & \textcolor{black}{0.30 $\pm$ 0.23}  & 3 \\
DC4/MC4  &  -7.509 &  -0.022 &  13.53 &  11.70  & 2.44 &  11.9  &  2.19 $\pm$  0.66 &    6.8 $\pm$  8.5 & \textcolor{black}{1.56 $\pm$ 0.13} & \textcolor{black}{0.70 $\pm$ 0.09}  & 3 \\
DC5  &  -7.210 &  0.447  &  13.61 &  11.86  & 2.35 &  12.6  &  3.12 $\pm$  0.56 &   37.4 $\pm$  5.1 & \textcolor{black}{1.76 $\pm$ 0.27} & \textcolor{black}{0.96 $\pm$ 0.13}  & 2  \\
DC6  &  -7.244 &  0.120  &  16.66 &  13.86  & 3.87 &  13.6  &  3.19 $\pm$  2.50 &  1.0 $\pm$  17.4 & \textcolor{black}{1.51 $\pm$ 0.24} & \textcolor{black}{1.38 $\pm$ 0.13}  & 1 \\
DC7/MC5  &  -7.260 &  0.064  &  15.94 &  13.93  & 2.75 &  13.6  &  3.99 $\pm$  1.82 &   27.3 $\pm$  11.8 & \textcolor{black}{1.46 $\pm$ 0.20} & \textcolor{black}{1.03 $\pm$ 0.10}  & 3 \\
DC8  &  -6.097 &  -0.119 &  15.36 &  13.41  & 2.67 &  11.9  &  1.87 $\pm$  2.91 &   47.1 $\pm$ 23.9 & \textcolor{black}{0.96 $\pm$ 0.26} & \textcolor{black}{1.23 $\pm$ 0.17} & 3 \\
DC9  &  -6.708 &  0.098  &  15.62 &  13.27  & 3.23 &  11.1  &  3.98 $\pm$  1.23 &  -40.5 $\pm$  8.5 & \textcolor{black}{0.35 $\pm$ 0.32} & \textcolor{black}{0.58 $\pm$ 0.16} & 3 \\
DC10 &  -3.175 &  -0.218 &  13.78 &  12.01  & 2.38 &  13.1  &  4.50 $\pm$  0.64 &    4.5 $\pm$  4.0 & \textcolor{black}{1.05 $\pm$ 0.25} & \textcolor{black}{0.76 $\pm$ 0.17} & 3 \\
DC11 &  -3.677 &  -0.040 &  15.83 &  13.09  & 3.79 &   8.8  &  2.96 $\pm$  1.11 &   -9.7 $\pm$  10.1 & \textcolor{black}{1.07 $\pm$ 0.30} & \textcolor{black}{0.84 $\pm$ 0.13} & 3 \\
DC12 &  -2.158 &  -0.078 &  13.96 &  12.13  & 2.47 &  12.3  &  4.33 $\pm$  1.19 &    3.5 $\pm$  7.6 & \textcolor{black}{0.00 $\pm$ 0.90} & \textcolor{black}{0.65 $\pm$ 0.22} & 3 \\
DC13 &  -2.416 &  -0.170 &  13.48 &  12.10  & 1.85 &   9.4  &  0.80 $\pm$  0.68 &  -19.8 $\pm$  18.1 & \textcolor{black}{1.23 $\pm$ 0.22} & \textcolor{black}{1.43 $\pm$ 0.16} &\textcolor{black}{1} \\
DC14 &  -1.491 &  -0.079 &  14.66 &  12.05  & 3.58 &  12.1  &  4.09 $\pm$  0.97 &  -15.2 $\pm$  6.6 & \textcolor{black}{0.71 $\pm$ 0.17} & \textcolor{black}{0.50 $\pm$ 0.11} & 3 \\
DC15 &  -1.508 &  0.186  &  13.41 &  11.62  & 2.41 &  11.6  &  4.68 $\pm$  0.78 &  -28.9 $\pm$  4.6 & \textcolor{black}{0.14 $\pm$ 0.32} & \textcolor{black}{0.13 $\pm$ 0.26} &3 \\
DC16 &  2.593  &  0.149  &  14.69 &  12.85  & 2.50 &  12.0  &  1.68 $\pm$  1.38 &  -80.1 $\pm$ 18.7 & \textcolor{black}{1.03 $\pm$ 0.37} & \textcolor{black}{0.95 $\pm$ 0.15} &3 \\
DC17/MC17 &  2.456  &  0.033  &  14.08 &  12.13  & 2.64 &  12.9  &  4.01 $\pm$  1.39 &    2.9 $\pm$ 9.2 & \textcolor{black}{1.02 $\pm$ 0.23} & \textcolor{black}{0.92 $\pm$ 0.13} & 3 \\
DC18/MC14 &  1.732  &  -0.001 &  12.91 &  11.48  & 1.90 &  11.7  &  2.54 $\pm$  0.85 &  -16.6 $\pm$ 9.0 & \textcolor{black}{0.40 $\pm$ 0.20} & \textcolor{black}{0.36 $\pm$ 0.11} & 3 \\
DC19/MC16 &  2.018  &  -0.041 &  14.62 &  12.30  & 3.17 &  10.8  &  1.57 $\pm$  0.77 &   -1.1 $\pm$  12.8 & \textcolor{black}{0.79 $\pm$ 0.21} & \textcolor{black}{0.86 $\pm$ 0.11} &3 \\
DC20 &  2.982  &  -0.185 &  15.33 &  13.08  & 3.09 &   9.8  &  4.66 $\pm$  1.50 &    5.1 $\pm$ 9.0 & \textcolor{black}{1.35 $\pm$ 0.42} & \textcolor{black}{1.20 $\pm$ 0.21} & \textcolor{black}{1} \\
DC21/MC18 &  2.838  &  -0.035 &  14.27 &  12.25  & 1.90 &  12.8  &  2.33 $\pm$  1.13 &  -38.7 $\pm$  12.4 & \textcolor{black}{-0.12 $\pm$ 0.33} & \textcolor{black}{1.28 $\pm$ 0.19} & 2 \\
DC22 &  2.767  &  0.088  &  15.47 &  13.25  & 2.76 &  11.4  &  4.58 $\pm$  2.44 &   -4.3 $\pm$  13.5 & \textcolor{black}{0.13 $\pm$ 0.49} & \textcolor{black}{0.31 $\pm$ 0.39} & 3 \\
DC23 &  4.528  &  0.075  &  14.37 &  12.64  & 2.34 &  12.3  &  2.62 $\pm$  1.06 &  -18.2 $\pm$  10.6 & \textcolor{black}{1.45 $\pm$ 0.26} & \textcolor{black}{1.19 $\pm$ 0.16} & 3 \\
DC24 &  5.822  &  0.115  &  15.31 &  13.37  & 2.65 &  11.4  &  0.95 $\pm$  0.78 &  -54.9 $\pm$  18.2 & \textcolor{black}{1.61 $\pm$ 0.17} & \textcolor{black}{1.29 $\pm$ 0.13} &1 \\
DC25 &  7.717  &  -0.147 &  14.67 &  12.45  & 3.03 &  11.2  &  4.03 $\pm$  1.03 &   -5.8 $\pm$  7.1 & \textcolor{black}{0.88 $\pm$ 0.22} & \textcolor{black}{0.89 $\pm$ 0.13} &3 \\
DC26 &  10.372 &  -0.189 &  16.52 &  13.00  & 4.89 &  10.5  &  2.27 $\pm$  1.03 &   64.8 $\pm$  11.7 &  \textcolor{black}{0.97 $\pm$ 0.65} & \textcolor{black}{0.63 $\pm$ 0.16} &3 \\
DC27 &  -9.222 &  -0.110 &  13.19 &  11.44  & 2.35 &  11.8  &  0.77 $\pm$  0.50 &   -2.9 $\pm$  15.8 & \textcolor{black}{0.17 $\pm$ 0.19} & \textcolor{black}{0.20 $\pm$ 0.12} & 3 \\
DC28 &  -7.559 &  0.086  &  12.91 &  11.09  & 2.45 &  11.0  &  1.50 $\pm$  0.30 &   27.5 $\pm$  5.5 & \textcolor{black}{0.60 $\pm$ 0.60} & \textcolor{black}{0.36 $\pm$ 0.11} & 3 \\
DC29/MC7 &  -5.302 &  0.039  &  12.09 &  11.01  & 1.40 &  11.5  &  3.83 $\pm$  0.80 &  -20.7 $\pm$  5.7 & \textcolor{black}{1.31 $\pm$ 0.19} & \textcolor{black}{1.41 $\pm$ 0.16} &1 \\
DC30 &  -1.801 &  0.103  &  13.05 &  11.72  & 1.76 &  11.6  &  2.55 $\pm$  0.43 &    6.4 $\pm$  4.7 &\textcolor{black}{1.05 $\pm$ 0.16} & \textcolor{black}{1.05 $\pm$ 0.16} & 3 \\
DC31 &  -2.371 &  -0.139 &  12.77 &  11.18  & 2.12 &  12.9  &  3.06 $\pm$  0.44 &   -2.2 $\pm$  4.1 &\textcolor{black}{1.41 $\pm$ 0.27} & \textcolor{black}{0.92 $\pm$ 0.29} &3 \\
DC32 &  2.073  &  0.047  &  13.90 &  12.35  & 2.08 &  12.4  &  2.88 $\pm$  0.88 &    6.4 $\pm$  8.8 & \textcolor{black}{1.35 $\pm$ 0.21} & \textcolor{black}{1.33 $\pm$ 0.11} & 1 \\
DC33 &  4.060  &  -0.109 &  13.16 &  11.46  & 2.28 &  12.6  &  5.03 $\pm$  0.56 &   -10.0 $\pm$  3.1 & \textcolor{black}{1.14 $\pm$ 0.21} & \textcolor{black}{1.15 $\pm$ 0.11} &\textcolor{black}{3} \\
DC34/MC21 &  3.915  &  -0.003 &  13.47 &  11.47  & 2.71 &  11.1  &  1.00 $\pm$  0.64 &    7.7 $\pm$  15.2 & \textcolor{black}{0.55 $\pm$ 0.26} & \textcolor{black}{0.90 $\pm$ 0.16} &3 \\
DC35 &  4.342  &  -0.109 &  13.03 &  11.42  & 2.15 &  12.0  &  4.39 $\pm$  0.71 &    1.7 $\pm$  4.6 & \textcolor{black}{1.80 $\pm$ 0.19} & \textcolor{black}{1.80 $\pm$ 0.14} & 1 \\
MC2  &  -9.634 &  0.058  &  12.83 &  11.54  & 1.70 &  12.8 &  1.46 $\pm$  0.54 &    0.1 $\pm$  9.9 & \textcolor{black}{0.47 $\pm$ 0.27} & \textcolor{black}{0.62 $\pm$ 0.22} & 3 \\
MC3  &  -8.405 &  -0.048 &  11.58 &  10.46  & 1.44 &  12.7 &  1.79 $\pm$  0.20 &   -1.4 $\pm$  3.1 & \textcolor{black}{1.14 $\pm$ 0.19} & \textcolor{black}{1.18 $\pm$ 0.17} &3 \\
MC6  &  -6.198 &  0.010  &  10.30 &  9.20   & 1.40 &  10.7 &  3.40 $\pm$  0.09 &  -13.2 $\pm$  0.8 & \textcolor{black}{2.39 $\pm$ 0.37} & \textcolor{black}{1.98 $\pm$ 0.27} &1 \\
MC8  &  -2.908 &  -0.062 &  12.31 &  10.77  & 2.04 &  14.1 &  3.65 $\pm$  0.34 &  -19.5 $\pm$  2.7 & \textcolor{black}{0.20 $\pm$ 0.25} & \textcolor{black}{0.46 $\pm$ 0.17} &3 \\
MC9  &  -2.058 &  -0.013 &  12.08 &  10.47  & 2.14 &  11.4 &  2.61 $\pm$  0.24 &   -9.6 $\pm$  2.6 & \textcolor{black}{0.74 $\pm$ 0.24} & \textcolor{black}{0.54 $\pm$ 0.16} &3 \\
MC10 &  -0.324 &  -0.026 &  12.14 &  10.32  & 2.44 &   7.9 &  5.52 $\pm$  0.27 &  -18.6 $\pm$  1.4 & \textcolor{black}{2.94 $\pm$ 1.30} & \textcolor{black}{0.98 $\pm$ 0.22} &3 \\
MC15 &  1.897  &  -0.031 &  12.58 &  10.95  & 2.17 &  10.9 &  2.29 $\pm$  0.30 &  -22.9 $\pm$  3.7 & \textcolor{black}{1.02 $\pm$ 0.17} & \textcolor{black}{1.00 $\pm$ 0.12} &3 \\
MC19 &  2.961  &  0.007  &  12.47 &  10.91  & 2.07 &  13.9 &  4.88 $\pm$  0.42 &    8.2 $\pm$  2.7 & - $\pm$ - & - $\pm$ - &- \\
MC20 &  3.035  &  0.010  &  11.22 &  9.86   & 1.77 &  13.9 &  2.81 $\pm$  0.16 &   -2.3 $\pm$  1.6 & - $\pm$ - & - $\pm$ - & - \\
MC22 &  5.165  &  -0.023 &  13.81 &  11.74  & 2.21 &  20.0 &  2.41 $\pm$  0.67 &  -17.1 $\pm$  7.6 & \textcolor{black}{1.30 $\pm$ 0.27} & \textcolor{black}{0.94 $\pm$ 0.16} &3 \\
MC23/DCC1 &  6.990  &  0.001  &  14.79 &  12.75  & 2.77 &  14.7 &  6.89 $\pm$  1.2 &  -13.6 $\pm$  3.5 & - $\pm$ - & - $\pm$ - & - \\
MC24/DCC1 &  6.996  &  0.001  &  14.85 &  12.73  & 2.89 &  13.8 &  3.67 $\pm$  2.3 &   -7.5 $\pm$  11.5 & - $\pm$ - & - $\pm$ - & - \\
MC25 &  7.486  &  0.059  &  13.90 &  11.99  & 2.58 &  15.0 &  2.55 $\pm$  0.92 &   17.4 $\pm$  9.9 & \textcolor{black}{0.26 $\pm$ 0.28} & \textcolor{black}{0.61 $\pm$ 0.16} &3 \\
MC26 &  7.973  &  0.003  &  11.68 &  10.35  & 1.72 &  18.2 &  2.19 $\pm$  0.29 &    5.0 $\pm$  3.6 & \textcolor{black}{0.89 $\pm$ 0.23} & \textcolor{black}{1.26 $\pm$ 0.18} &3 \\
MC27 &  8.094  &  -0.035 &  11.91 &  10.23  & 2.25 &  14.9 &  2.48 $\pm$  0.19 &  -19.8 $\pm$  2.1 & \textcolor{black}{1.43 $\pm$ 0.24} & \textcolor{black}{1.64 $\pm$ 0.16} &1 \\
MC28 &  9.068  &  -0.016 &  13.10 &  11.44  & 2.21 &  16.0 &  4.03 $\pm$  0.44 &   -9.6 $\pm$  3.1 & \textcolor{black}{1.38 $\pm$ 0.21} & \textcolor{black}{1.79 $\pm$ 0.17} &1 \\
MC29 &  9.674  &  -0.012 &  12.74 &  10.94  & 2.42 &  10.7 &  1.52 $\pm$  0.30 &   29.0 $\pm$  5.5 & \textcolor{black}{1.24 $\pm$ 0.40} & \textcolor{black}{1.14 $\pm$ 0.25} &3 \\
\hline
\label{tb:table1}
\end{longtable}
%(善光_191127)表のDflagとG1およびG2の修正を行った

\vspace{-0.5cm}

\section{The differential analysis}
Now we try to sort the stars in each field of view into three regions 
according to their distances.  
First, the %majorities 
\textcolor{black}{majority} %長田1220
of field stars belong to the bulge
\textcolor{black}{(e.g., \cite{Hatano})} %長田1223
, and 
are concentrated around the distance of 
the Galactic center $\sim \, 8$ kpc.  
These are ``bulge'' stars.  
Second, the Cepheids are in the far side of the Galaxy, and their distances are well determined.  
The field stars which suffer interstellar extinction similar to the Cepheids go into the ``background'' category.   
Last, the field stars close to us can be detected in the optical wavelengths, and %(善光191105_2)usではなくsunにするのはありだろうか？
are listed in the $Gaia$ DR2 catalog.  
They are ``foreground'' stars.   
Therefore, we %divides 
will divide each field %of view ●
into the three regions %●
a) foreground, b) bulge, and c) background, 
and %●calculates 
calculate the differential polarization between them. %(善光191105_2)among themはどうか？
%%%%

%%&&&&&&&&&&&&&&&&&&&&&&&&&&&&&&&&&
%%&&&&&&&&&&&&&&&&&&&&&&&&&&&&&&&&&
%\subsection{The method of differential analysis}

%
%●第2段落 foreground
%The Galactic plane ($|b|$ $\leq$ $1^{\circ}$) has the significant dust extinction, and Gaia parallax catalog lists almost all of the star with D $\leq$ 4 kpc. 
Our Cepheid fields are close to the Galactic plane, and 
most of the stars in the $Gaia$ DR2 catalog is confined to the distance 
$D$ $\leq$ 4 kpc.  
%We define the stars matching the Gaia parallax catalog as foreground stars.
We define the stars as the ``foreground'' stars whose parallax ranges 
determined in the $Gaia$ DR2 catalog cover the distance 4~kpc or closer. %(善光191105_2)受け身では？
%We set a color range in order to prevent the distant stars from crossmatching the Gaia DR2 stars.
%
%\cite{Hatano} constructed CMD for the field stars which are located at the Galactic center ($\sim$ $3^{\circ}$ $\times$ $2^{\circ}$) and found two weak concentrations of the stars around ($H$ - $K_{\mathrm S}$, $J$ - $H$) $\sim$ (0.1, 0.3) mag and $\sim$ (0.2, 0.7) mag. 
%
%Based on the model by \citet{Wainscoat}, the two concentrations predominantly consist of A/F dwarfs and G/K giants with small extinction. 
%
%Therefore, \cite{Hatano} defined the field stars with $H$ - $K_{\mathrm S}$ $<$ 0.4 mag as the disk stars. 
Following \citet{Hatano}, who regarded stars with $H$ $-$ $K_{\mathrm S}$ $<$ 0.4 mag as the disk stars, we set a color range of 0 mag $\leq$ $H$ $-$ $K_{\mathrm S}$ $\leq$ 0.5 mag and matched these stars with the $Gaia$ DR2 catalog, using a match radius of 1 arcsec.  
%
%We set a color range with 0 mag $\leq$ $H$ - $K_{\mathrm S}$ $\leq$ 0.5, because we can determine whether the foreground stars or not using the Gaia DR2 catalog. 
%
%The field stars with 0 mag $\leq$ $H$ - $K_{\mathrm S}$ $\leq$ 0.5 mag are matched to the Gaia DR2 sources, using a match radius of 1 arcsec. 
%
%Distances of the field stars are derived using the relation $d$ = $(1000/\omega)$ pc (where $\omega$ is the parallax in milliarcseconds).
If we confine the match to the $Gaia$ 10$\sigma$ stars 
whose ratios of parallaxes to uncertainties $\omega / \sigma_\omega$ are greater than 10, 
very few matches are found.  
%
%If we provide the field stars 10 $\sigma$ criterion (i.e. ratios of parallaxes to their uncertainties, $\omega/\sigma_{\omega} \geq 10$), they have the accuracy distance but do not have the $K_{\mathrm{S}}$ band polarization. 
%
Therefore we use all the $Gaia$ stars 
whose ratios of parallaxes to uncertainties are greater than 1.  
%We provide the foreground stars with the $1\sigma_{\mathrm{\omega}}$ criterion ($\omega/\sigma_{\mathrm{\omega}} > 1$) in order to prove the polarization as a function of distance, although this criterion has the large distance error. 
%
We remove the field stars whose distance are more than 4 kpc 
even if we assume $\omega + \sigma_\omega$ as their parallaxes. 
%In addition, we remove field stars satisfied with d-$1000/(\omega + \sigma_{\omega})$ $\geq$ 4000 pc criterion, and we choose these having the potential to be foreground at least. (8/17)

\begin{figure*}[h]
\begin{minipage}{0.3\hsize}
\begin{center}
\includegraphics[scale=0.55]{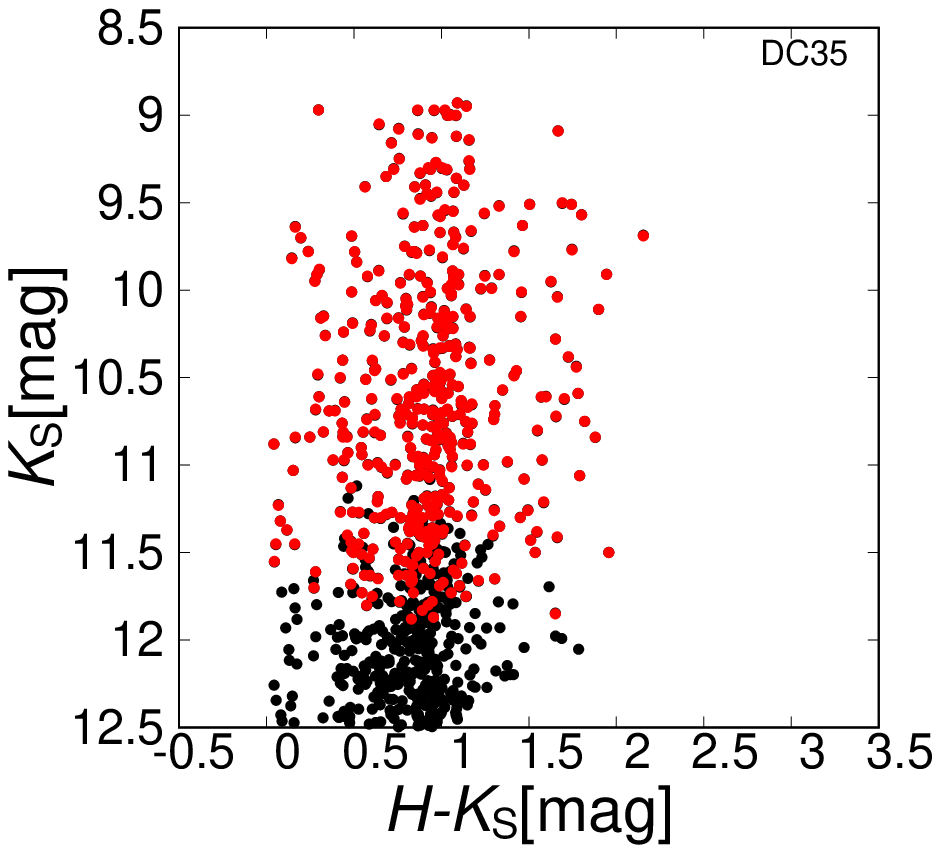}
\end{center}
\end{minipage}
\begin{minipage}{0.3\hsize}
\begin{center}
\includegraphics[scale=0.55]{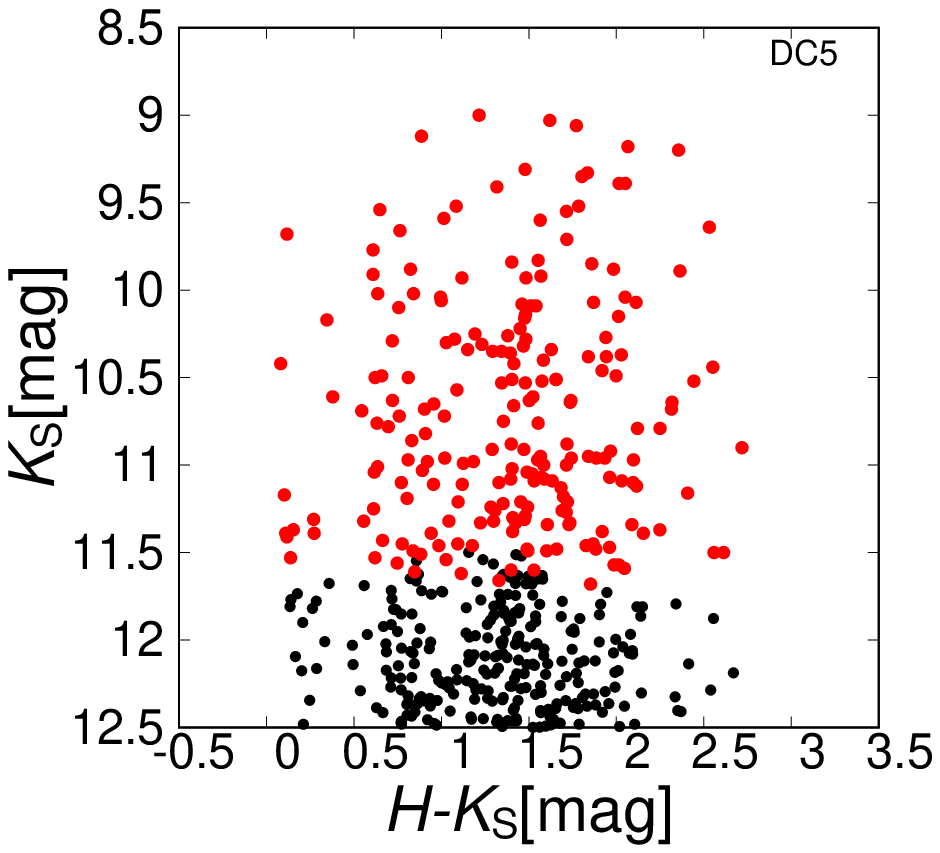}
\end{center}
\end{minipage}
\begin{minipage}{0.3\hsize}
\begin{center}
\includegraphics[scale=0.55]{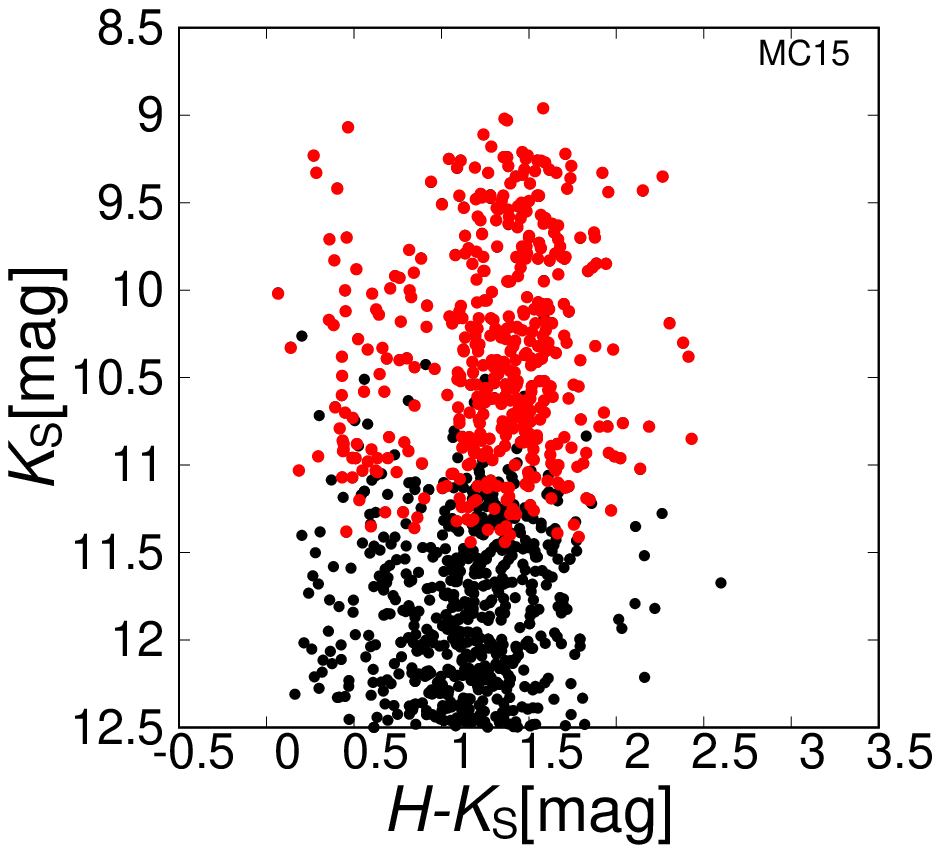}
\end{center}
\end{minipage}
\vspace{1.5cm}
\caption{$K_{\mathrm{S}}$ vs. $H$ $-$ $K_{\mathrm S}$ %●δですか？
color magnitude diagram for stars with %$K_{\mathrm S} < 0.06$ mag. Field names are shown in upper right
$\delta K_{\mathrm S} < 0.06$~mag. 
Red dots are stars whose polarization is measured with $\delta P_{K_{\mathrm{S}}} < $ 0.5 \%,  
and black dots are those with $\delta P_{K_{\mathrm{S}}} > $ 0.5 \%.
%Stars whose polarization is measured with  $\delta P_{K_{\mathrm{S}}} < $ 0.5 \% (red dots) are on the whole %brighter than 
%$K_{\mathrm S} \lesssim 11.5$~mag. %▲いや、これはこれで正しい文かもしれないけど、不自然ですよねえ●その情報がないとあんまり意味がない気がする。で、何で８月中頃に（ヒストグラムの赤をやめるとともに）その情報をやめちゃったんでしたっけ。 
Field names are shown in upper right}
\label{fig:Fig4}
\end{figure*}

%●第3段落 background
%We define the field stars which have similar distance to the Cepheids as background stars, because these stars and Cepheids are located on the far side of the Galactic bulge. 
We regard the field stars whose distances are similar to the Cepheids 
as the ``background'' stars. 
First, we compare the observed color magnitude diagrams 
of the field stars in figure~\ref{fig:Fig4} %(善光191105_2)どれと比べるのかを明示する必要があるか？
with the Galaxy stellar model by \citet{Wainscoat}. 
The stellar types whose $H$ $-$ $K_{\mathrm{S}}$ color and $K_{\mathrm{S}}$ magnitude 
are compatible with the observed ranges, and also the distance to the Cepheids are 
M5$\rm I\hspace{-.1em}I \hspace{-.1em}I$, 
M6$\rm I\hspace{-.1em}I \hspace{-.1em}I$ and 
A-G $\rm I$-$\rm I\hspace{-.1em}\rm I$.  
%
%However, since we can not directly calculate the distance of background stars like the foreground stars, we have to use physical parameter instead of the distance. 
%The $H$ - $K_{\mathrm{S}}$ color is used by the distance indicator (e.g. \cite{bNishiyama}), and we use this. 
%Our method uses the distance modulus and the extinction of the Cepheid as a base point, and we estimate the typical mean $H$ - $K_{\mathrm{S}}$ color of the background stars. 
%We calculate the $H$ and $K_{\mathrm S}$ magnitude of each stellar type combining distance modulus of the Cepheid, extinction $A_{K_{\mathrm{S}}}$ of the Cepheid and absolute magnitudes of each stellar type referenced by \citet{Wainscoat}. We plot the $K_{\mathrm S}$ magnitude and the $H$ - $K_{\mathrm S}$ color of each stellar type on the color-magnitude diagram (CMD) of the observed stars, and we choose the stellar types having the similar $K_{\mathrm S}$ magnitude and $H$ - $K_{\mathrm S}$ color to the observed sources. In this result, M5$\rm I\hspace{-.1em}I \hspace{-.1em}I$, M6$\rm I\hspace{-.1em}I \hspace{-.1em}I$ and A-G$\rm I\hspace{-.1em}I$ have $K_{\mathrm{S}}$ band and $H$ - $K_{\mathrm{S}}$ matching the observation result in the all field of views. M4$\rm I\hspace{-.1em}I \hspace{-.1em}I$ have $K_{\mathrm{S}}$ band and $H$ - $K_{\mathrm{S}}$ matching the observation result in several field of views. (8/17)
%
We calculate the mean $H$ $-$ $K_{\mathrm{S}}$ color using these stellar types 
and stellar densities in \citet{Wainscoat}.  
The %typical 
mean background $H$ $-$ $K_{\mathrm S}$ color of the 
DC35 field is 1.91 mag, 
DC5 is 2.03 mag, and 
MC15 is 1.88 mag;  
we regard all the stars in the $H$ $-$ $K_{\mathrm S}$ color range of 0.5 mag 
around them as 
the ``background'' stars.  
%We calculate the mean $H$ - $K_{\mathrm{S}}$ color using the aforementioned stellar types. M4$\rm I\hspace{-.1em}I \hspace{-.1em}I$, M5$\rm I\hspace{-.1em}I \hspace{-.1em}I$, and M6$\rm I\hspace{-.1em}I \hspace{-.1em}I$ have the small different $H$ - $K_{\mathrm{S}}$ color from each other, but A-G$\rm I\hspace{-.1em}I$ is bluer about 0.3 mag than those. A-G$\rm I\hspace{-.1em}I$ type star is approximately hundredfold less stellar dense than the other types, and we defined the mean $H$ - $K_{\mathrm{S}}$ wighted by the stellar dense as the typical $H$ - $K_{\mathrm{S}}$. In this paper, We choose the field stars within plus or minus 0.25 mag from the typical mean $H$ - $K_{\mathrm{S}}$ as the background stars, and this color wide corresponds to the wide of foreground $H$ - $K_{\mathrm S}$ color (0.0 mag $\leq$ $H$ - $K_{\mathrm{S}}$ $\leq$ 0.5 mag). The typical background $H$ - $K_{\mathrm S}$ color of DC35 is 1.91 mag, DC5 is 2.03 mag, and MC15 is 1.88 mag. (8/17)

%%%%%%%%%%%%%%%%%%%%%%%%%%%%%%%%%%%%%%%%%%%%%%%%%%%%

\begin{figure*}[h]
\begin{minipage}{0.5\hsize}
\begin{center}
\includegraphics[scale=0.7]{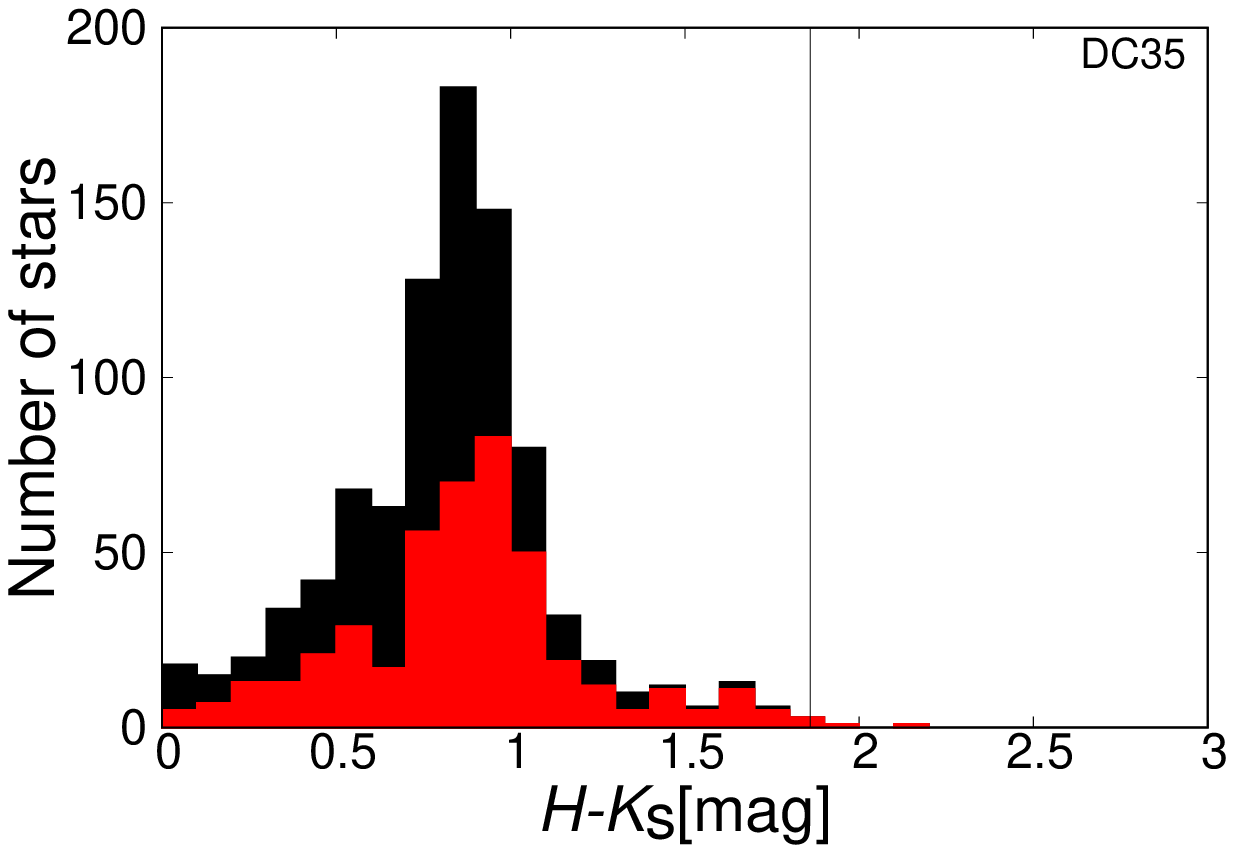}
\end{center}
\end{minipage}
\begin{minipage}{0.5\hsize}
\begin{center}
\includegraphics[scale=0.7]{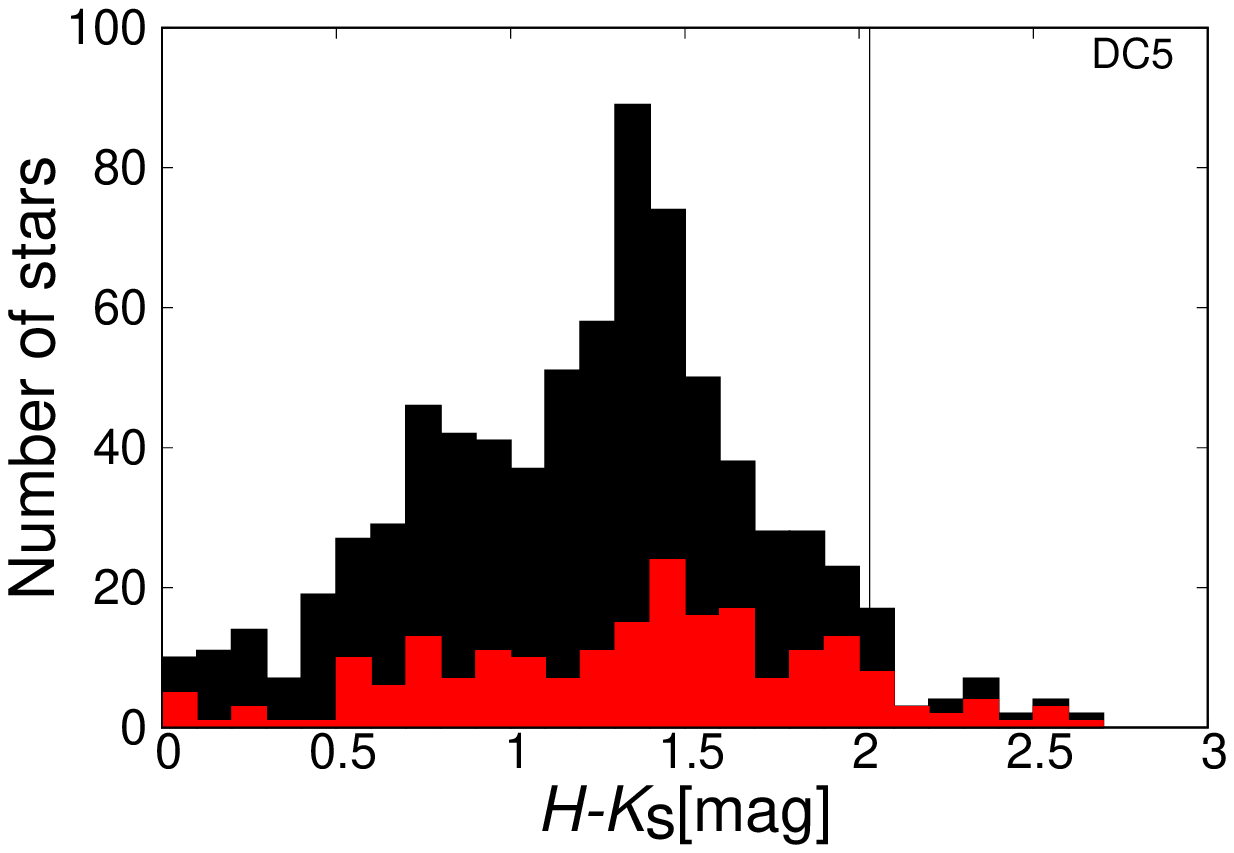}
\end{center}
\end{minipage}
\begin{minipage}{0.5\hsize}
\begin{center}
\includegraphics[scale=0.7]{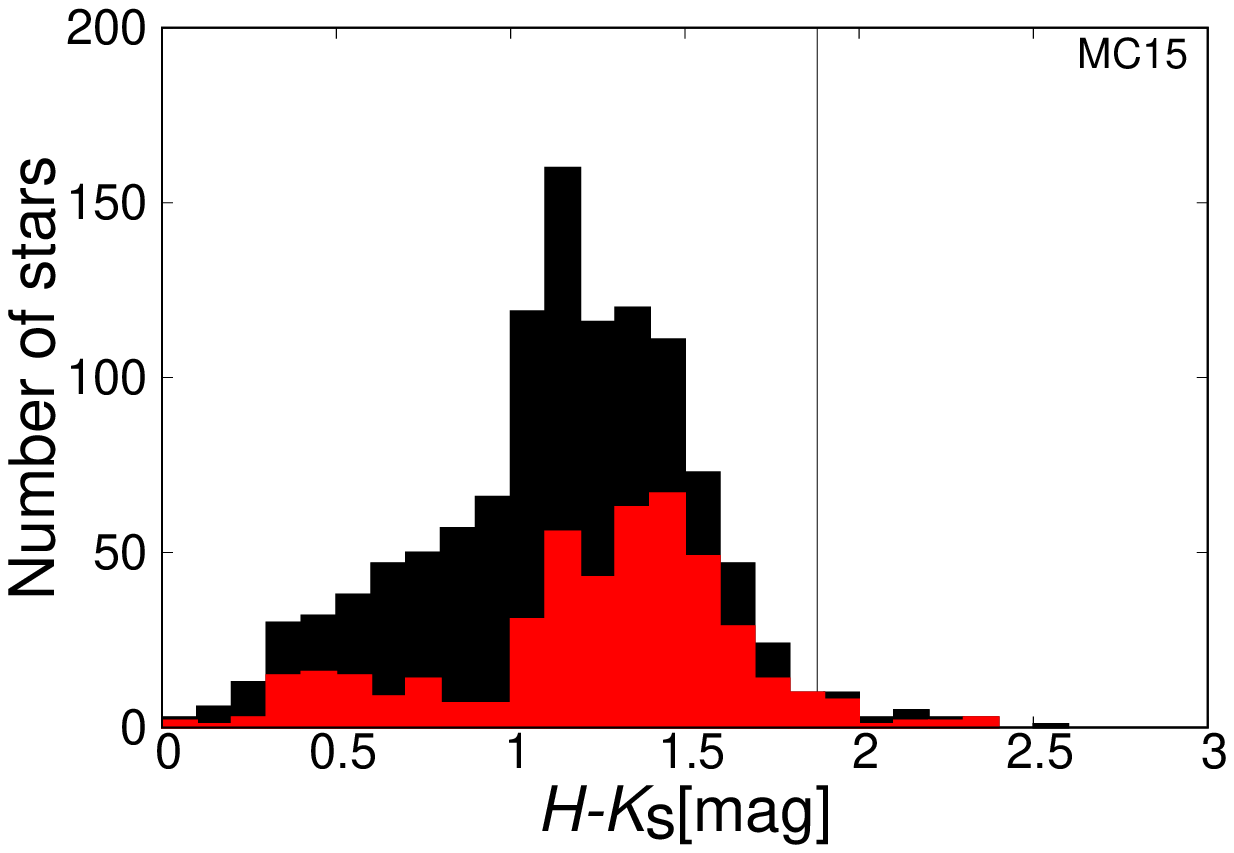}
\end{center}
\end{minipage}
\vspace{1.5cm}
\caption{Histograms of $H$ - $K_{\mathrm S}$ color for stars with $\delta K_{\mathrm S} < 0.06$ mag. Field names are shown in upper right. 
The red histograms %▼なんでインクルードだっけ？include 
are stars whose polarization is measured with  $\delta P_{K_{\mathrm{S}}} < $ 0.5 \%, 
and the black ones are those with  $\delta P_{K_{\mathrm{S}}} > $ 0.5 \%. %(善光191105) histogram が大文字になっていた
Vertical lines are the \textcolor{black}{mean background $H$ - $K_{\mathrm S}$ color of each field}: 
1.86~mag for DC35, 2.03~mag for DC5, and 1.88~mag for MC15.}
%{Black histograms of degree of $H$ - $K_{\mathrm S}$, for stars with $K_{\mathrm S} < 0.06$ mag. Field names are shown in upper right. Black vertical lines indicate the $H$ - $K_{\mathrm S}$ color of each background, DC35 from 1.86 mag, DC5 from 2.03 mag, and MC15 from 1.88 mag.}
\label{fig:Fig3}
\end{figure*}

Near-infrared observations toward the Galactic center produced 
a double peak structure in the $H$ $-$ $K_{\mathrm S}$ histogram; 
the small peak on the bluer side of the histogram corresponds to the foreground stars, and 
the main peak in the redder side of the histogram corresponds to the bulge stars (e.g.,  \cite{bNishiyama}). 
In the $H$ $-$ $K_{\mathrm S}$ histograms of the 48 Cepheid fields, 
many of them show a similar structure.  
%The Galactic bulge ($\sim$ 8 kpc) is located between the foreground and the background. Since we conduct the differential analysis among the foreground, the bulge and the background, we choose the field stars located in the Galactic bulge. Near-infrared observation toward the Galactic center proved that there is a double peak on the $H$ - $K_{\mathrm S}$ histogram. The small peak with the bluer side of the histogram correspond to the foreground stars, and the strong peak with the redder side of the histogram correspond to the bulge stars (e.g. \cite{bNishiyama}). We draw the $H$ - $K_{\mathrm S}$ histograms in each 48 Cepheid fields, and most of them show the similar to the $H$ - $K_{\mathrm S}$ histogram of the Galactic center. Therefore, we define the the strong peak value as the typical $H$ - $K_{\mathrm S}$ of bulge stars. 
%
In the color magnitude diagram (figure~\ref{fig:Fig4}) and 
$H$ $-$ $K_{\mathrm {S}}$ histogram (figure~\ref{fig:Fig3}), 
stars (black dots) with photometric errors 
$\delta H_{}$ $<$ 0.06 mag and 
$\delta K_{\mathrm{S}}$ $<$ 0.06 mag and stars (red dots) with photometric errors 
$\delta H_{}$ $<$ 0.06 mag, $\delta K_{\mathrm{S}}$ $<$ 0.06 mag and polarimetric errors $\delta P_{K_{\mathrm{S}}} < $ 0.5 \% are plotted.
%The red histogram of the peak $H$ $-$ $K_{\mathrm {S}}$ color of the DC35 field is 0.95 mag,
%0.95ではなく0.85 
The 
\textcolor{black}{peak of the red hisogram 
} %長田1220
$H$ $-$ $K_{\mathrm {S}}$ color of the DC35 field is 0.95 mag,
DC5  1.45 mag, and MC15  1.45 mag. %>>(善光)isはなし？(191106:疑問解消)
We basically regard the field stars whose colors are within %plus or minus 
$\pm 0.25$ mag from the main peak of $H$ $-$ $K_{\mathrm S}$ as the bulge star candidates.
However, if the bulge color range overlaps with the foreground or the background color range, we limit the color range of the bulge so that they do not overlap.
The color range of the DC35 and DC5 fields is within $\pm 0.25$ mag, and that of the MC15 field is within $\pm 0.18$ mag. %like that of MC15.%(善光191105)解析の仕方が変わったのを反映

%%&&&&&&&&&&&&&&&&&&&&&&&&&&&&&&&&&
%%&&&&&&&&&&&&&&&&&&&&&&&&&&&&&&&&&
\subsection{DC35}
\begin{figure*}
\begin{minipage}{0.5\hsize}
\begin{center}
\includegraphics[scale=0.65]{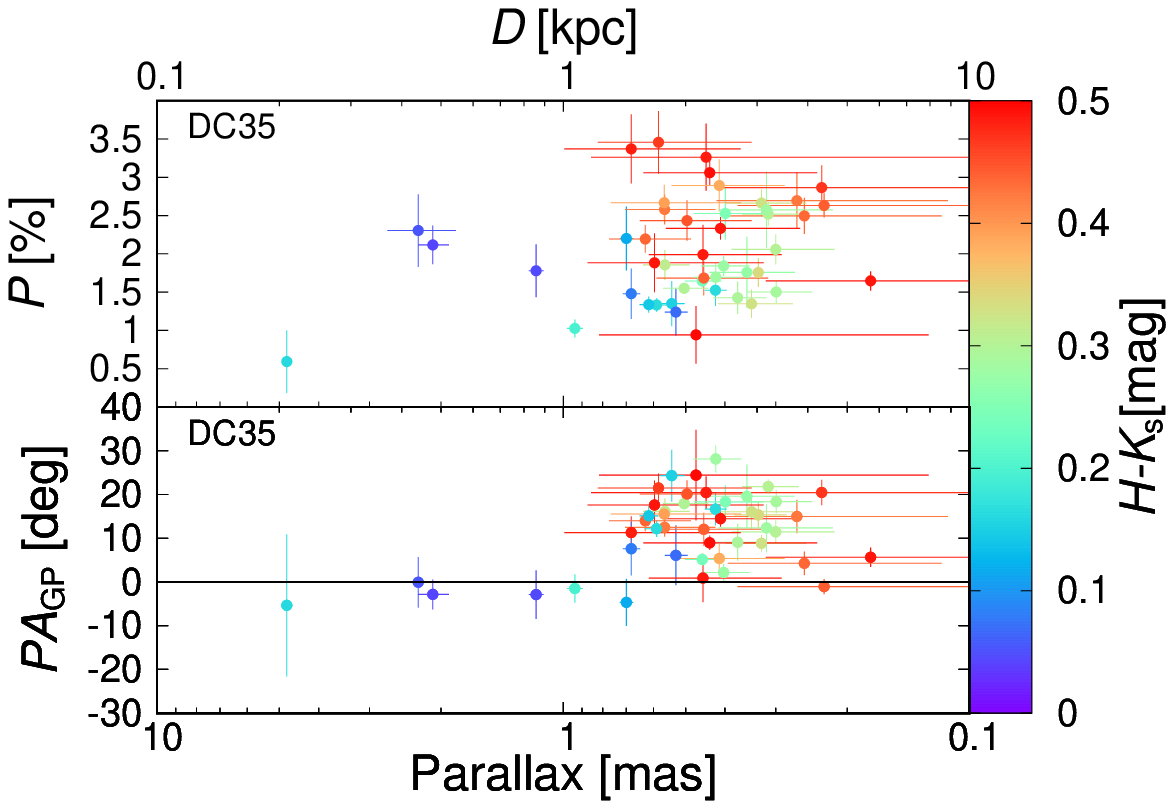}
\end{center}
\end{minipage}
\begin{minipage}{0.5\hsize}
\begin{center}
\includegraphics[scale=0.65]{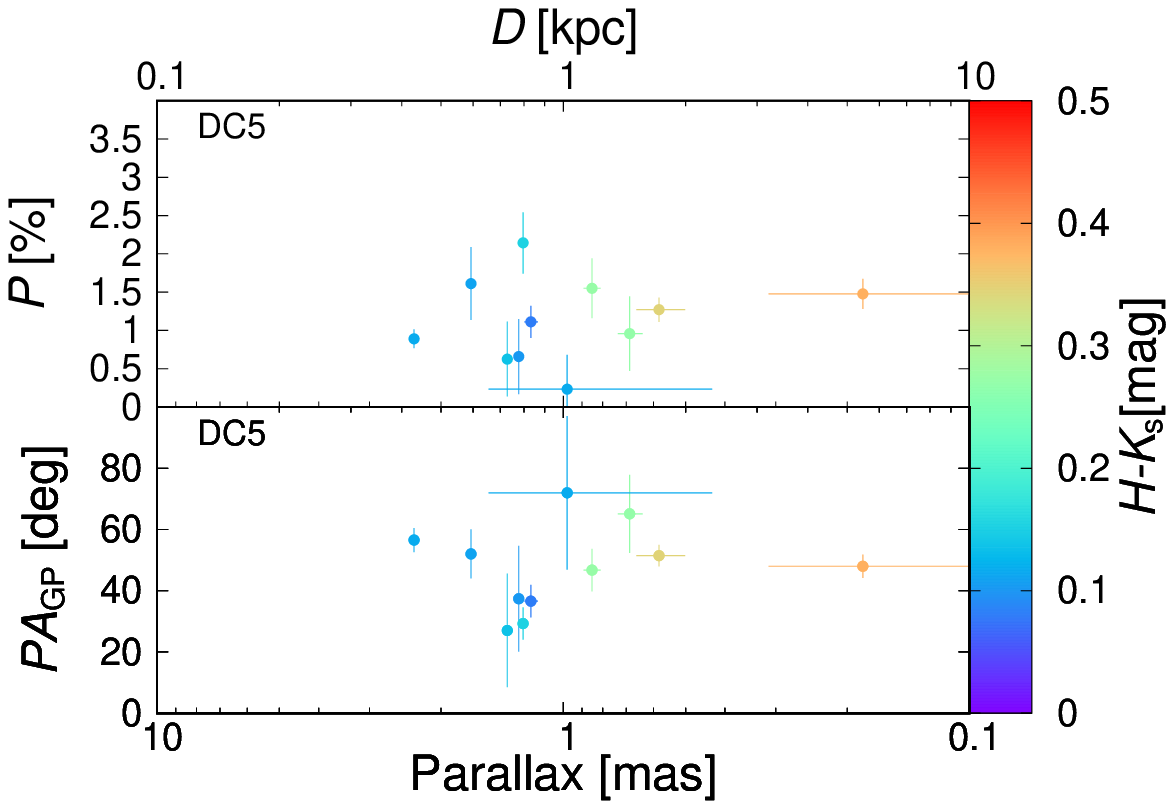}
\end{center}
\end{minipage}\\
\begin{minipage}{0.5\hsize}
\begin{center}
\includegraphics[scale=0.65]{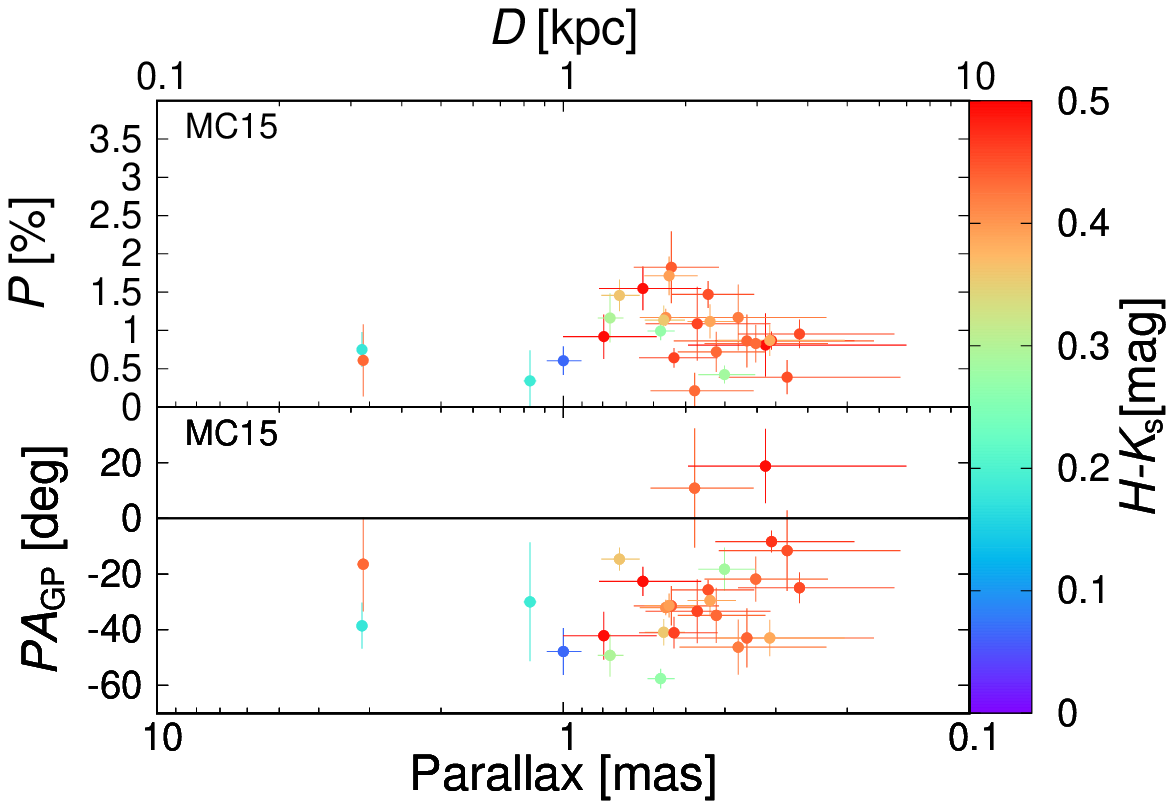}
\end{center}
\end{minipage}
\vspace{1.2cm}
\caption{Parallax vs. Polarization degree $P$ and position angle $PA_{\mathrm{GP}}$ of foreground stars cross-matched with the Gaia DR2 catalog. 
Field names are shown in upper left. }
%▲これが要らなくなった　Note that the $PA_{\mathrm{GP}}$ range for the DC35 field is only from $-30^\circ$ to $+40^\circ$ 
%because the position angles are in this small range.}
%{Parallax vs. Polarization degree P or position angle $PA_{\mathrm{GP}}$ of foreground stars cross-matched with the catalog Gaia DR2. Field names are shown in upper left. The error bar size is 1 $\sigma$. ●これは視差もPも±1シグマですよね？ならば自明だから不要と思います。Black vertical lines represent the Galactic plane.これも。}
\label{fig:fig2}
\end{figure*}

\begin{figure*}
\begin{minipage}{0.3\hsize}
\begin{center}
\includegraphics[scale=0.55]{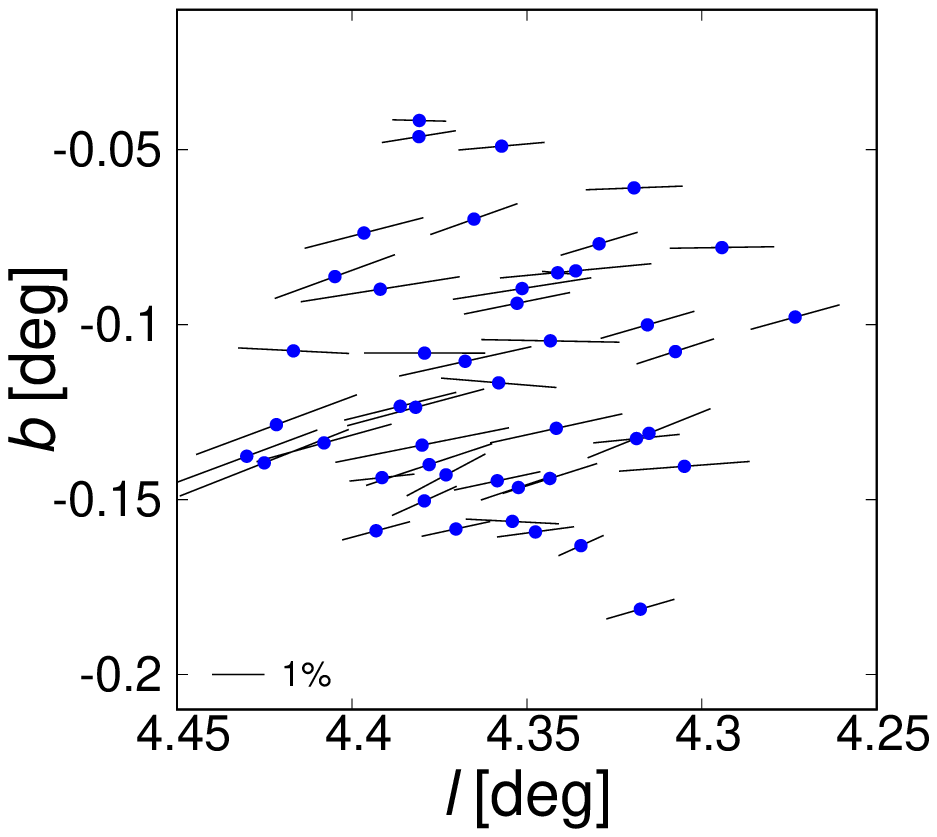}
\end{center}
\end{minipage}
\begin{minipage}{0.3\hsize}
\begin{center}
\includegraphics[scale=0.55]{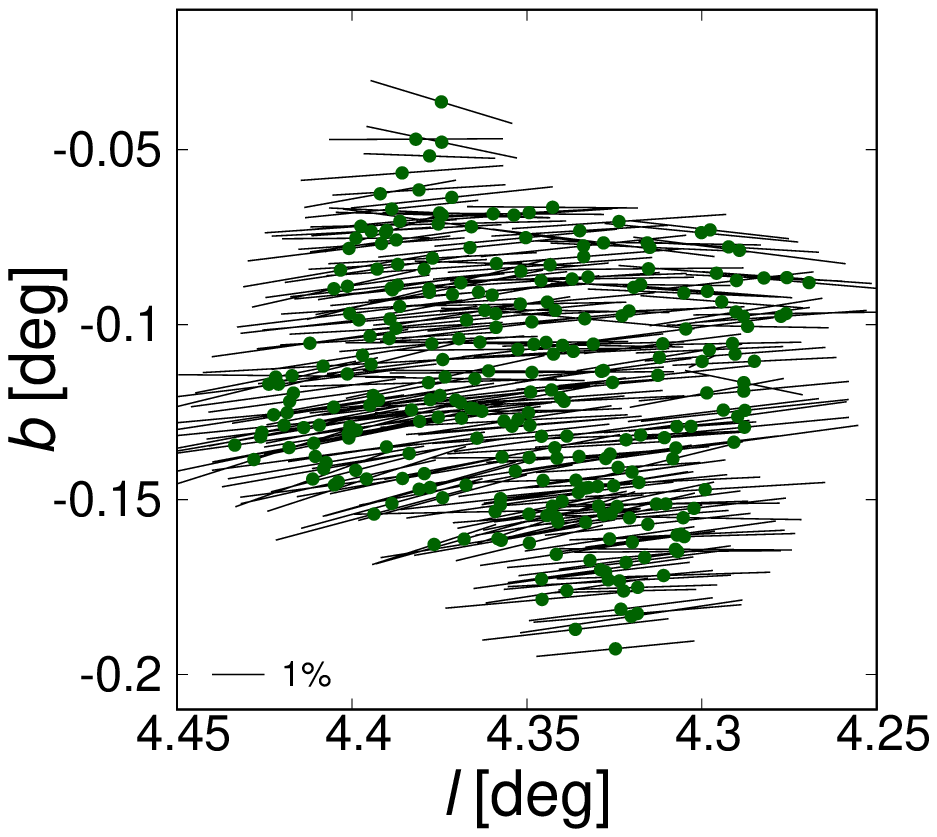}
\end{center}
\end{minipage}
\begin{minipage}{0.3\hsize}
\begin{center}
\includegraphics[scale=0.55]{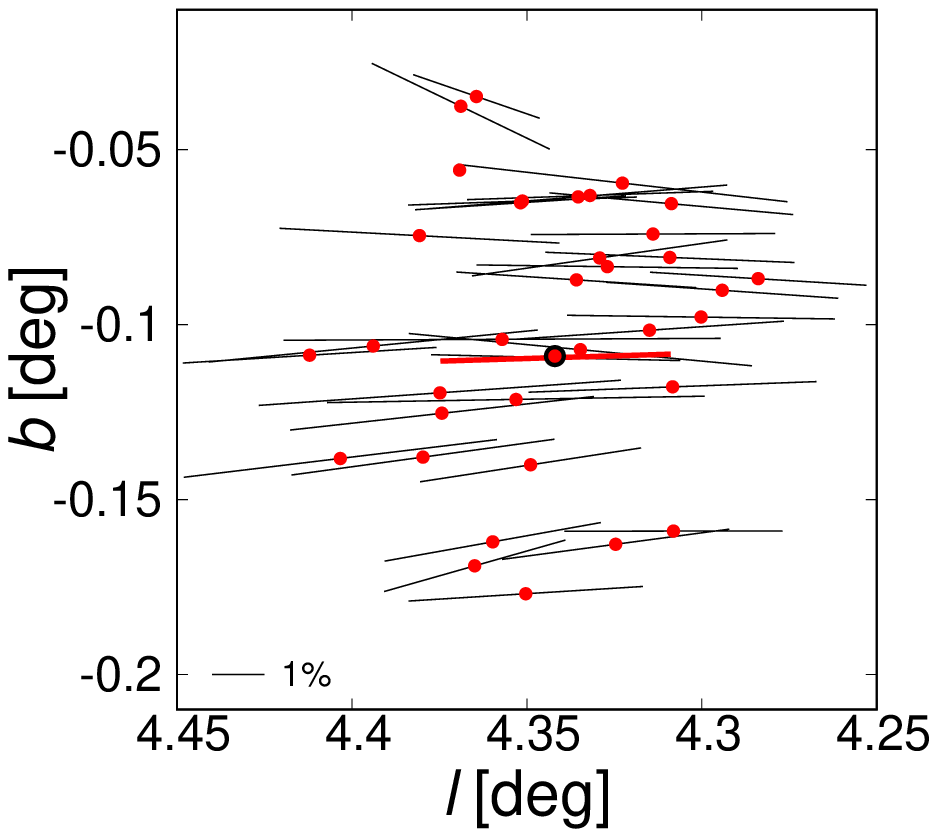}
\end{center}
\end{minipage}\\
\vspace{1.5cm}
\caption{Polarization maps of DC35 field %are ●題名だから文にしない
divided into foreground (left) in blue, bulge (middle) in green, and background (right) in red.
The red filled circle surrounded by a black frame is the Cepheid position.
}
%{DC35 field. Polarization for significantly detected stars (polarization degree error $P$ $\leq$ 0.5\%) are shown as black vectors centered on their stars. Vector lengths represent percentage linear polarization $-$ a 1\% reference bar is shown in lower left. Color dots represent the regions: blue is foreground, green is bulge, and red id background.}
\label{fig:Fig10}
\end{figure*}

%In the upper left panel of figure \ref{fig:fig2}, we show the relation between the distance and the polarization of the foreground stars matched to Gaia DR2.The polarization of the foreground stars indicates significantly disturbed magnetic field. Most of foreground stars are less than 4 kpc and the median distance is 2.21 kpc. 5 field stars with 0.9 mas $\leq$ $\omega$ have the position angles $PA_{\mathrm{GP}}$ $\sim$ $0^{\circ}$, the other foreground stars have the position angles $PA_{\mathrm{GP}}$ $\sim$ $+10^{\circ}$.However, we cannot suggest evidence for clear change in the orientation of magnetic field from this observed feature, because almost all of the foreground stars have the large distance errors.The mean $H$ - $K_{\mathrm S}$ of the foreground star is 0.31 mag. 
%In the upper left panel of figure~\ref{fig:fig2}, ●●順番変更  DC35, DC5, MC15のサブセクションのほとんどの段落が、In the right panel of...のたぐいで始まっていて、まあ偉大なるマンネリでも良い場合もあるのだけど、それでは「ほとんどの段落は図の説明に終始している」ように見えるので、少なくともIn the figureなんとか  というフレーズで始まらない段落をたくさん作りました。
We show the relation between the distance and the polarization of the foreground stars 
in the upper left panel of figure~\ref{fig:fig2}.  
%In the upper left panel of figure \ref{fig:fig2}, we show the relation between the distance and the polarization of the foreground stars matched to Gaia DR2.
%The polarization of the foreground stars indicates significantly disturbed magnetic field. ●これっていったいどういう意味ですか？　私には正反対に思えるのですが。
%Most of foreground stars are less than 4 kpc and the median distance is 2.21 kpc. 後ろへ
All the five stars close to us with $\omega$ $\geq$ 0.9 mas have the position angles $PA_{\mathrm{GP}}$ $\sim$ $0^{\circ}$ or slightly smaller, 
while the position angles of the other more distant foreground stars are 
distributed around $PA_{\mathrm{GP}}$ $\sim \, 15^{\circ}$.  %◎10じゃなく15にしました。良いよね？
%5 field stars with 0.9 mas $\leq$ $\omega$ have the position angles $PA_{\mathrm{GP}}$ $\sim$ $0^{\circ}$, the other foreground stars have the position angles $PA_{\mathrm{GP}}$ $\sim$ $+10^{\circ}$.
%However, we cannot suggest evidence for clear change in the orientation of magnetic field from this observed feature, because almost all of the foreground stars have the large distance errors.●これもよくわからんです。私の意見は以下のとおり。
Therefore, a clear change in the magnetic field direction seems to exist 
at a distance slightly more than 1~kpc.  
Such a small change is difficult to discern in the $H$ $-$ $K_{\mathrm S}$ color discussion 
in the previous section.  %●●これはホントですかね。H-Kでもこれら5つは小さいのかな。>>(善光_191026)図を変更しH-Kがわかるようにしました。◎discernにした
The median distance of these 46 foreground stars is 2.21 kpc, and 
their mean $H$ $-$ $K_{\mathrm S}$ color is 0.31 mag. %●xxに数字を入れて。
%% the median distance is 3.10 => the median distance is 2.21
%%the foreground stars is 0.35 mag => the foreground stars is 0.31 mag
If we calculate the mean polarization degree, $P$ is $\sim \, 2$\% and 
the mean position angle $PA_{\mathrm{GP}}$ is $\sim \, 10^\circ$ (table~\ref{tb:table2}). 
%
%
%In the left panel of figure~\ref{fig:Fig10}, ●●ここも順番変更と同様
The polarization map (the left panel of figure~\ref{fig:Fig10}) also demonstrates that %◎demonstrateにした
the vectors of the foreground stars are the mixture of the 
two coherent patterns of 
$PA_{\mathrm{GP}}$ $\sim$ $0^{\circ}$ and around $\sim$ $15^{\circ}$. 
%mentioned above.  
%◎下の文はよろしくないよねえ、上に移しました。
%The mean polarization degree $P$ is $\sim 2$~\% and 
%the mean position angle $PA_{\mathrm{GP}}$ is nearly parallel to the Galactic plane (table~\ref{tb:table2}). 
%
%
%The polarization vectors of the foreground stars show an overlap of two coherent patters (in the left panel of figure\ref{fig:Fig10}). The mean polarization degree is more than 1.5\% and the mean position angle $PA_{\mathrm{GP}}$ is parallel (less than $20^{\circ}$) to the Galactic plane (table \ref{tb:table2}). 

%In the upper left panel of figure~\ref{fig:Fig11}, ●●順番変更
We plot the foreground stars with the blue error bars 
in the Stokes $q_{\mathrm{GP}}- u_{\mathrm{GP}}$ plane 
in the Galactic coordinates (the upper left panel of figure~\ref{fig:Fig11}), 
where 
the rightmost points are large polarization parallel to the Galactic plane,  
the uppermost points are polarization $45^{\circ}$ clockwise (east from the increasing $l$ direction) from the Galactic plane, 
the leftmost points are perpendicular to the Galactic plane, and so on.  
The foreground stars are located in a small region on the right side of the origin in this figure, so 
the magnetic field along the line of sight aligns in the same direction between the observer and the foreground stars. 
Comparing this  Stokes $q_{\mathrm{GP}}- u_{\mathrm{GP}}$ figure with %%◎◎これを加えた！！！
the upper left panel of figure~\ref{fig:fig2} in detail, 
we can say that the magnetic field in this line of sight is 
in the direction of slightly negative $PA_{\mathrm{GP}}$
(close to the $q_{\mathrm{GP}}$ axis but in the the fourth quadrant of 
the $q_{\mathrm{GP}}- u_{\mathrm{GP}}$ plane) up to 1.1~kpc from us, and then 
in the poistive $PA_{\mathrm{GP}}$ direction (the first quadrant of 
the $q_{\mathrm{GP}}- u_{\mathrm{GP}}$ plane) to show the polarization of 
$PA_{\mathrm{GP}} \sim \, 15^{\circ}$ around the distance of $D \, \sim \, 2-3$~kpc.  
%In the upper left panel of figure~\ref{fig:Fig11}, the group of the foreground (the blue bars) is located on the right side of the origin, and this feature means that the magnetic field along the line of sight aligns in the same direction well between the observer and the foreground. 
%●ここは段落を分けますかねえ。DC35は最初なので、前景３段落、バルジ、後景。他のフィールドは前景１、バルジ１、後景１。

We have checked the dark cloud catalog by \citet{Dobashi}, and 
only one dark cloud No. 216 on the upper boundary of the DC35 field exists, with the center coordinates of ($l,\, b$) = ($4\fdg33$, $-0\fdg03$) and the surface are of 18 $\mathrm{arcmin^{2}}$; %>>(善光_191026)面積の値を追加
this cloud does not seem to affect the polarization of the field.  
%We check the dark cloud catalog (\cite{Dobashi}), because the magnetic field orientation inside a dense dust region like dark cloud differs from the global interstellar magnetic field (e.g. \cite{Messinger}). 
%There is a dark cloud No. 216 on the upper side of the DC35 field, and the center galactic coordinates of this cloud is (l, b) = ($4.33^{\circ}$, $-0.02^{\circ}$).
%There are no polarized stars around this cloud (figure \ref{fig:Fig7}), and this cloud does not affect our differential analysis.●ん？？　少なくともcenterはこの視野にかかってないのでは？　どれぐらいの大きさ？　大きさによっては上記の書き方を変えないと。
We have also checked the $V$ band polarimetry catalog (\cite{bHeiles}), but there are no stars located in the DC35 field of view. 
%In the upper left panel of figure\ref{fig:Fig11}, the group of the foreground (the blue bars) is located on the right side of the origin, and this feature means that the magnetic field along the line of sight aligns in the same direction well between the observer and the foreground. We check the dark cloud catalog (\cite{Dobashi}), because the magnetic field orientation inside a dense dust region like dark cloud differs from the global interstellar magnetic field (e.g. \cite{Messinger}). There is a dark cloud No. 216 on the upper side of the DC35 field, and the center galactic coordinates of this cloud is (l, b) = ($4.33^{\circ}$, $-0.03^{\circ}$).There are no polarized stars around this cloud (figure \ref{fig:Fig7}), and this cloud does not affect our differential analysis.We also check the $V$ band polarimetry catalog (\cite{bHeiles}), but there are no stars located in the DC35 field of view. 
%2kpc以内において、Lallement et al. (2019)の3次元ダストマップは約1.2kpcではっきりとした赤化の増加を示している。これはω\UTF{2265}0.9あたりで偏光の位置角が変わった距離に近い。それゆえ、この視線方向において、約1.2kpcの前後で磁場の向きが変化していると考えられる。
\textcolor{black}{
However, the web 3D dust map mentioned in \citet{Lallement} indicates a clear increase in reddening 
$E(B-V)$ at a distance around 1.2~kpc, which might correspond to the change in the position angle $PA_{\mathrm{GP}}$ at $\omega \sim 0.9$~mas.  
} %長田1222 長田1223 長田1223

In the middle panel of figure~\ref{fig:Fig10}, 
most of the vectors of the bulge stars seem to be aligned in the Galactic plane direction. 
In the bulge, the mean polarization degree $P$ increases and the position angle becomes closer to parallel to the Galactic plane than the forground (table~\ref{tb:table2}).
%In the middle panel of figure\ref{fig:Fig10}, most of vectors approximately look to be aligned with the Galactic plane direction. The mean polarization (table \ref{tb:table2}) indicates that the bulge position angles $PA_{\mathrm{GP}}$ are approximately parallel to the Galactic plane. Comparing the foreground polarization map and the bulge polarization map, the polarization degrees P seem to increase and the position angles $PA_{\mathrm{GP}}$ show the directional coherence. 
Therefore, the differential polarization between the foreground and the bulge (table~\ref{tb:table3}) shows polarization increase of more than 1\% in the direction parallel to the Galactic plane. 
The polarization increase per color change $\Delta P$$/$$\Delta (H-K_{\mathrm{S}})$ is  as high as 2.9~$\% / \mathrm{mag}$, indicating that 
a uniform magnetic component dominates between the foreground and the bulge.
%The differential polarization between the foreground and the bulge (table \ref{tb:table3}) indicates the large polarization degree (more than 1 \%) and the direction parallel to the Galactic plane. We calculate the polarization efficiency $\Delta P$ $/$ $\Delta (H-K_{\mathrm{S}})$ = 3.19 $\% / mag$. This value is high and means that the energy density of the uniform component is higher than that of the random component of the magnetic field between the foreground and the bulge. 
In the upper left panel of figure~\ref{fig:Fig11}, 
the bulge stars (the green error bars) are located further right from the foreground stars, 
indicating that 
the magnetic field between the foreground and the bulge along the line of sight aligns 
parallel to the Galactic plane very well.  
%In the upper left panel of figure\ref{fig:Fig11}, the group of the bulge (the green bars) is located around the boundary between first quadrant and quadrant 4. The group of the bulge is distributed on the right side of the group of the foreground, and this result is consistent with the differential polarization between the foreground and the bulge. 

The polarization vectors of the background stars are almost parallel to the Galactic plane in the entire field, %◎◇
%which is quite consistent with the polarization of DC35 itself at the distance $D$ of 12.0~kpc 
and are quite consistent with the polarization of DC35 itself, which is at the distance $D$ of 12.0~kpc 
(the right panel of figure~\ref{fig:Fig10}). %◇
The mean polarization degree $P$ of the background stars further increases and their position angle becomes even closer to parallel to the Galactic plane than the bulge (table~\ref{tb:table2}). 
%In the background, the mean polarization degree $P$ further increases and the position angle becomes even closer to parallel to the Galactic plane than the bulge (table~\ref{tb:table2}). 
%The number of background stars are fewer than the bugle, because the background stars suffer from the large dust extinction and is located in the far side of the Galactic center. 
The differential polarization between the bulge and the background stars 
(table \ref{tb:table3}) shows further polarization increase 
of more than 1\% in the position angle $PA_{\mathrm{GP}}$ 
nearly parallel to the Galactic plane. 
%◇The differential polarization between the bulge and the background (table \ref{tb:table3}) shows further polarization increase of more than 1~\% in the position angle $PA_{\mathrm{GP}}$ nearly parallel to the Galactic plane. 
The polarization increase per color change 
$\Delta P$$/$$\Delta (H-K_{\mathrm{S}})$ is \textcolor{black}{large} (\textcolor{black}{1.7$\% / \mathrm{mag}$}),%(善光_191127)  
indicating again that 
a uniform magnetic component dominates between the bulge and the background. 
%The mean polarization in the background (table \ref{tb:table2}) indicates that the bulge position angles $PA_{\mathrm{GP}}$ are approximately parallel to the Galactic plane. Since the standard deviation of the position angles $PA_{\mathrm{GP}}$ is small, the background polarization vectors look to be well aligned with the Galactic plane direction. Comparing to the bulge polarization map and the background polarization map, almost all of the polarization degrees P get up to more than 1\% and the position angles $PA_{\mathrm{GP}}$ clearly show the directional coherence. 
%
%●ここになぜ段落変更がまた入ることになったのか私にはわかりません。
%The differential polarization between the bulge and the background (table \ref{tb:table3}) indicates the large polarization degree (more than 1\%) and position angle $PA_{\mathrm{GP}}$ parallel to the Galactic plane. We calculate the polarization efficiency $\delta P$ $/$ $\delta (H-K_{\mathrm{S}})$ = 1.91 $\% / mag$. This value is relatively high and means that the energy density of the uniform component is higher than that of the random component of the magnetic field between the bulge and the background. 
%◇ここは強いメッセージがなかった
In the $q_\mathrm{GP} - u_\mathrm{GP}$ diagram (the upper left panel of figure~\ref{fig:Fig11}), %(善光191105)括弧閉じ
DC35 itself is located near the extension of the distribution of 
foreground stars (in blue) and 
bulge stars (in green).  
Most of the background stars (in red) are distributed close to DC35. %, but
Although some of them are far from it, 
such field stars are located in the uppermost part in the field of view.  
A few bulge stars there also have polarization like them; 
thus this seems to be due to variation in the field of view.  
%the is broad distribution might indicate 
%the presence of somewhat disturbed magnetic field in the far side of the Galactic center. 
%some of them are far from it; 
%this broad distribution might indicate 
%the presence of somewhat disturbed magnetic field in the far side of the Galactic center. 
%% ◇いや、そもそも(q=2.2,u=-3)とか(1.8, -1.5)とか(0,0)とかってどの星？　正しいのん？
%>>(善光)Fig. 14.の後景の偏光マップでいうと視野の上部にある2天体です。バルジにも似たようなものがあるので正しいと思います。
%※え？？、それ（バルジにも似たようなものがある）なら、英語の文章が正しくないよね。
%In (the upper left panel of figure~\ref{fig:Fig11}, 
%the background stars (the red error bars) are distributed broadly on the right side of the group of the bulge, 
%but this broad distribution might indicate 
%the presence of somewhat disturbed magnetic field.  
%In the upper left panel of figure\ref{fig:Fig11}, the group of the background stars (the red error bars) is distributed on the right side of the group of the bulge.These results indicate the differential polarization between the bulge and the background traces the large magnetic field structure. 

\subsection{DC5}  
\begin{figure*}
\begin{minipage}{0.3\hsize}
\begin{center}
\includegraphics[scale=0.55]{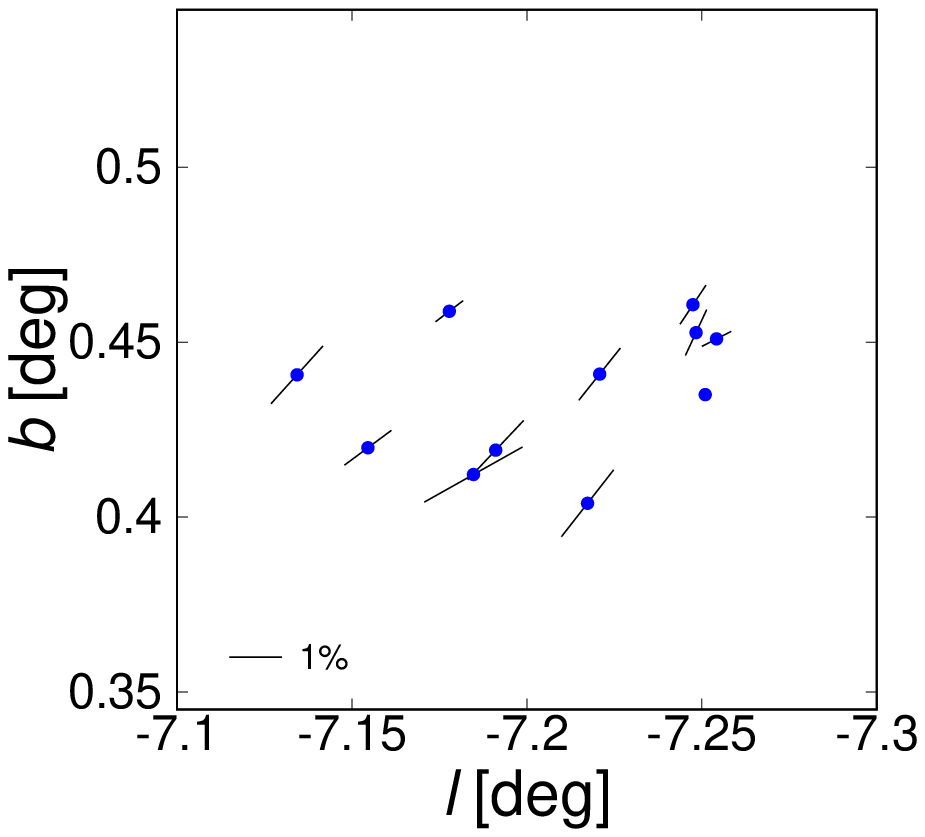}
\end{center}
\end{minipage}
\begin{minipage}{0.3\hsize}
\begin{center}
\includegraphics[scale=0.55]{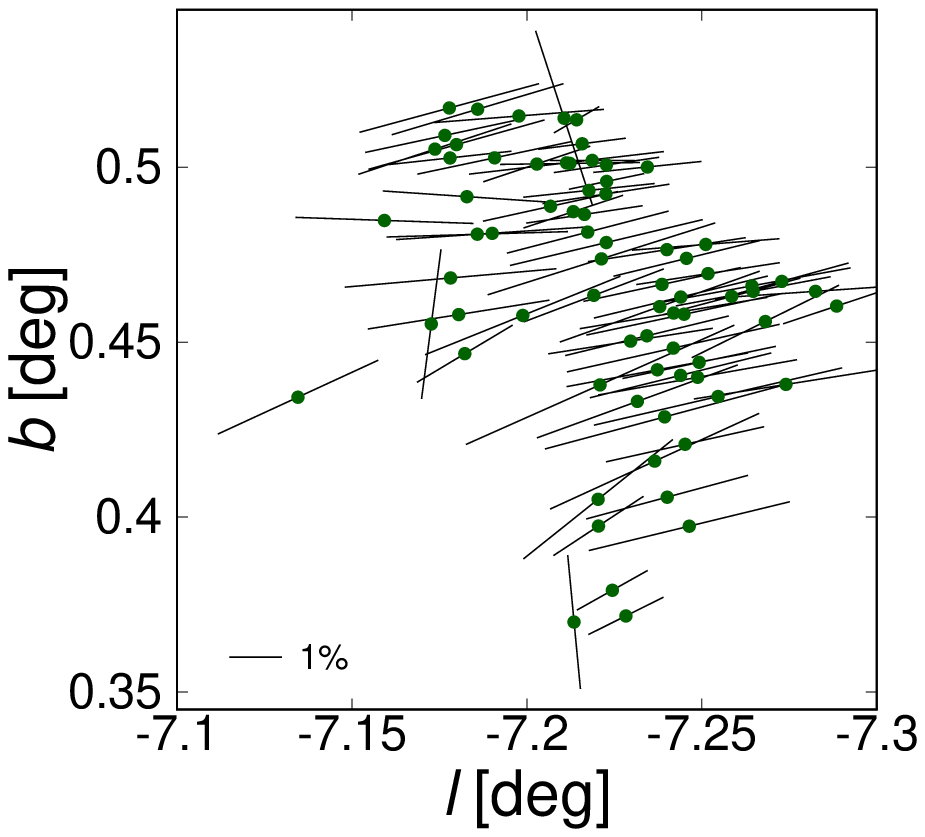}
\end{center}
\end{minipage}
\begin{minipage}{0.3\hsize}
\begin{center}
\includegraphics[scale=0.55]{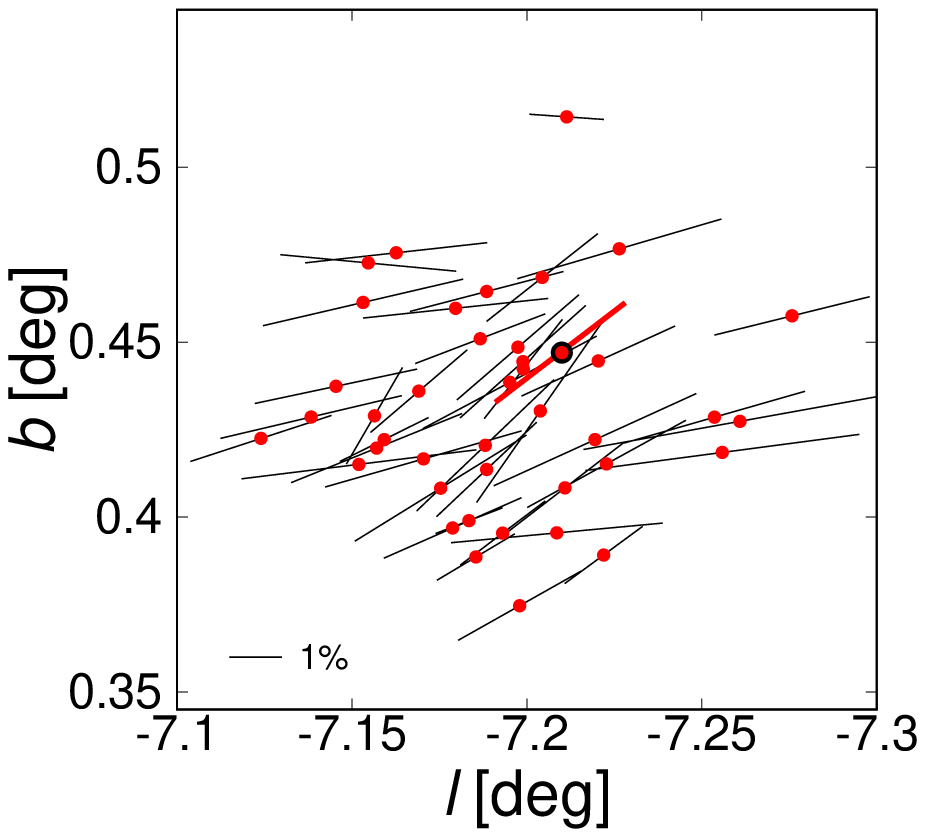}
\end{center}
\end{minipage}\\
\vspace{1.5cm}
\caption{Same as figure~\ref{fig:Fig10}, but for DC5.}
%{DC5 field. See figure\ref{fig:Fig10} for vector descriptions}
\label{fig:Fig8}
\end{figure*}

%In the upper right panel of figure~\ref{fig:fig2}, %●●順番変更
%we show the relation between the distance and the polarization of the foreground stars matched to Gaia DR2 in the DC5 field of view. 
All the \textcolor{black}{11} foreground stars have the position angles $PA_{\mathrm{GP}}$ $\sim \, 50^{\circ}$, 
and no foreground stars are 
polarized parallel to the Galactic plane 
$\sim \, 0^{\circ}$ %
in the upper right panel of figure~\ref{fig:fig2}.  
%while there are no foreground stars around $\sim$ $0^{\circ}$.●これだと、0度付近に前景の星がないという意味ですよね、何が0度？　銀緯が？ >>(善光)位置角が0度である天体が無いです●いや、それはわかってまっせ・・・英語を読んだ人が上記のように思う、という意味です。
%Almost all of the foreground stars are less than 2 kpc, and the weighted mean distance is 0.92 kpc. The range of polarization degrees are 0\% to 2.5\%, and the position angles $PA_{\mathrm{GP}}$ is $\sim$ $40^{\circ}$, almost oblique to the Galactic plane. Almost all of the foreground stars have the large distance errors, and it is not clear how the polarization degrees and the position angles $PA_{\mathrm{GP}}$ change against the distance. 
The median distance of these \textcolor{black}{11} stars is \textcolor{black}{0.83} kpc, and their mean $H$ $-$ $K_{\mathrm S}$ color is 0.19 mag. 
%同じ段落で良いような。
The polarization map (in the left panel of figure~\ref{fig:Fig8}) thus shows %few detectable polarizations and ってどういう意味？ >>(善光) 数が少ないといいたいけど、蛇足ですかね●蛇足じゃないけど、そういう意味には読めなかった。
a coherent pattern of $PA_{\mathrm{GP}}$ $\sim$ $50^{\circ}$. 
%as noted above. 
%The foreground field shows few detectable polarizations and an orientation angle coherence (in the left panel of figure\ref{fig:Fig8}). 
%All vectors look to be aligned with the oblique direction to the Galactic plane. 
%The mean position angle $PA_{\mathrm{GP}}$ (table \ref{tb:table2}) indicates more than $45^{\circ}$, and the global magnetic field direction is oblique to the Galactic plane. There are no field stars in this upper side of the field, and we check the presence or absence of the field stars to use the 2MASS catalog (\cite{Skrutskie}). There are some field stars in the upper side of this field, but they have large polarization error. 
The mean polarization degree $P$ is $\sim$ 1\% and the mean position angle $PA_{\mathrm{GP}}$ is oblique to the Galactic plane (table \ref{tb:table2}).
%不要のような。Therefore, the magnetic field direction is oblique to the Galactic plane. >>(善光) 4.4の議論で書くので不要
%Therefore, we consider that there are no field stars in this upper side of the field due to the citation with $P_{\mathrm err}$ $\leq$ 0.5\%.
%同じ段落で良いような。
%
In the upper right panel of figure~\ref{fig:Fig11}, the foreground stars are located in a region on the upper side of the origin (small $u_\mathrm{GP} >\, 0$).  %◎
This is the magnetic field up to the distance $D\, \sim$ 1.5~kpc in this line of sight.   
%so the magnetic field along the line of sight aligns in the same direction between the observer and the foreground stars. >>(善光)くどいか? 上のa coherent pattern of $PA_{\mathrm{GP}}$ $\sim$ $50^{\circ}$で事足りるという判断
%In the upper right panel of figure\ref{fig:Fig11}, the Stokes qu values of the foreground is located near the origin on the second quadrant, and this feature is consistent with the mean polarization. 
%◎改行してみた

We have also checked the dark cloud catalog %◎
by \citet{Dobashi}, 
and two dark clouds, No. 7204 and 7208, 
exist near this field, with the center coordinates of ($l,\, b$) = ($-7\fdg13$, $0\fdg283$) and ($l,\, b$) = ($-7\fdg05$, $0\fdg367$) and the surface areas of 143 $\mathrm{arcmin^{2}}$ and 141 $\mathrm{arcmin^{2}}$; %>>(善光_191026)面積も追加。
these clouds %
do not seem to affect the polarization of the field.
%We confirm whether a dark cloud is there or not as well as DC35 field of view, two dark cloud No. 7204 and 7208 (Dobashi et al. 2011) are located near this field. 
%However, since the center of the dark clouds are away from this field, we conclude that two dark clouds have little effect on the field stars.
The $V$ band polarimetry catalog by \citet{bHeiles} has no stars in the DC5 field of view, either.  
%We have also check the $V$ band polarimetry catalog, but there are no stars located in the DC5 field of view.

%DC5のバルジ
In the middle panel of figure~\ref{fig:Fig8}, the polarization vectors of the bulge stars seem to have broad distribution 
from $\sim \, 50^{\circ}$, which is similar to foreground stars, 
to nearly $\sim \, 0^{\circ}$, parallel to the Galactic plane. 
%●ではないでしょうか. >>(善光) シームレスに接続している感じに見えませんかね？微妙ですね。●接続しているという意味だったのか？！　　だけど、そうだとしても、前景・バルジ・後景と3つに分けてしまうと良くわからないきらいはあるけど、H-Ksで見たのが正しければ、バルジって前景の50度から0度に落ちてくるんだよね、0度から40度の後景へシームレスにつながるのではなく。In the middle panel of figure\ref{fig:Fig8}, the polarization vectors of the bulge stars are the connection of the two coherence patterns of $PA_{\mathrm{GP}}$ $\sim$ $0^{\circ}$ and $\sim$ $40^{\circ}$. 
%Some polarization vectors are almost parallel to the Galactic plane, and the other polarization vectors are oblique or orthogonal to the Galactic plane. 
The mean polarization degree $P$ 
of the bulge stars increases and the position angle $PA_{\mathrm{GP}}$ becomes clearly closer to parallel to the Galactic plane than the foreground stars (table~\ref{tb:table2}).
%The mean polarization of the bulge is accumulated by two different components, the foreground and the bulge (table \ref{tb:table2}). 
%The mean position angle $PA_{\mathrm{GP}}$ of the bulge is about $30^{\circ}$ smaller than that of the foreground, and the parallel component to the Galactic plane is higher than the perpendicular component to the Galactic plane between the foreground and the bulge. 
%Therefore,
The differential polarization between the foreground and the bulge (table~\ref{tb:table3}) shows polarization increase of more than 2\% in the direction parallel to the Galactic plane.
%Comparing the bulge polarization map and the foreground polarization map, the polarization degrees P seem to increase, and the position angles $PA_{\mathrm{GP}}$ show a different coherent patterns in the bulge field. 
The polarization increase per color change $\Delta P$$/$$\Delta (H-K_{\mathrm{S}})$ is %relatively 
large (2.0 $\% / \mathrm{mag}$)%%(善光191105)計算間違い(1.2 $\% / \mathrm{mag}$)
, indicating that a uniform magnetic component dominates between the foreground and the bulge.
%The differential polarization degree between the foreground and the bulge is more than 1.5 \%, and the polarization efficiency is 2.04 $\% / mag$. 
%This value is relatively high and means that the energy density of the uniform component is higher than that of the random component of the magnetic field between the foreground and the bulge. 
%Since the differential position angle $PA_{\mathrm{GP}}$ is $\sim 0^{\circ}$, the global magnetic field direction between the foreground and the bulge is parallel to the Galactic plane. 
%同じ段落で良いような。
%
%In 
This is clear also in the upper right panel of figure~\ref{fig:Fig11}, almost all of the bulge stars are located %further 前景が右じゃないのだからfurtherではない
right of the foreground stars, %, indicating that the magnetic field between the foreground and the bulge along the line of sight aligns parallel to the Galactic plane.
which means that the magnetic field between the foreground and the bulge is in the direction parallel to the Galactic plane.  
%In the upper right panel of figure\ref{fig:Fig11}, most of the bulge stars are distribute in the right side of the group of the foreground stars, and some bulge stars are distributed in the same place as the foreground stars. 
%Two bulge stars, (l, b) = ($-7.211^{\circ}$, $-0.514^{\circ}$) and (l, b) = ($-7.144^{\circ}$, $0.424^{\circ}$), clearly have different polarization from the other bulge stars. 
%We consider that two bulge stars are young stellar object (YSO) candidates, because many YSOs are intrinsically polarized due to the scattering of the stellar light by dust grains in their circumstellar disk (e.g., \cite{Whitney}, \cite{Tamura}, \cite{Yoshikawa}). 
%However, since the identification of stellar type is different from the purpose of this paper, we do not consider two bulge stars as YSO in this paper. 

%DC5の後景
%The ◇
If we proceed to the position of the Cepheid DC5 at the distance $D\, =\, 12.6$~kpc, 
the $PA_\mathrm{GP}$ of polarization becomes oblique to the Galactic plane again.  
%However, the Cepheid DC5 at the distance $D = 12.6$~kpc has the polarization 
%whose $PA_\mathrm{GP}$ is 37$^\circ$, rather oblique to the Galactic plane.  
In the right panel of figure~\ref{fig:Fig8},
the background stars 
%◇again 
show 
broad distribution 
from nearly $\sim \, 0^{\circ}$, parallel to the Galactic plane,  
to $\sim \, 50^{\circ}$, which is as indicated in the $H$ $-$ $K_{\mathrm S}$ vs. polarization angle $PA_{\mathrm{GP}}$ of the bottom panel of figure~\ref{fig:Fig5}, in the more reddened and probably more distant stars.  
%a overlap of two coherent patterns of $PA_{\mathrm{GP}}$ $\sim$ $+45^{\circ}$ and around $\sim$ $+10^{\circ}$. 
In the background, the mean polarization degree $P$ %slightly increases and the position angle becomes slightly inclined than the bulge (table \ref{tb:table2}).●ん？これは私には理解できないのですが。>> (善光) '偏光度はちょっとしか変わっていないけど、向きは大きく変化していますよ'ということがいいたかった。書き直した文章が意図した内容になっています。
is almost unchanged from the bulge, and 
the position angle becomes tilted again from the Galactic plane (table~\ref{tb:table2}). 
% ◎The polarization 
%change 
However, the 
differential polarization 
between the bulge and the background (table~\ref{tb:table3}) 
is 
%shows polarization %increase of 
%◎more than %less than %◎
nearly 1\%
%0.5 \% 
in the direction oblique to the Galactic plane.
%◎◎うーん、以下の文、特に「今やっている解析のやりかたをすると」およびメール文の「解析手法を少し変えた影響で4.2章の解釈に変更」があんまり理解できていないのですが、以上および以下の英語であらわせてます？◎(善光)今やっている解析のやりかたをすると$\Delta P$ $/$ $\Delta (H-K_{\mathrm{S}})$の値で逆転が生じた。ここのでのポイントはバルジと後景の平均の偏光度は約0.2\%ぐらいしか違わないのに、差分の偏光度は0.9\%あることが注目すること。乱流が強いとかではなくて、はっきりと磁場の向きが変わっていること主張できる。
%
%The polarization per color %ここでの表記変えはなぜ？●もともとはincreasedと動詞？それとも動詞の過去分詞で形容詞？になってたので、それをやめたのでした。そもそもincrease（偏光度上昇を示唆）というよりは変化なのでchangeとしたかったけど上と合わせてくどいかと。だけど、意味不明瞭よりはくどい方がよっぽど良いね。変更します。
%This polarization change per reddening 
The polarization change per reddening 
$\Delta P$$/$$\Delta (H-K_{\mathrm{S}})$ is
%, however, 
\textcolor{black}{relatively high} (\textcolor{black}{1.5$\% / \mathrm{mag}$}). %(善光_191127)
Therefore, 
%random magnetic components dominate, but 
%on the whole, 
the magnetic field %can be 
is 
tilted as large as $\sim \, 60^{\circ}$ (table~\ref{tb:table3}, upper right panel of figure~\ref{fig:Fig11}), 
and this change should exist 
between the bulge at the distance of $D \, \sim$ 8~kpc and the position of DC5 at 
$D\,=\, 12.6$~kpc.   
% indicating that a random magnetic component dominates between the foreground and the bulge. >> (善光)磁場の向きが変わったからという内容は4.4に入れるつもりでここでは乱流が強いことを示唆しているとした。
%10/18善光(変更)
%change in the magnetic field orientation exist between the bulge and background a uniform magnetic component dominates between the bulge and the background.
%◇
The location of DC5 in the $q_\mathrm{GP} - u_\mathrm{GP}$ diagram %(the upper right panel of figure~\ref{fig:Fig11} 
is near the upper left end of the distribution of 
background stars (in red).
\subsection{MC15}
\begin{figure*}
\begin{minipage}{0.3\hsize}
\begin{center}
\includegraphics[scale=0.55]{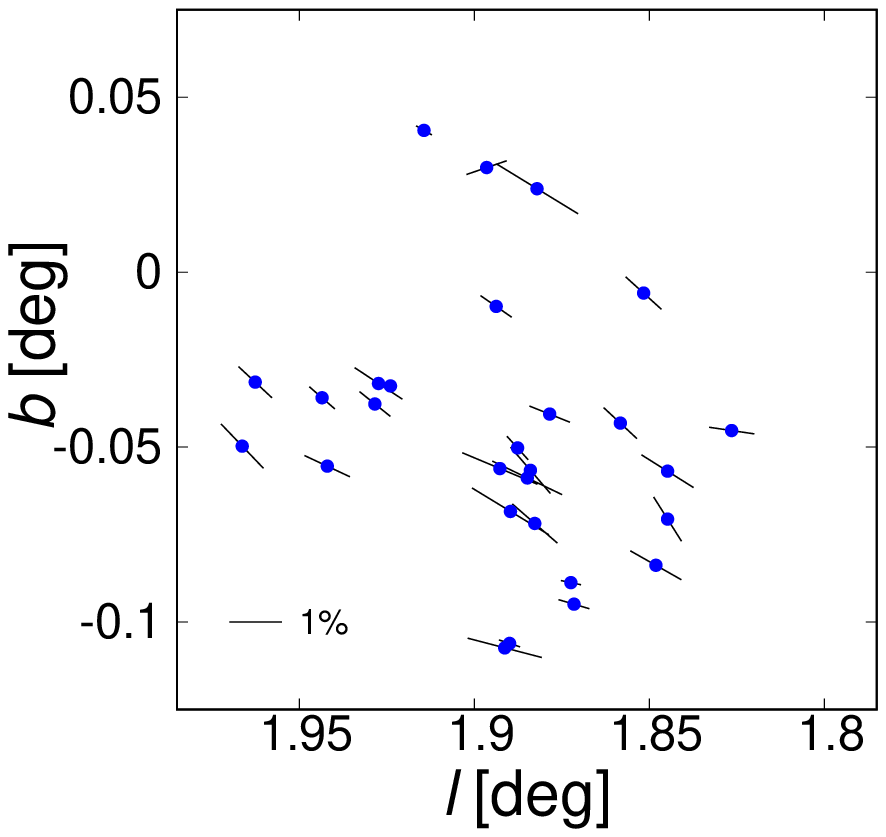}
\end{center}
\end{minipage}
\begin{minipage}{0.3\hsize}
\begin{center}
\includegraphics[scale=0.55]{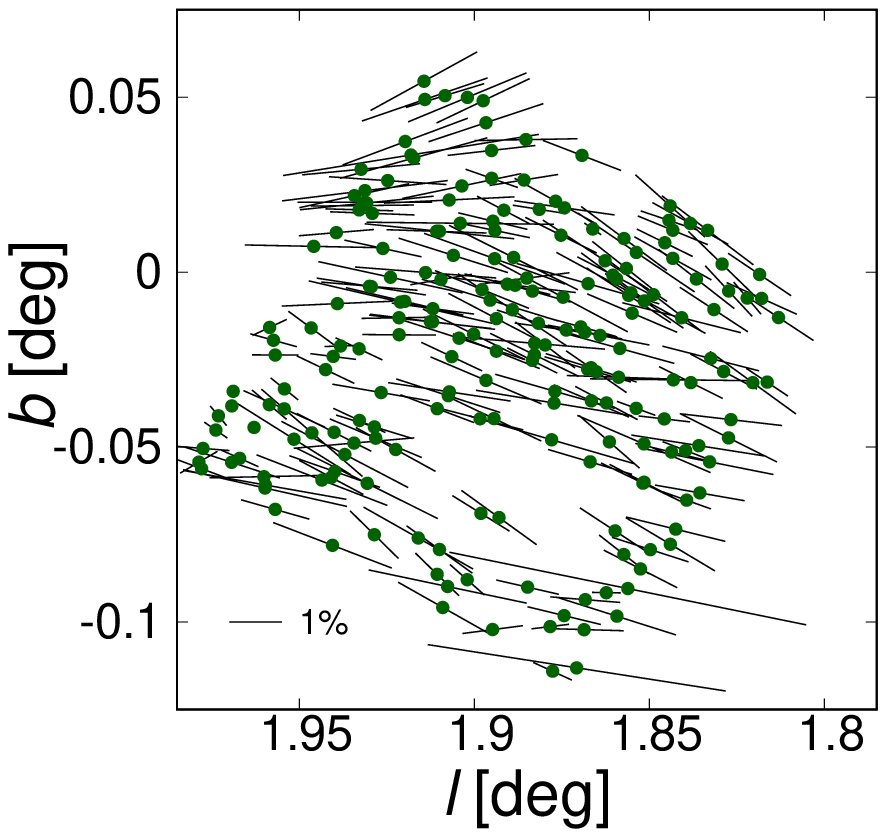}
\end{center}
\end{minipage}
\begin{minipage}{0.3\hsize}
\begin{center}
\includegraphics[scale=0.55]{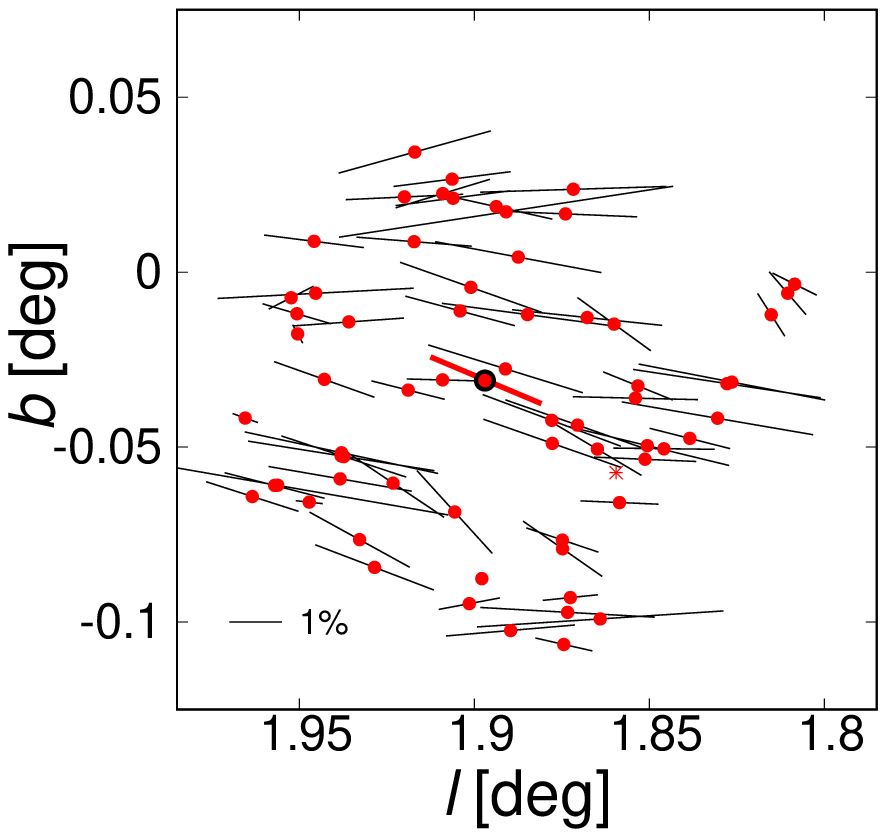}
\end{center}
\end{minipage}\\
\vspace{1.5cm}
\caption{Same as figure~\ref{fig:Fig10}, but for MC15.}
%{MC15 field. See figure\ref{fig:Fig10} for vector descriptions}
\label{fig:Fig9}
\end{figure*}

\begin{figure*}
\begin{minipage}{0.5\hsize}
\begin{center}
\includegraphics[scale=0.8]{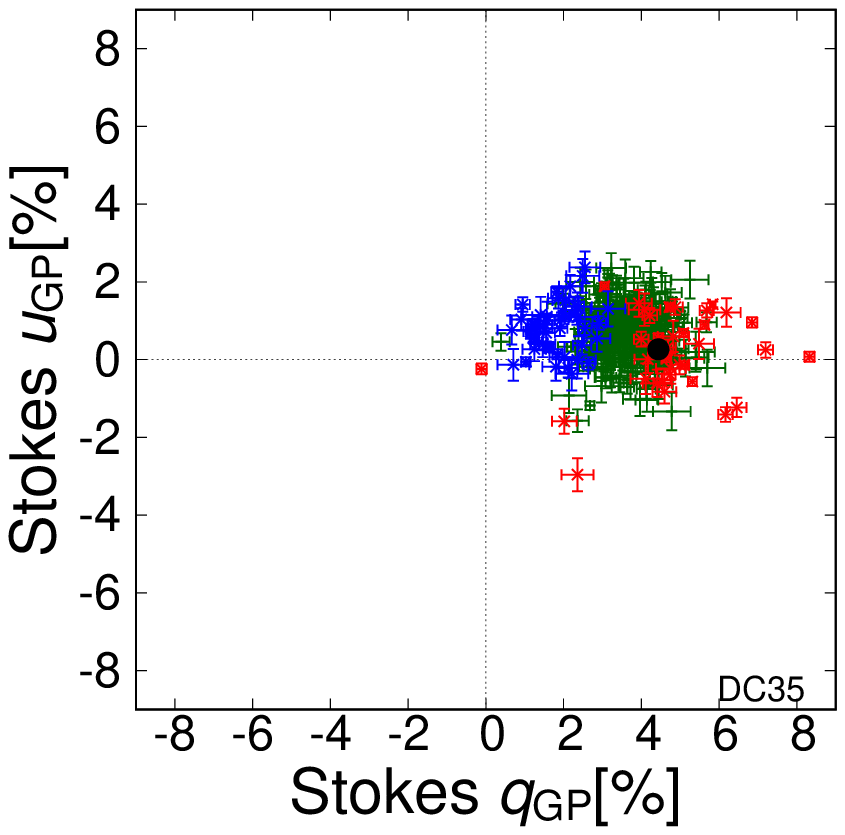}
\end{center}
\end{minipage}
\begin{minipage}{0.5\hsize}
\begin{center}
\includegraphics[scale=0.8]{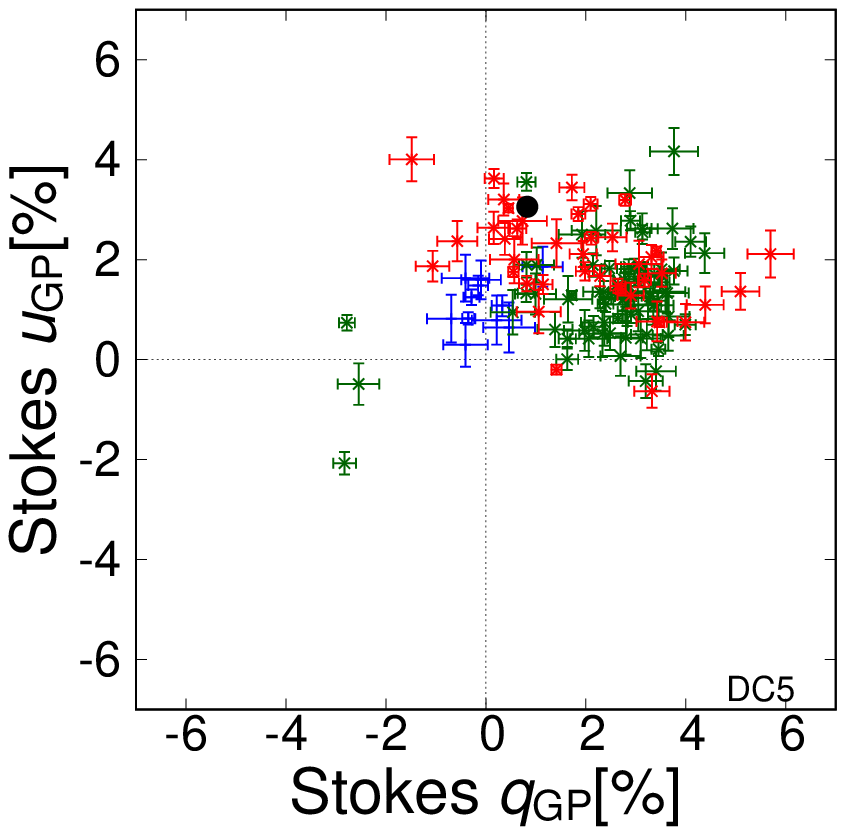}
\end{center}
\end{minipage}\\
\begin{minipage}{0.5\hsize}
\begin{center}
\includegraphics[scale=0.8]{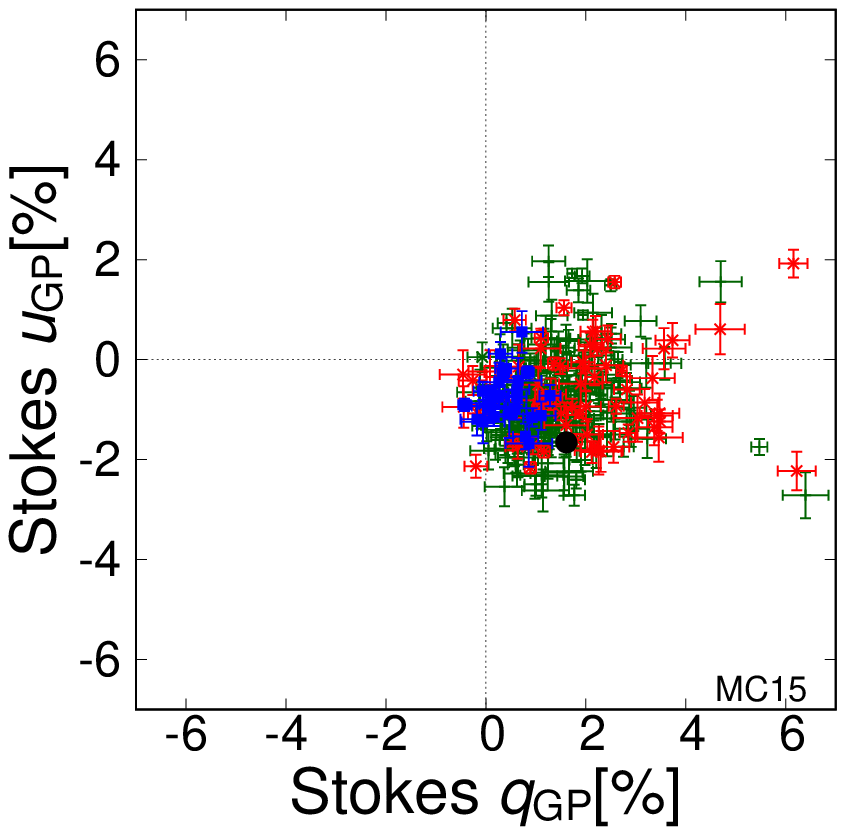}
\end{center}
\end{minipage}
\vspace{1.5cm}
\caption{Stokes parameters $q_{\mathrm{GP}}$ and $u_{\mathrm{GP}}$ of field stars, foreground (blue), bulge (green), and background (red). 
Field names are shown in lower right. 
Black filled circles are
Stokes parameters of each Cepheid. } %▲eachは単数Cepheids, $(q_{\mathrm{GP}},u_{\mathrm{GP}})_{\mathrm{DC35}}=(4.44\pm0.72, 0.27\pm0.71)$, $ 
%▲いや、この情報は要らないと思うのです。これは一度消していたのにドッペルゲンガーでよみがえったのでした。で、むしろ、誤差の伝播の大きな問題点（松永氏の指摘）を示している気がする。中島氏も含めて、Ｄ論では考え直すべきかもしれない。ちょっと違う話題だけど。(q_{\mathrm{GP}},u_{\mathrm{GP}})_{\mathrm{DC5}}=(0.83\pm0.56, 3.06\pm0.56)$, and $(q_{\mathrm{GP}},u_{\mathrm{GP}})_{\mathrm{MC15}}=(1.61\pm0.30, -1.66\pm0.30)$.
%{Stokes $q_{\mathrm{GP}}$ as a function of Stokes $u_{\mathrm{GP}}$●これは違うよねえ、べつにuの関数としてqを表しているわけではなく、2つの独立変数quでしょ, colored by the foreground (blue), the bulge (dark-green) and the background (red) stars. Field names are shown in lower right. Error bars show $\pm 1\sigma$ uncertainties●これも自明だよね.
\label{fig:Fig11}
\end{figure*}

Almost all %of 
the foreground stars have %the position angle
polarization of $PA_{\mathrm{GP}}$ $\sim \, -30^{\circ}$ 
in the bottom left panel of figure~\ref{fig:fig2}.  %●●順番変更.  
%polarization of $PA_{\mathrm{GP}}$ $\sim-25^{\circ}$. ●●なんで２４日の昼に-２５に変わったんだっけ。-３０の方がtableとも矛盾しないような気がするのだけど。 %, and two field stars with $\omega$ $\leq$ 1~mas have the position angles $PA_{\mathrm{GP}}$ $\sim$ $-20^{\circ}$.●いや、プラス20度だよね。だけど、1つはsignificantじゃないし、もう1つの星に関してこれを言う必要がある？>> (善光)ないです。%Therefore, the magnetic field direction seem slightly oblique to the Galactic plane. >> 下の文章と組み合わせる
%Therefore, the magnetic field direction seem slightly oblique to the Galactic plane. >> 下の文章と組み合わせる
%これは3章へ戻すべきでしょ。>> (善光)レファレンス間違いなので消さない ●というのはどういう意味？ >>(善光) figure~\ref{fig:fig2}をfigure~\ref{fig:Fig2}としていた。ここは年周視差vs偏光度および位置角のところなので、3章に書いてしまうと意味不明になるので>>(善光_192026)まだ3章のものは消していない
%In the bottom left panel of figure~\ref{fig:Fig2}, 
%most of field stars have %the 
%position angles $PA_{\mathrm{GP}}$ $\sim -30^{\circ}$. 
%Therefore, the magnetic field direction seem slightly oblique to the Galactic plane.
The median distance of these \textcolor{black}{28} foreground stars is \textcolor{black}{1.98} kpc, and their mean $H$ $-$ $K_{\mathrm S}$ is 0.39~mag.
%DC13の名残: The median distance of these 17 foreground stars is 1.86 kpc, and their mean $H$ - $K_{\mathrm S}$ is 0.26 mag.
%We show the relation between the distance and the polarization of the foreground stars matched to Gaia DR2 in the MC15 field in the bottom left panel of figure\ref{fig:Fig2}. 
%Most of the foreground stars are less than 2 kpc, and the weighted mean distance is 2.10 kpc. 
%The position angles $PA_{\mathrm{GP}}$ of the foreground stars show $\sim$ $30^{\circ}$, and the range of polarization degrees are 0 \% to 2 \%. 
%In this field of view, most of the foreground stars have the large distance errors, and it is not clear how the polarization degree and the position angle $PA_{\mathrm{GP}}$ change against the distance. 
%The mean $H$ - $K_{\mathrm S}$ of the foreground star is 0.39 mag. 
The polarization vectors of the foreground stars (in the left panel of figure~\ref{fig:Fig9}) thus show a coherent pattern of $PA_{\mathrm{GP}}$ $\sim \, -30^{\circ}$, 
oblique to the Galactic plane; 
%In the left panel of figure\ref{fig:Fig9}, the polarization vectors of the foreground stars seem to be oblique to the galactic plane.
the mean polarization degree is $\sim$ 0.9\% (table~\ref{tb:table2}). %and the mean position angle $PA_{\mathrm{GP}}$ is oblique to the Galactic plane (table \ref{tb:table2}).●文を分けると、くどいだけと思う
%In the foreground polarization map (in the left panel of figure\ref{fig:Fig9}), the polarization vectors show a coherent pattern that is oblique to the Galactic plane. 
%The mean position angle $PA_{\mathrm{GP}}$ (table \ref{tb:table2}) is $\sim$ $30^{\circ}$, and this indicates the global magnetic field is oblique to the Galactic plane.
%In the 
The bottom left panel of figure~\ref{fig:Fig11} also shows that the foreground stars are located %bottom 
%slightly lower right ($q_{\mathrm{GP}}>0, u_{\mathrm{GP}}<0$) from the origin.◎変更
in the fourth quadrant ($q_{\mathrm{GP}}>0,\, u_{\mathrm{GP}}<0$) with small distances.  
Thus, the magnetic field up to $D \, \sim$ 3~kpc is aligned reasonably well in this line of sight. %(善光191105)theeをtheに修正 
%in this figure, so the magnetic field along the line of sight aligns in the same direction between the observer and the foreground stars.
%We also have checked the 
The dark cloud catalog 
by \citet{Dobashi} 
lists no dark cloud in the neighborhood of the MC15 field.
%and there is no dark cloud nearby the MC15 field.   ●「赤外暗黒星雲」は3.3に移しました。
The $V$ band polarimetry catalog by \citet{bHeiles} has no stars in the MC15 field of view, either.
%We also have checked the $V$ band polarimetry catalog, but there are no stars located in the MC15 field of view.
%In the bottom left panel of figure\ref{fig:Fig11}, the Stokes qu values of the foreground (the blue error bars) is distributed near the origin on the quadrant 4. 
%Therefore, the magnetic field direction between the foreground and the background extends diagonally from lower right to upper left.
%We confirm whether a dark cloud is there or not as well as DC35 and DC5 field of view. 
%There is no dark cloud, but there is the large dust extinction region in the galactic coordinate (l, b) = ($1.9^{\circ}$, $-0.05^{\circ}$). 
%There is no polarized stars around this large dust extinction region, and we conclude that this region does not affect the observed stars.
%We also search the $V$ band polarimetry catalog (Heiles 2000), but there are no stars located in the MC15 field of view. 

%MC15のバルジ
%そもそもfieldで一番上の場所については、３章で言っておく。それを受けて
%The polarization vectors of the 
The bulge stars (in the middle panel of figure~\ref{fig:Fig9}) show a coherent pattern of 
polarization, 
$PA_{\mathrm{GP}}$ $\sim \, -20^{\circ}$, except for the uppermost small area, as we notice in the section~3.3.  
The angle from the Galactic plane is smaller negative, and the  
%In the middle panel of figure~\ref{fig:Fig9}, most of vectors of the bulge stars seem to be oblique to the Galactic plane around $PA_{\mathrm{GP}}$ $\sim$ $-20^{\circ}$, while other vectors of bulge stars which are located on the upper side of the field seem to be parallel to the Galactic plane.
%In the middle panel of figure~\ref{fig:Fig9}, the polarization vectors of the bulge stars are almost parallel to the galactic plane in almost all field.
%In the bulge, the 
mean polarization degree $P$ %increased 
increases from 0.9 to \textcolor{black}{1.6}\% %and the position angle %become 
%becomes closer to 
%more parallel to the Galactic plane than foreground
 (table~\ref{tb:table2}).  
%>>(善光)前景よりは平行という主張●平行にmoreは無いよん。more correctのたぐい。
%Therefore, t
This is evident in 
the bottom left panel of figure~\ref{fig:Fig11}, 
where the bulge stars (the green error bars) are distributed broadly in the right side of the foreground; 
$q_{\mathrm{GP}}$ increases and $u_{\mathrm{GP}}$ does not change.  
%indicating that 
Therefore, the magnetic field between the foreground and the bulge %along the line of sight 
aligns parallel to the Galactic plane.
However, 
its differential polarization %between the foreground and the bulge くどいかと思った
(table~\ref{tb:table3}) shows polarization increase of %less than 1\% 
only $\sim$ \textcolor{black}{1.0}\% in the direction parallel to the Galactic plane.
The polarization increase per color change $\Delta P$$/$$\Delta (H-K_{\mathrm{S}})$ is rather small (0.9 $\% / \mathrm{mag}$), indicating that a random magnetic component dominates between the foreground and the bulge. %(善光_191127)長田1221青字を消した
The Cepheid MC15 at $D\, =\, 10.9$~kpc has polarization whose $P\, =\, 2.3\%$ and $PA_\mathrm{GP}\,= \,-23^\circ$, and 
the polarization vectors of the background stars (in the right panel of figure~\ref{fig:Fig9}) show patterns 
%◎ even 
slightly closer to parallel to the Galactic plane, with %
the marginally increased mean polarization degree $P$ 
%◇ from 1.5 to 1.9\% %increased  %●そして誤差を考えるとmarginalとした
of \textcolor{black}{2.0}\% (table~\ref{tb:table2}).  
%In the background, the mean polarization degree slightly increase and the position angle become closer to parallel to the Galactic plane than the bulge (table \ref{tb:table2}).
The differential polarization degree between the bulge and the background (table \ref{tb:table3}) is less than \textcolor{black}{0.6}\% in the direction nearly parallel to the Galactic plane.
%The differential polarization between the bulge and the background (table \ref{tb:table3}) show polarization increase of less than 0.5 \% in the position angle $PA_{\mathrm{GP}}$ nearly parallel to the Galactic plane.
The polarization increase per color change $\Delta P$$/$$\Delta (H-K_{\mathrm{S}})$ is small (\textcolor{black}{1.4$\% / \mathrm{mag}$}), and the background stars %◇%(善光_191127) %長田1223
and MC15 itself
(the red error bars %◇
and the black filled circle, respectively,  
in the bottom left panel of figure~\ref{fig:Fig11}) overlap the bulge stars. 
Therefore, random magnetic components seem to dominate more here than %●抜けてました
%>>(善光) 簡潔に書くことを主眼に●だけど、「どこで」は書かないと仕方がないでしょ。
between the foreground and the bulge. %The polarization increase per color change $\Delta P$ $/$ $\Delta (H-K_{\mathrm{S}})$ is very small 0.57 $\% / mag$, indicating that a random magnetic opponent dominates between the foreground and the bulge.%MC15用に計算し直した(10/23)
%In the bottom left panel of figure\ref{fig:Fig11}, the background stars overlap in the distribution of the bulge stars, and this overlap distribution might indicate the presence of violently disturbed magnetic field.
%In the right panel of figure\ref{fig:Fig9}, the background field shows some detectable polarizations and several orientation angle coherences. 
%Most of vectors look to be aligned with the Galactic plane direction, and others are oblique to the Galactic plane.
%The mean position angle $PA_{\mathrm{GP}}$ is $\sim$ $-10^{\circ}$, and this $PA_{\mathrm{GP}}$ value indicates the magnetic fields along the line of sight are parallel to the Galactic plane. 
%In background polarization map, there are also no polarized stars around the galactic coordinate (l, b) = ($1.9^{\circ}$, $-0.05^{\circ}$), and the background stars also are not suffered from the large dust extinction region. 
%Comparing the bulge polarization map and the background polarization map, we cannot discern a change in the polarization degrees P, and the polarization vectors of the background stars more parallel than that of the bulge stars.
%The differential position angle $PA_{\mathrm{GP}}$ is almost parallel to the Galactic plane, and this indicates that the magnetic field orientation is parallel to the Galactic plane. 
%The polarization efficiency = 0.29 is very low, and this value indicates that random components has a higher strength than the uniform component of the magnetic field.

%\subsection{Discussion}◇いや、ここは4章(differential解析が良いかどうか)の、というよりも、全体でのdiscussionにした方が良いと思います。

\section{Discussion}
%
%◇この2つを問題にするという善光君の意識は賛成です。
%Three Cepheids are located in front and back Far-3kpc arm.
%◇まず、Cepheidsとして3つを問題にするんだからもちろんTheでしょ。それからin front なんかbackなんかハッキリせえよ、論文なら。で、論文ならどれがfrontかbackかをちゃんと書くなり、そんなんどうでも良いなら（実際これまで3kpc armなんてあとにも先にも出てきてないからどうでも良いと私は思うけど）そういう書き方をしない。あるいはちょっとだけ「the Far-3kpc arm（armなんだからtheがついて）」に関して述べる。ただ、Far-3kpc armなんて、誰が見たん？　証拠proofなんて無い気がする、まあええとこevidenceの断片がいくつか。
The three Cepheids DC35, DC5, and MC15 are at 12.0, 12.6, and 10.9~kpc from us, respectively, 
and located in the far side of the Galactic center, possibly in a spiral arm (e.g., the Far-3kpc arm; \cite{Dame}; \cite{dHan}). %(善光191105)括弧の中に括弧が入っていたので修正
Many of the field stars observed in the near-infrared are in the bulge, which is around 8~kpc from us; 
the polarization changes from the bulge stars to the Cepheids 
provide us with the information of the magnetic field in the distance range of 8$-$12~kpc.   
%◇The polarization of background (table~\ref{tb:table2}) is similar to the polarization of the Cepheid in three fields.
%◇Therefore, these differential polarization from the bulge to background clearly represent the magnetic field orientation between the bulge and Far-3kpc arm.
%◇実は私は、これは論理的には今ひとつと思っているので、下のような書き方をしました。
The polarization of background stars (table~\ref{tb:table2}) is generally similar to the polarization of the Cepheids, 
so the classification of the field stars into the background 
on the basis of their $H - K_\mathrm{S}$ color, and therefore 
the differential polarization in table~\ref{tb:table3} seems reasonable.  
%
%MC15 and DC35 fields are consistent with magnetic field orientation predicted from the model, but DC5 field is inconsistent with this.
%◇これはmodelが何かを言っていないので、ダメな英文。それに、モデルに合ってますという言い方はヘンですよねえ
The magnetic field in the DC35 and MC15 sightlines runs parallel to the Galactic plane 
in the far side of the Galactic center, but 
in the DC5 sightline, the magnetic field seems to be oblique to the Galactic plane.  
%
%(新しいこと:セファイドの距離がわかるからこそ、どこからどこまでという指定ができる。単にH-Ksでは銀河系中心よりも遠方としかわからない)
%今回選んだ3つのセファイドの距離は10kpc以上でFar-3kpc armの前後に位置している
%それゆえバルジー後景間の差分の偏光はFar-3kpc armあたりの磁場の向きを示している
%(あと何か言いたいが、うまく言えない)
%(これを書くかは微妙なところ)DC5はバルジー後景間で磁場の向きがはっきりと変化しているが、Dflag=2に分類された視野がDC5を含めて3つしかないことから、銀河系中心の奥の磁場構造も大局的に平行であることを示唆している。

%We classified the Cepheid fields as Dflag = 1 or not, by using the slope of polarization per color.
%◇Discussionでは少し引いたところから議論するのが通例でもある。
Dflag of 1 is assigned to the lines of sight 
where the slope of polarization increase per the $H - K_\mathrm{S}$ change 
is large all the way to the $H - K_\mathrm{S}$ color of 3.0~mag.  
%◇うーん、以下ではJonesのAlfven waveの方のモデルを言ってる？　だけど、彼らはどっちが良いとは何も決めてないよねえ、それをこう言うのは良くないのでは？
%◇This slope represents relation between extinction and polarization degree $P$.
%For the relation, \cite{Jones} constructed the wave model depending on the geometry of magnetic fields along the line of sight.
%\cite{Jones} fitted the model to the data, and concluded that the relation between the polarization degree $P$ and the extinction represents the ratio of the energy density of the magnetic field to the kinematic energy density of moving clouds.
%If the kinematic energy density of moving clouds is higher than the energy density of the magnetic field, the slope of polarization per color is small.
According to the model by \citet{Jones}, this slope corresponds to 
the ratio of constant to random magnetic components. 
The magnetic field in the line of sight to DC35 (Dflag = 1) seems to have greater constant components than MC15 (Dflag = 3), both in the far side and in our neighborhood ($D \, \sim$ 1~kpc).  
More lines of sight have been assigned to Dflag = 3, which means that 
random magnetic field components are often dominant in some regions 
in the range of $\sim$ 12~kpc 
to the far side of the Galactic center.  

\section{SUMMARY}
We have measured near-infrared polarization of 52 Cepheid fields in the Galactic plane, 
toward the Galactic center. 
The magnetic field orientation between the Sun and the far side of the Galactic center seems to be close to parallel to the Galactic plane in most cases.
We classify %▲the 52 
48 Cepheid fields into three types on the basis of the polarization characteristics. %偏光の特徴から 
We have chosen a field of view from each of the three types.
%We chose a field of view from each of the three types, and discussed the change of magnetic field structure in detail. 
%>>(善光_191026)DC35 field shows the magnetic field orientation is parallel to the Galactic plane between the observer and the background.
%>>(善光_191026)DC5 field indicates the magnetic field orientation changes more than $45^{\circ}$ between the bulge and background.
%>>(善光_191026)MC15 fields show the random magnetic field component is dominant.
%各視野の結果を端的に●うん、これを復活させれば良いと思いますよ
The DC35 field shows the magnetic field nearly parallel to the Galactic plane, 
well aligned all the way of $\sim$ 12~kpc from the Sun to the Cepheid position 
in the other side of the Galactic center, %◎
with a very small change in the position angle we can detect at the distance of 1.1~kpc. %(善光191105)ここは年周視差0.9 masの方がいいような気がする。すぐさま変換できなかった(いや、書いてあったのでこれは無視)
However, sightlines which show such well aligned magnetic fields in the Galactic plane is rather small in number.   
%>>(善光_191026)DC5 field indicates the magnetic field orientation changes more than $45^{\circ}$ between the bulge and background.
%>>(善光_191026)MC15 fields show the random magnetic field component is dominant.
%
%
%The field of views which have the magnetic field along the line of sight aligns well in the same direction are in the minority, and 32 fields do not show any significant pattern.
The MC15 field, along with %▲theつけましょう
the other \textcolor{black}{36} Cepheid fields, indicates that random magnetic field components are %dominant.    ◆◎◎ここは30になるのかな%(善光191105)
significant.  
The DC5 field and %▲
the other field indicate that the magnetic field orientation changes more than $45^{\circ}$ in the line of sight. 
%The DC5 field and other few fields indicate that the magnetic field orientation changes more than $45^{\circ}$ in the line of sight. 
%In three fields, the Galactic magnetic field orientation between the observer and the far side of Galactic center is almost parallel to the Galactic plane in most cases.
The polarization increase per color change $\Delta P$$/$$\Delta (H-K_{\mathrm{S}})$ 
varies from region to region, 
reflecting the change in the ratio of the magnetic field strength and the turbulence strength.  
%さらに何か？>>(善光_191027)追加 将来的な話
%In the future application, $J$ and $H$ band starlight polarimetry will allow the foreground Galactic magnetic field to be unambiguously.Also, new classical Cepheids and Type $I\hspace{-.1em}I$ Cepheids catalog (\cite{cDekany}) allow us to analyze magnetic field structure widely and finely
%The polarization increase per color change $\Delta P$ $/$ $\Delta (H-K_{\mathrm{S}})$ among the fields have shown that the ratio of the magnetic field and turbulence varies from region to region in three fields.%>>(善光_191026)下の文章はdiscussionを省く関係上削除。その代わりに上の3つの文を入れる。また、各視野結果を端的に書くよりも全体的なことを言うことにする
%The differential position angles $PA_{\mathrm{GP}}$ of 48 fields usually are $\sim$ $0^{\circ}$, and the galactic magnetic field orientation is parallel to the Galactic plane.
%位置角~0度の視野が多い 大局的な磁場の向きは銀河面に対して平行 
%However, the most of the 48 fields show the polarization vectors like DC15, indicating that the turbulence energy dominates the magnetic field energy.
%乱流が磁場よりも強い

\bigskip
\begin{ack}
%The author would like to thank Tetsuya Nagata, Mikio Kurita, Kino Maseru, Shogo Nishiyama, and Noriyuki Matsunaga for useful discussions. 
%The author thanks Yasushi Nakajima for our analysis program comments.
We thank the staff of the South Africa Astronomical Observatory (SAAO) for %there ◎
their support during our observations. 
The IRSF/SIRIUS project was initiated and supported by Nagoya University and the National Astronomical Observatory of Japan in collaboration with the SAAO.
%◎科研費?
This work is supported by JSPS KAKENHI grants 18H03720, 18H05441, and 19H00695.
%宇宙ユニット?
TZ thanks Kyoto University Unit of Synergetic Studies for Space for the support for the provision of 
overseas observation trip expenses.   
This publication makes use of data from the Two Micron All Sky Survey, a joint project of the University of Massachusetts, the Infrared Processing and Analysis Center, NASA, and NSF.
%Gaia
This publication makes use of data from the $Gaia$ processed by the $Gaia$ Data Process-
ing and Analysis Consortium. %>>(善光_191107)最後まで書ききった
\end{ack}

\begin{table*}[h]
\tbl{The mean polarization of foreground, bulge, and background stars in each field of view.}{%
%\tbl{The mean polarization}{%●もう少しだけ詳しく。そしてIDに
  \begin{tabular}{lrrrrrr}
  \hline
ID &  $ \overline{P_{\mathrm{fore}}} $ & $ \overline{PA_{\mathrm{(GP,fore)}}} $ & $ \overline{P_{\mathrm{bulge}}} $  & $ \overline{PA_{\mathrm{( GP,bulge)}}} $ & $ \overline{P_{\mathrm{back}}} $ & $ \overline{PA_{\mathrm{( GP,back)}}} $  \\
  \hline
 &  \%  & deg  &  \% & deg  &  \% & deg \\
  \hline
DC35 & 1.99 $\pm$ 0.61 & 11.84 $\pm$ 9.12 & 3.72 $\pm$ 0.70 & \textcolor{black}{5.42 $\pm$ 5.05} & \textcolor{black}{5.12 $\pm$ 1.67} & \textcolor{black}{0.45 $\pm$ 5.92}\\
DC5 & \textcolor{black}{1.12} $\pm$ 0.49 &  \textcolor{black}{45.58 $\pm$ 12.74} & \textcolor{black}{2.78 $\pm$ 1.29} & \textcolor{black}{13.64 $\pm$ 10.67} & 2.79 $\pm$ 1.33 & 22.72 $\pm$ 13.75\\
MC15 & 0.87 $\pm$ \textcolor{black}{0.46} & \textcolor{black}{-30.97 $\pm$ 13.71} & \textcolor{black}{1.61 $\pm$ 0.91} & \textcolor{black}{-15.31 $\pm$ 15.99} & \textcolor{black}{2.06 $\pm$ 1.31} & \textcolor{black}{-9.39 $\pm$ 13.49}\\
\hline
  \end{tabular}}
  \label{tb:table2}
  \begin{tabnote}
  \footnotemark[$*$] These $P$ values have not been corrected for positive statistical bias.\\
  \end{tabnote}
\end{table*}
%(善光_191127)

\begin{table*}[h]
\tbl{Differential polarization. \\  
From observer (0) to foreground (1), \\
from foreground (1) to bulge (2), and \\
from bulge (2) to background (3) in each field of view.}{%
%\tbl{Differential polarization of three field of views.}{%●で、やっぱり縦と横を逆にするべきだと思います！●そしてさらに少し変えてみました◎◎◎qとuは要らんのでは？と思い直し、消してみました。いかが？
  \begin{tabular}{llrrrr}
  \hline 
ID & Diff  & unit & 0 $\rightarrow$ 1 & 1 $\rightarrow$ 2 & 2 $\rightarrow$ 3\\
\hline
DC35 & $ \Delta P$ & \% & 1.99 & \textcolor{black}{1.86}  & \textcolor{black}{1.58} \\
         & $ \Delta PA_{\mathrm{GP}} $  & deg & 11.84 & \textcolor{black}{-1.50} & \textcolor{black}{-11.55} \\
%         & $ \Delta q_{\mathrm{GP}} $  & \% & 1.81 & 1.83  & 1.22 \\
%         & $ \Delta u_{\mathrm{GP}} $ & \% & 0.80 & -0.10 & -0.55\\
  \hline 
DC5   & $ \Delta P$  & \% & \textcolor{black}{1.12}  &  \textcolor{black}{2.50}  & \textcolor{black}{0.88} \\
         & $ \Delta PA_{\mathrm{GP}} $ & deg & \textcolor{black}{45.58} & \textcolor{black}{1.84} & \textcolor{black}{63.13} \\
%         & $ \Delta q_{\mathrm{GP}} $ & \% &-0.07  &2.46  &-0.43  \\
%         &$ \Delta u_{\mathrm{GP}} $ & \% &1.16 & -0.01 & 0.82 \\
  \hline 
MC15 & $ \Delta P$ & \% &\textcolor{black}{0.87}  &\textcolor{black}{0.98}  &\textcolor{black}{0.59}\\
         & $ \Delta PA_{\mathrm{GP}} $ & deg &\textcolor{black}{-30.97} &\textcolor{black}{-1.47} & \textcolor{black}{7.75}\\
%         & $ \Delta q_{\mathrm{GP}} $ & \% &0.42  &0.99  &0.40\\
%         & $ \Delta u_{\mathrm{GP}} $ & \% &-0.77&-0.06 & 0.12\\
\hline
  \end{tabular}}
  \label{tb:table3}
    \begin{tabnote}
  %\footnotemark[$*$] Interval numbers refer to the region, zero being the observer, one being the foreground, two being the bulge and three being the background.\\●逆にするのも含め、このfootnoteは要らなくなるかも。ところで、DC5のforegroundの1.16（この表）と1.17（前の表）はどっちが正しい？
  \end{tabnote}
\end{table*}
%(善光_191127)

\vspace{-0.2cm}

\newpage

\end{document}